\documentclass[twoside,a4paper,fleqn,titlepage,preprintnumbers,nofootinbib,%
               prd,12pt,showpacs,showkeys]{revtex4-1}
\usepackage{url}
\usepackage{subfig}
\usepackage{graphicx}
\begin{document}
%
%
\title{The Golden Channel at a Neutrino Factory revisited: improved sensitivities from a Magnetised Iron Neutrino Detector}
%
\author{R.~Bayes}
\author{A.~Laing}
\author{F.J.P.~Soler}
\affiliation{School of Physics \& Astronomy, University of Glasgow,
  Glasgow, UK.}
\author{A. Cervera Villanueva}
\author{J.J.~G\'omez Cadenas}
\author{P.~Hern\'andez}
\author{J.~Mart\'in-Albo}
\affiliation{IFIC, CSIC \& Universidad de Valencia, Valencia, Spain.}
\author{J.~Burguet-Castell}
\affiliation{Universitat de les Illes Balears, Spain.}
\begin{abstract}
This paper describes the performance and sensitivity to neutrino mixing
parameters of a Magnetised
Iron Neutrino Detector (MIND) at a Neutrino Factory with a neutrino
beam created from the decay of 10~GeV muons. Specifically, it is
concerned with the ability of such a detector to detect muons of the
opposite sign to those stored (wrong-sign muons) while suppressing
contamination of the signal from the interactions of other neutrino
species in the beam. 
A new more realistic simulation and analysis, which
improves the efficiency of this detector at low energies, has been
developed using the GENIE neutrino event generator and the GEANT4
simulation toolkit. Low energy neutrino events down to 1 GeV were selected, while reducing
backgrounds to the $10^{-4}$ level. Signal efficiency plateaus of 
$\sim$60\% for $\nu_\mu$ and $\sim$70\% for $\overline{\nu}_\mu$ events were achieved starting at $\sim$5~GeV. Contamination from the $\nu_\mu\rightarrow \nu_\tau$
oscillation channel was studied for the first time and was found to be at the level between 1\% and 4\%. 
Full response matrices are supplied for all the signal and background channels from 1~GeV to 10~GeV.
The sensitivity of an experiment involving a  MIND detector of 100~ktonnes at
2000 km from 
the Neutrino Factory is calculated for the case of $\sin^2 2\theta_{13}\sim 10^{-1}$.  For this value
of $\theta_{13}$, the accuracy in the measurement of the CP violating phase is estimated to be
$\Delta \delta_{CP}\sim 3^\circ - 5^\circ$, depending on the value of  $\delta_{CP}$, the
CP coverage at 5$\sigma$ is 85\% and the mass hierarchy would be determined with better 
than 5$\sigma$ level for all values of $\delta_{CP}$.
\end{abstract}%
\maketitle

\section{Introduction}
\label{Intro}
The Neutrino Factory, a new type of accelerator facility in which a
neutrino beam is created from the decay of muons in flight in a
storage ring, is perhaps the most promising facility design to
resolve the problem of CP violation in the neutrino sector. The
physics potential of this facility was first described by Geer
\cite{Geer:1997iz}. The expected absolute flux and spectrum of
neutrinos from such a facility can be calculated with smaller
systematic errors than those associated with the beams of alternate
facilities due to the ability to measure the muon beam flux and the
highly accurate measurement of muon decay kinematics \cite{TWIST:2011aa}. 
Since, in principle, both $\mu^+$ and $\mu^-$ can be
created with the same systematic uncertainties on the flux, any
oscillation channel can be studied with both neutrinos and
antineutrinos, improving sensitivity to CP
violation. Table~\ref{tab:oscFluxes} shows the oscillation channels
that will contribute to the flux at any far site due to the decay of
$\mu^+$.
\begin{table}[!ht]
  \caption{\emph{Oscillation channels contributing to flux from the decay of $\mu^+$.}}
  \label{tab:oscFluxes}
  \begin{center}
    \begin{tabular}{cc|cc}
      \multicolumn{2}{c}{$\nu_e$ origin} & \multicolumn{2}{c}{$\overline{\nu}_\mu$ origin}\\
      \hline
        & $\nu_e \rightarrow \nu_e \mbox{ (}\nu_e\mbox{ disappearance channel)}$ & & $\overline{\nu}_\mu \rightarrow \overline{\nu}_\mu \mbox{ (}\overline{\nu}_\mu\mbox{ disappearance channel)}$\\
       & $\nu_e$ $\rightarrow \nu_\mu \mbox{ (Golden channel)}$ & & $\overline{\nu}_\mu$ $\rightarrow \overline{\nu}_\tau \mbox{ (Dominant oscillation)}$\\
        & $\nu_e \rightarrow \nu_\tau \mbox{ (Silver channel)}$ & & $\overline{\nu}_\mu \rightarrow \overline{\nu}_e \mbox{ (Platinum channel)} $
    \end{tabular}
  \end{center}
\end{table}

The sub-dominant $\nu_e \rightarrow \nu_\mu$
oscillation~\cite{DeRujula:1998hd} was identified as the
most promising channel to explore CP violation at a Neutrino
Factory. The charged current interactions of the ``Golden Channel''
$\nu_\mu$ produce muons of the opposite charge to those stored in the
storage ring (wrong-sign muons) and these can be detected with a large
magnetised iron detector~\cite{Cervera:2000kp}. The original analyses
were carried out assuming a Neutrino Factory storing 50~GeV muons and,
as such, were optimised for high energy using a detector with 4~cm
thick iron plates and 1~cm scintillator planes. However, subsequent
phenomenological studies carried out as part of the International
Scoping Study (ISS) for future neutrino facilities \cite{Bandyopadhyay:2007kx,Apollonio:2009} favoured a stored muon
energy of 25~GeV and showed the importance of neutrinos with energies
below 5~GeV.  The Magnetised Iron Neutrino Detector (MIND) is a large
scale iron and scintillator sampling calorimeter, similar to
MINOS~\cite{Michael:2008bc}, which was re-optimized from the original
studies motivated by these
findings~\cite{Cervera:2000vy,CerveraVillanueva:2008zz,Abe:2007bi}. The
performance obtained indicated that the combination of two Magnetised
Iron Neutrino Detectors at 4000~km and 7500~km would give optimum
sensitivity to the mixing parameters \cite{Agarwalla:2010hk}.

The studies of MIND mentioned above evaluated the performance of the
detector using deep inelastic scattering events only, with a
simplified simulation, reconstruction and kinematic analysis. 
The performance needed to be evaluated and improved using a full
simulation and analysis of all physical processes. As part of the
International Design Study for a Neutrino
Factory~\cite{IDS-NF,EUROnu}, a software framework to perform these
studies has been developed. Pattern recognition and analysis
algorithms were developed and first applied to data generated using
the same simulation as was used in the ISS studies. The development of
the algorithm and the results of its application were described
in~\cite{Cervera:2008nf} and~\cite{Cervera:2010rz}, where it was shown
that under these conditions the efficiency and background could be
maintained at a similar level to that achieved in the ISS
studies. This paper introduces the full spectrum of possible neutrino
interactions generated using
the neutrino event generator GENIE~\cite{Andreopoulos:2009rq} and a comparison with another event generator, NUANCE~\cite{Casper:2002sd}. These interactions were tracked
through a new GEANT4
simulation~\cite{Agostinelli:2002hh,Apostolakis:2007zz} with full
hadron shower development and a new detector digitisation not present
in previous studies. The events were then subject to the pattern
recognition algorithm presented in~\cite{Cervera:2010rz}, reoptimised
for the new simulation. Finally a likelihood based analysis was used
to further suppress backgrounds. A preliminary version of this
analysis using the NUANCE package has been published in the Interim
Design Report of the IDS-NF \cite{NF:2011aa}. This paper includes
the full GENIE simulation, a comparison to NUANCE, an estimate of
systematic errors, and sensitivity calculations for $\theta_{13}$, the
neutrino mass hierarchy (sign of $\Delta m^2_{13}=m^2_1-m^2_3$) 
and the CP violating phase $\delta_{CP}$.

Recent results from the reactor experiments Daya Bay, RENO and Double Chooz
\cite{An:2012eh,Ahn:2012nd,Abe:2011fz}, as well as evidence from T2K \cite{Abe:2011sj} and MINOS \cite{Adamson:2011qu}, have demonstrated that the value
of $\theta_{13}$ is large (with a combined average of $\sin^22\theta_{13}= 0.097\pm 0.012$).
These results increase the likelihood of a discovery of 
CP violation and the determination of the mass hierarchy in neutrinos.
It was shown in the Interim Design Report of the IDS-NF \cite{NF:2011aa} that at a value
of $\sin^22\theta_{13}\sim 0.1$ the optimum Neutrino Factory configuration is achieved
with a muon energy of 10~GeV and with a far detector at a distance of 2000 km.
While the Neutrino Factory was designed to discover
CP violation for a large range of values of $\theta_{13}$ (down to values of $\sin^22\theta_{13}\sim 10^{-4}$), it will be shown in this paper that it also offers
the best chance to discover CP violation and the mass hierarchy at large values of 
$\theta_{13}$, regardless of whether $\Delta m^2_{13}$ is positive or negative
(inverted or normal mass hierarchy).

This paper is organised as follows. Section~\ref{sec:theo} introduces
the relevant backgrounds and contaminations of the golden signal and
describes the required suppressions. Section~\ref{sec:simulation}
describes the simulation tools and gives a description of MIND and the
assumptions still made for this study. The analysis is described in
Section~\ref{sec:analysis}, with a detailed demonstration of all
variables and functions used to identify signal from background. The
results from this analysis, including signal efficiencies, background
rejection capabilities and performance to the $\nu_\mu\rightarrow
\nu_\tau$ oscillation signal, are presented in
Section~\ref{sec:response}. A discussion of some of the systematic
errors of the analysis is described in
Section~\ref{sec:syst}. Finally, full sensitivity to $\theta_{13}$,
$\delta_{CP}$ and the neutrino mass hierarchy will be presented in
Section~\ref{sec:MINDsens}. Response (migration) matrices of this
detector system for all signal and background will be shown in the
Appendix.

\section{Sources of impurity in the golden sample\label{sec:theo}}
The primary sources of background to the wrong-sign muon search come
from the Charged Current (CC) and Neutral Current (NC) interactions of
the non-oscillating neutrinos present in the beam. Specifically, the
CC interactions of $\nu_\mu~(\overline{\nu}_\mu)$ being reconstructed
as $\overline{\nu}_\mu~(\nu_\mu)$, NC from all neutrino types in the
beam being reconsructed as $\overline{\nu}_\mu~(\nu_\mu)$ and the CC
interactions of $\overline{\nu}_e~(\nu_e)$ being reconstructed as
$\overline{\nu}_\mu~(\nu_\mu)$. Since these interactions are in far
greater abundance than those of the signal channel and contain little
or no discernible information about the key parameters $\theta_{13}$
and $\delta_{CP}$, they must be suppressed sufficiently so that the
statistical error on the background is smaller than the expected
signal level. This corresponds to a suppression of at least 10$^{-3}$
for each channel in the signal region.

In addition to the golden channel appearance oscillation there are
three other appearance channels that will introduce neutrinos to the
flux incident on the far detectors (shown in table~\ref{tab:oscFluxes}). 
The dominant oscillation, $\nu_{\mu}(\bar{\nu}_{\mu})\to
\nu_{\tau}(\bar{\nu}_{\tau})$, which must be considered when fitting
for the $\nu_\mu~(\overline{\nu}_\mu)$ disappearance signal, should not pose
a problem for fitting the golden channel since, at large $\theta_{13}$, 
the dependence of this channel on this mixing
angle is very small and the interaction would have to be reconstructed
with the opposite charge to that of the true primary
lepton. See~\cite{Indumathi:2009hg} for a detailed discussion of tau
contamination in the disappearance channel. The platinum channel,
$\nu_{\mu}(\bar{\nu}_{\mu})\to\nu_{e}(\bar{\nu}_{e})$, 
should pose no problem since the number of interactions should be
similar to that produced by the golden channel and $\nu_e$
interactions produce a penetrating muon-like track in only a small
fraction of cases. However, the silver channel oscillation,
$\nu_{e}(\bar{\nu}_{e})\to\nu_{\tau}(\bar{\nu}_{\tau})$,
would be expected to contribute a similar amount of $\nu_\tau$ to the flux as
the golden channel does to $\nu_\mu$, and since the primary $\tau$ decays with a $\sim
17.65\%$ probablility via channels containing muons, a significant
proportion of these interactions would be expected to pass the
analysis cuts. As discussed in~\cite{Donini:2010xk}, fitting the
observed spectrum without accounting for the presence of these
$\nu_\tau$ interactions leads to significantly reduced accuracy in the
fits. However, since this oscillation contains complimentary
information about both $\theta_{13}$ and $\delta_{CP}$, handling this
`contamination' correctly has the potential to perhaps improve the fit
accuracy compared to using an analysis which attempts to remove it
from the sample.

\section{Simulation and reconstruction of MIND}
\label{sec:simulation}
In previous studies \cite{Cervera:2000kp,Abe:2007bi,Cervera:2010rz},
only deep inelastic scattering (DIS) events generated with LEPTO 6.1
\cite{Ingelman:1997Cp} were considered. However, at energies below
5~GeV there are large contributions from quasi-elastic (QE), single
pion production (1$\pi$) and other resonant production (RES)
events. QE and 1$\pi$ events are expected to exhibit lower
multiplicity in the detector output, which makes candidate muon
reconstruction simpler. This should improve reconstruction efficiency
in low energy CC interactions but also potentially increase low energy
backgrounds, particularly from NC 1$\pi$ interactions. Other nuclear
resonant events, producing two or three pions, as well as diffractive and
coherent production, have much smaller contributions. Moreover, the
presence of QE interactions allows for the calculation of neutrino
energy without hadron shower reconstruction, improving neutrino energy
resolution.

\subsection{Neutrino event generation and detector simulation}
\label{Sec:event_generation}
\begin{figure}
  \begin{center}$
  \begin{array}{cc}
    \subfloat[$\bar{\nu}_{\mu}$ CC]{
      \includegraphics[width=0.45\textwidth]{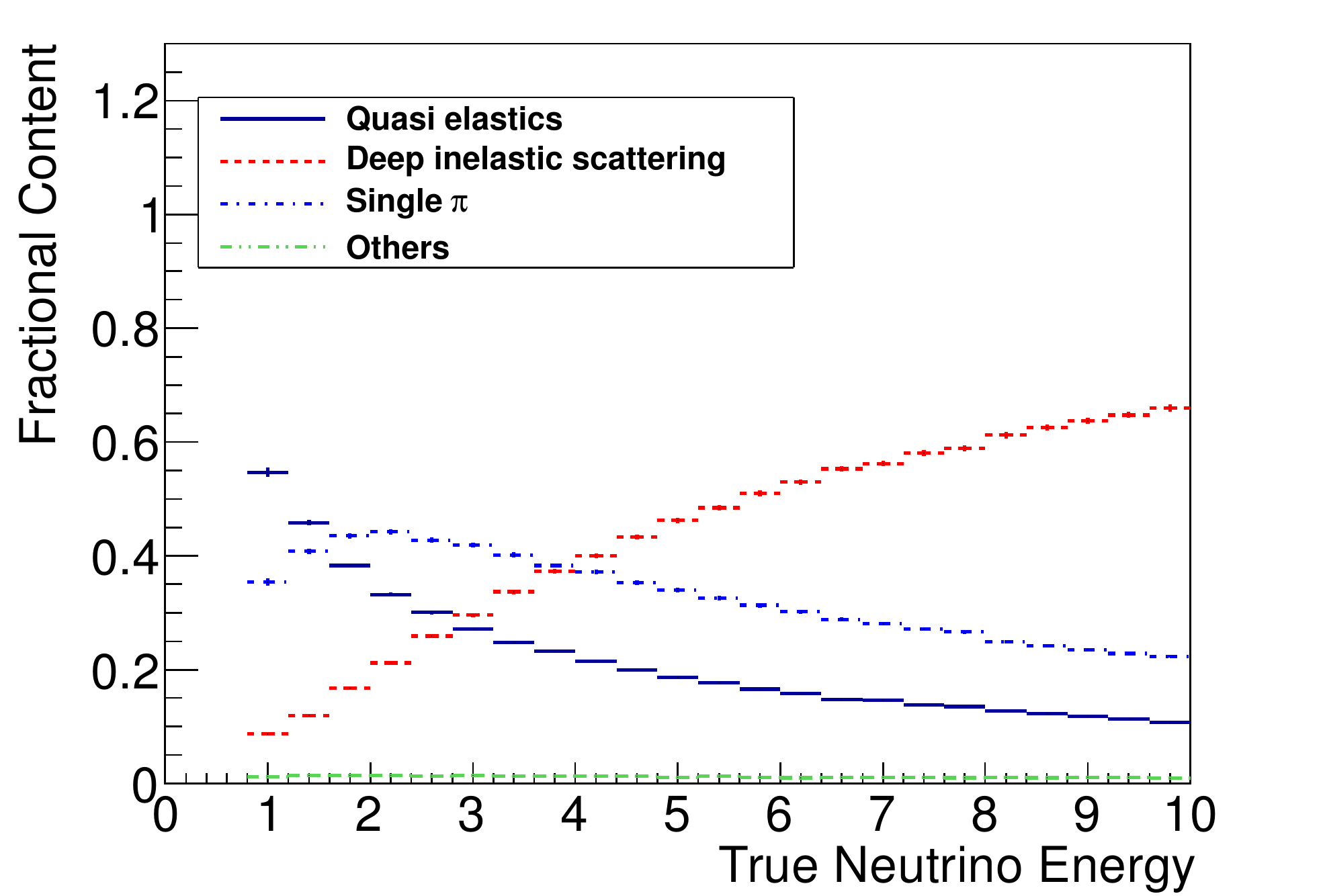}
    } & 
    \subfloat[$\nu_{e}$ CC]{
      \includegraphics[width=0.45\textwidth]{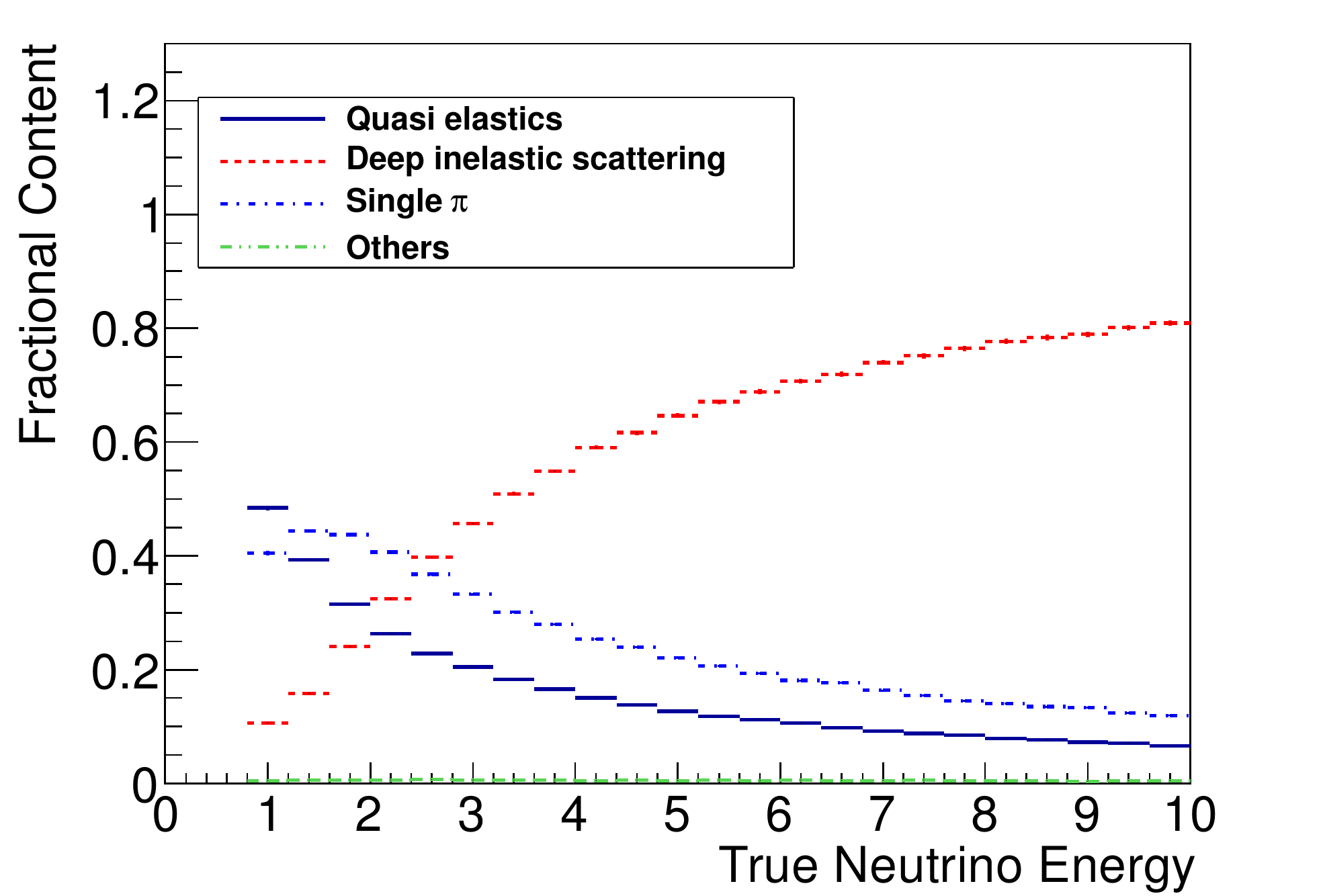}
    }\\%
    \subfloat[$\nu_{\mu}$ CC]{
      \includegraphics[width=0.45\textwidth]{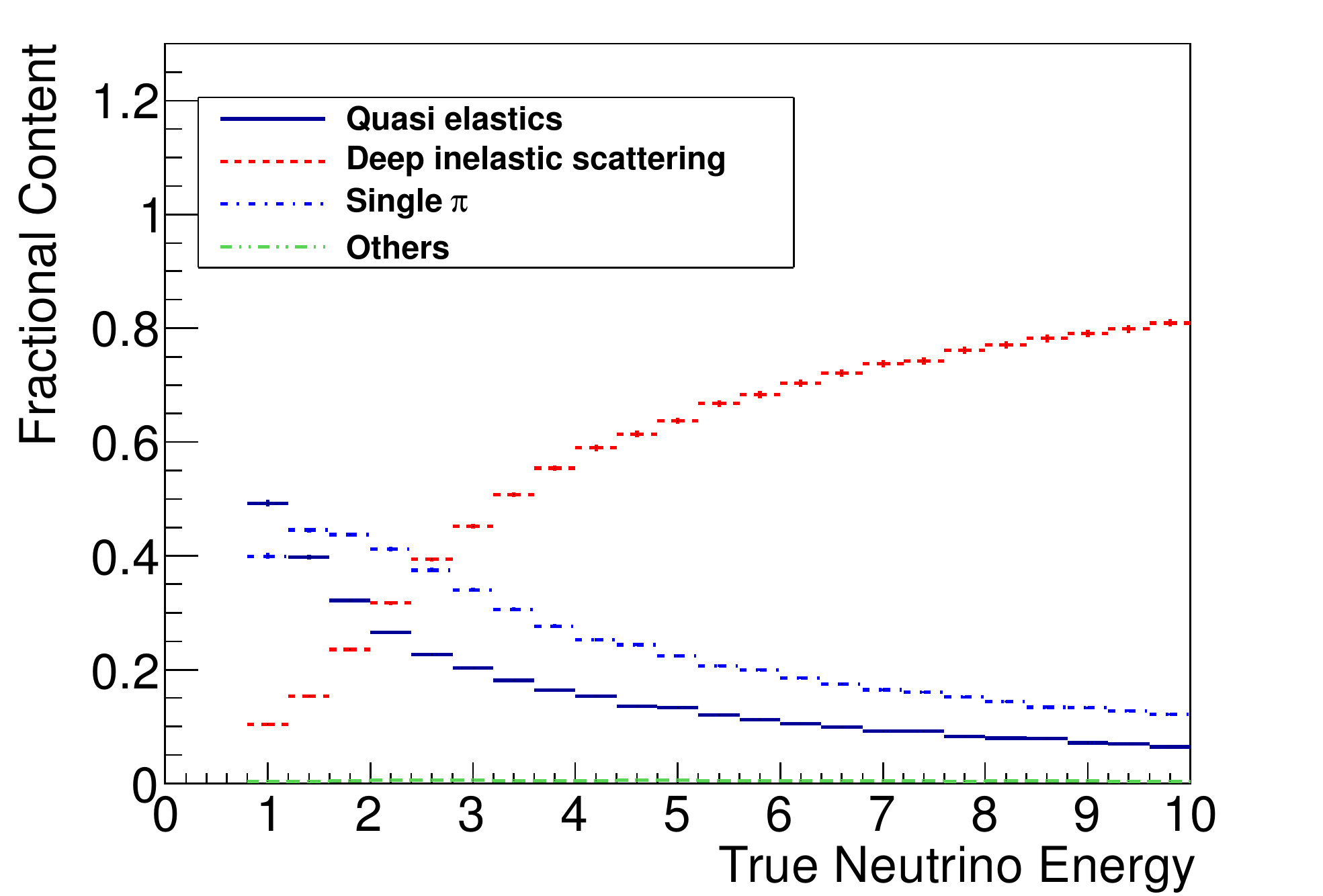}
    }& %
    \subfloat[$\bar{\nu}_{e}$ CC]{
      \includegraphics[width=0.45\textwidth]{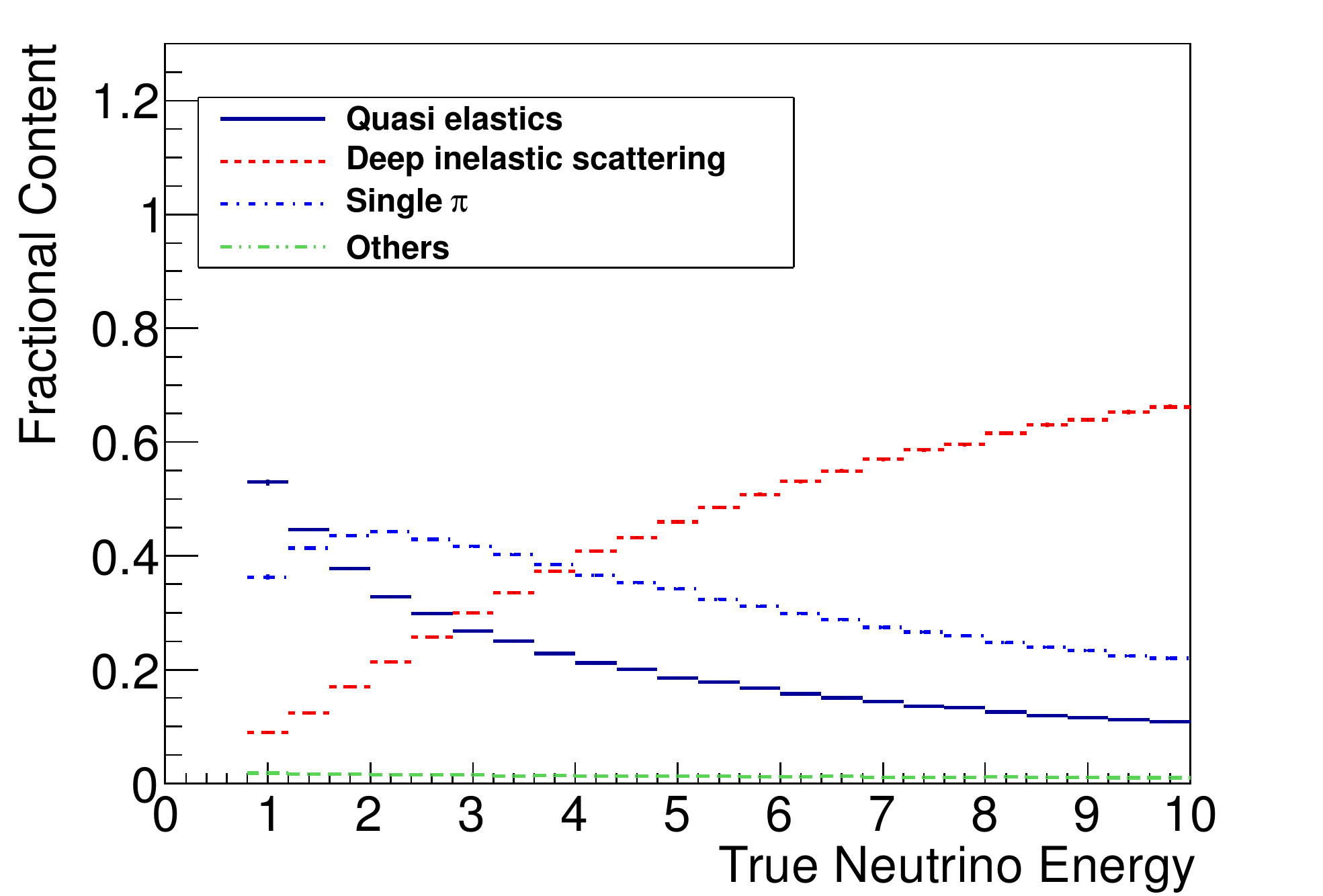}
    }\\%
    \subfloat[$\bar{\nu}_{\mu}$ NC]{
      \includegraphics[width=0.45\textwidth]{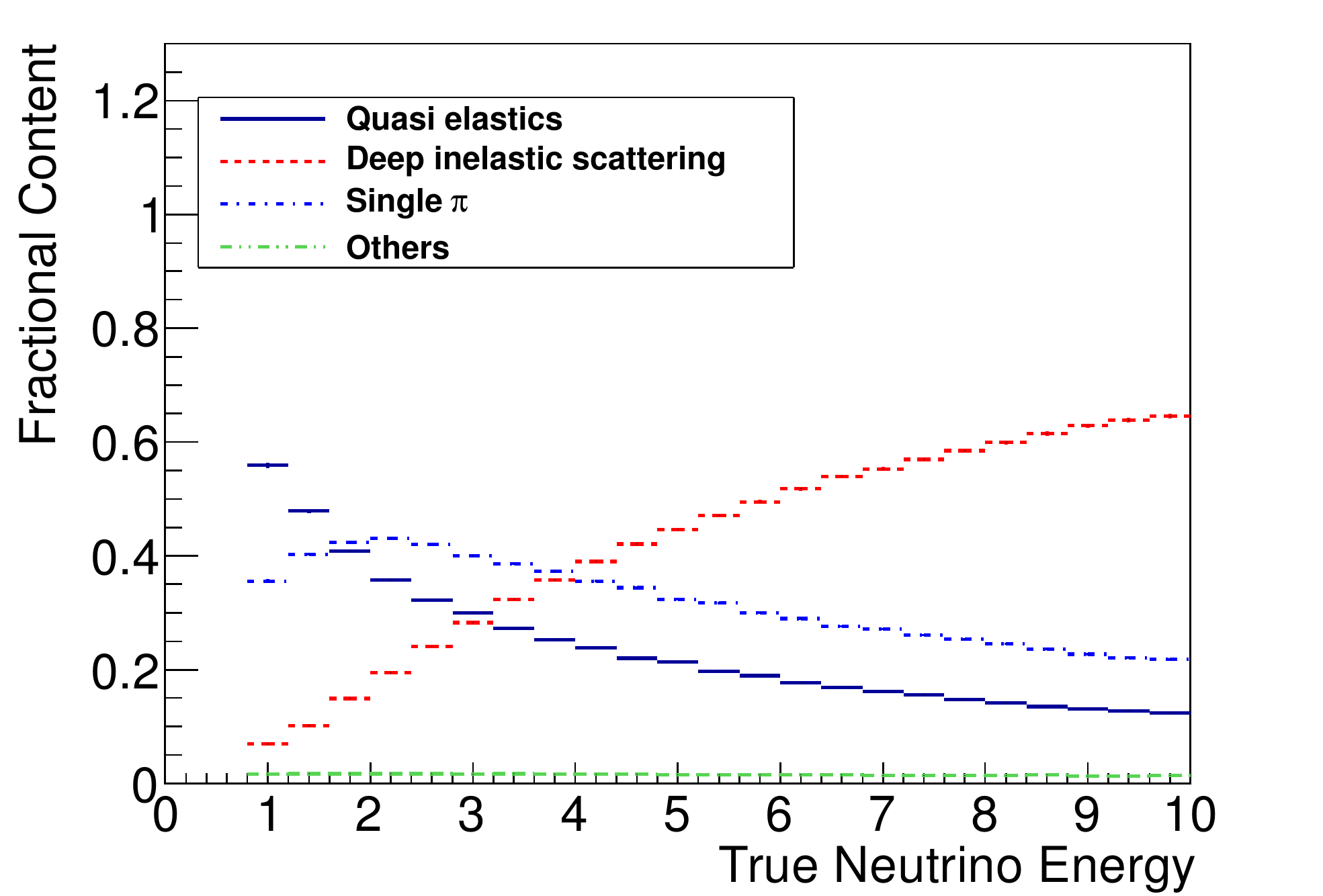}
    }& %
    \subfloat[$\nu_{\mu}$ NC]{
      \includegraphics[width=0.45\textwidth]{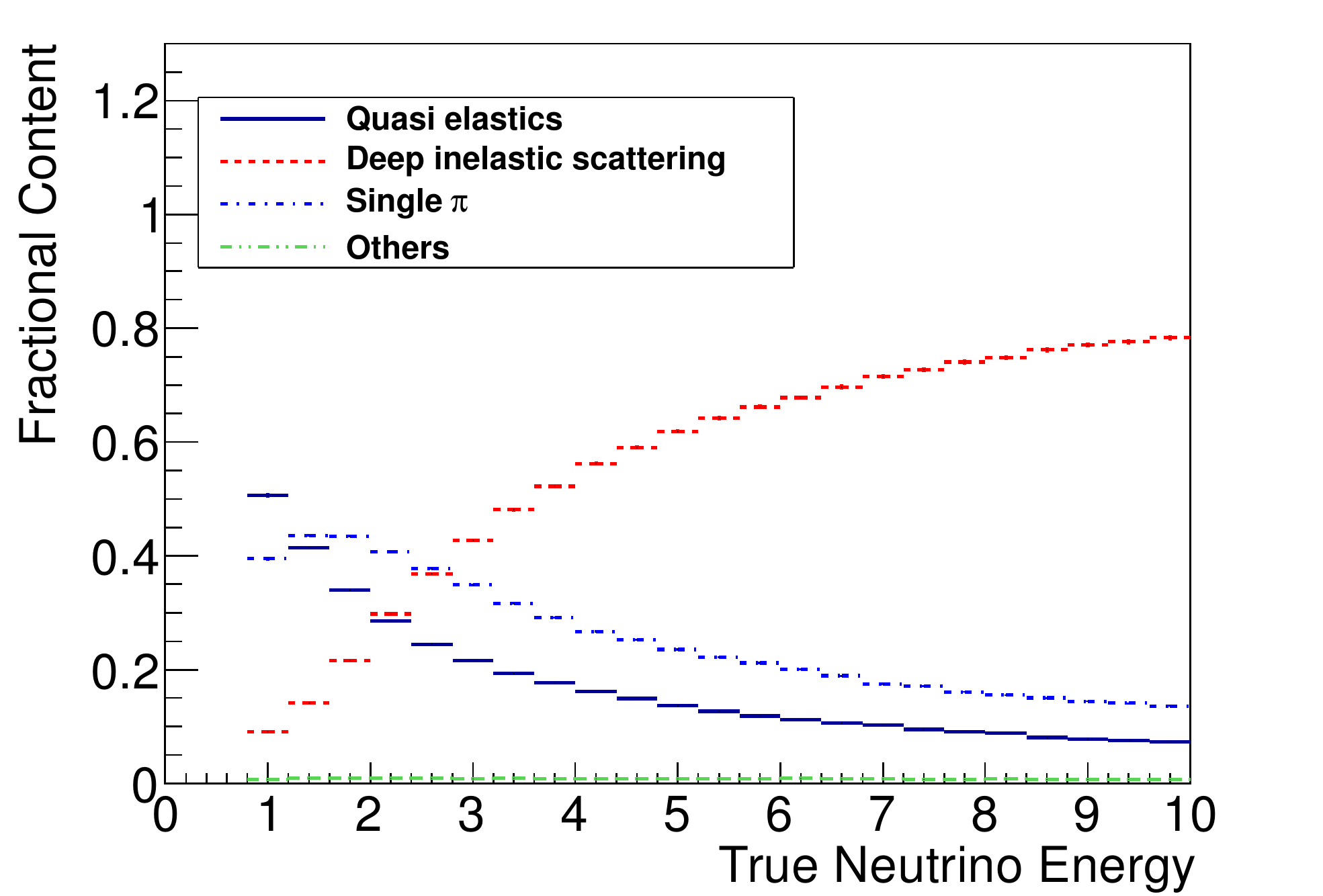}
    }\\%
  \end{array}
  $
  \end{center}
  \caption{Proportion of total number of interactions of different
    $\nu$ interaction processes for events generated using GENIE and
    passed to the G4MIND simulation.}
  \label{fig:intprops}
\end{figure}

Generation of all types of interaction was performed using the GENIE
framework~\cite{Casper:2002sd}. The exclusive event samples generated
by GENIE are shown in figure \ref{fig:intprops}, where `other'
interactions include the resonant, coherent and diffractive processes
other than single pion production. The relative rates below 1~GeV are
included for completeness, but will have negligible effect at a
Neutrino Factory. GENIE also includes a treatment to simulate the
effect of re-interaction within the participant nucleon, which is
particularly important for low energy interactions in high-$Z$ targets
such as iron.

A new simulation of MIND using the GEANT4
toolkit~\cite{Apostolakis:2007zz} (G4MIND) was developed to provide
flexibility to the definition of the geometry, to carry out full
hadron shower development and to perform a proper digitisation of the
events. This allows optimisation of all aspects of the detector, such
as the dimensions and spacing of all scintillator and iron pieces,
external dimensions of the detector and detector readout
considerations.

The detector transverse dimensions ($x$ and $y$ axes) and length in
the beam direction ($z$ axis), transverse to the detector face, are
controlled from a parameter file. A fiducial cross section of
14~m$\times$14~m, including 3~cm of iron for every 2~cm of polystyrene
extruded plastic scintillator (1~cm of scintillator per view), was
assumed. A constant magnetic field of 1~T is oriented in the positive
$y$ direction throughout the detector volume. Events generated for
iron and scintillator nuclei are selected according to their relative
weights in the detector and the resultant particles are tracked from a
vertex randomly positioned in three dimensions within a randomly
selected piece of the appropriate material. Physics processes are
modelled using the QGSP\_BERT physics lists provided by GEANT4
\cite{geant4phys}.

Secondary particles are required to travel at least 30~mm from their
production point or to cross a material boundary between the detector
sub-volumes to have their trajectory fully tracked. Generally,
particles are only tracked down to a kinetic energy of
100~MeV. However, gammas and muons are excluded from this cut. The
end-point of a muon track is important for muon pattern recognition.

\begin{figure}
  \begin{center}$
    \begin{array}{cc}
      \includegraphics[width=0.45\textwidth]{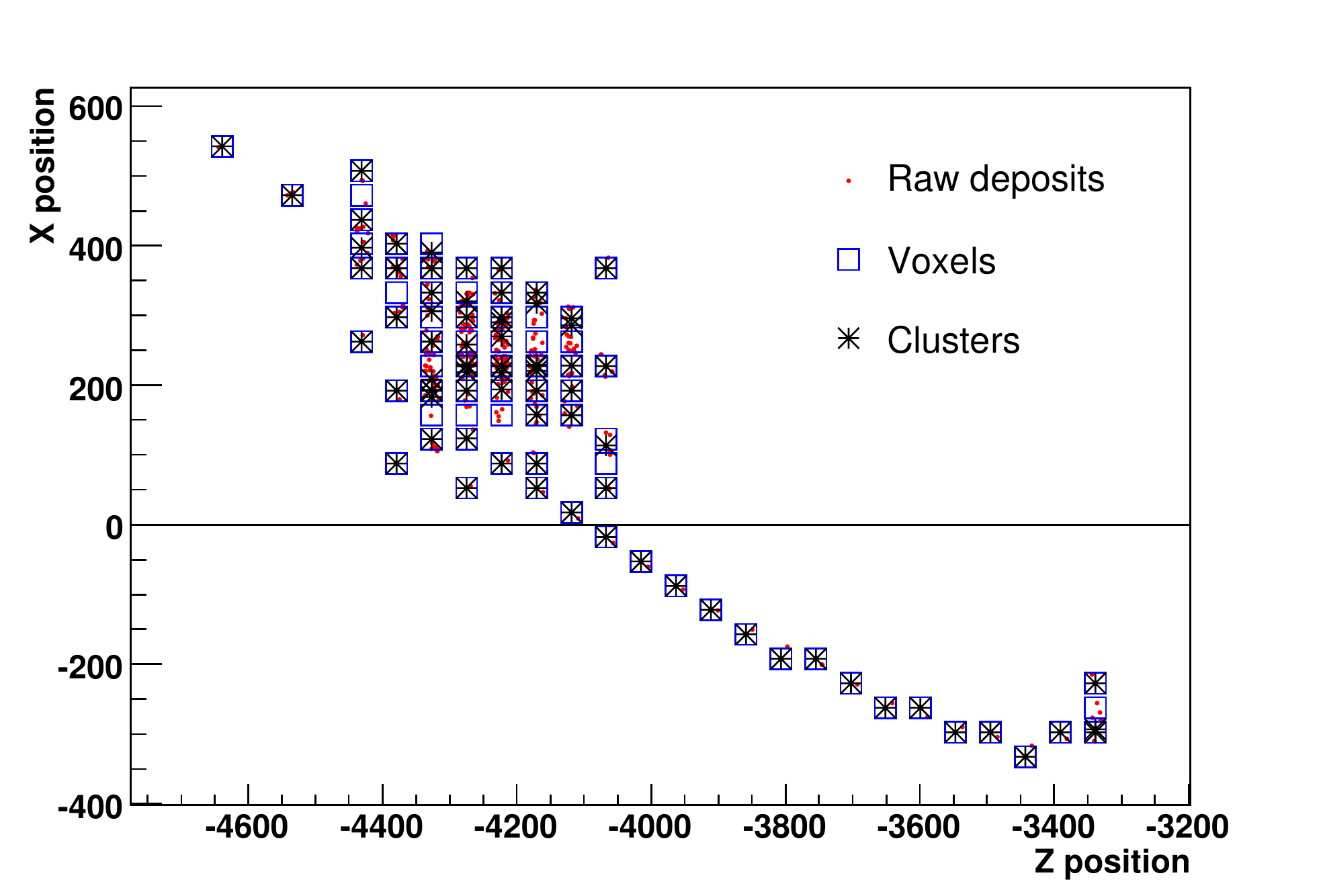} &
      \includegraphics[width=0.45\textwidth]{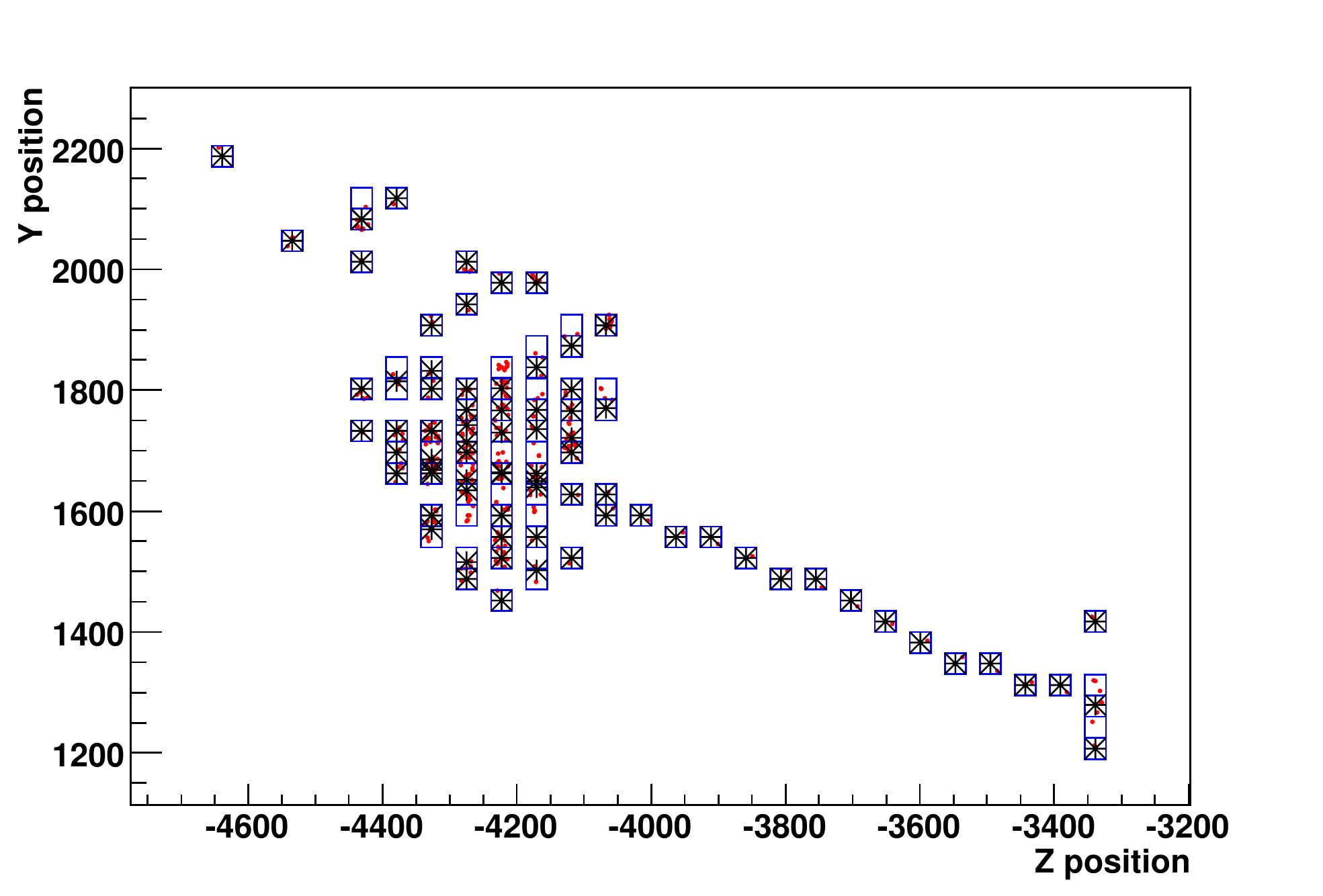} \\
    \end{array}$
    \includegraphics[width=0.45\textwidth]{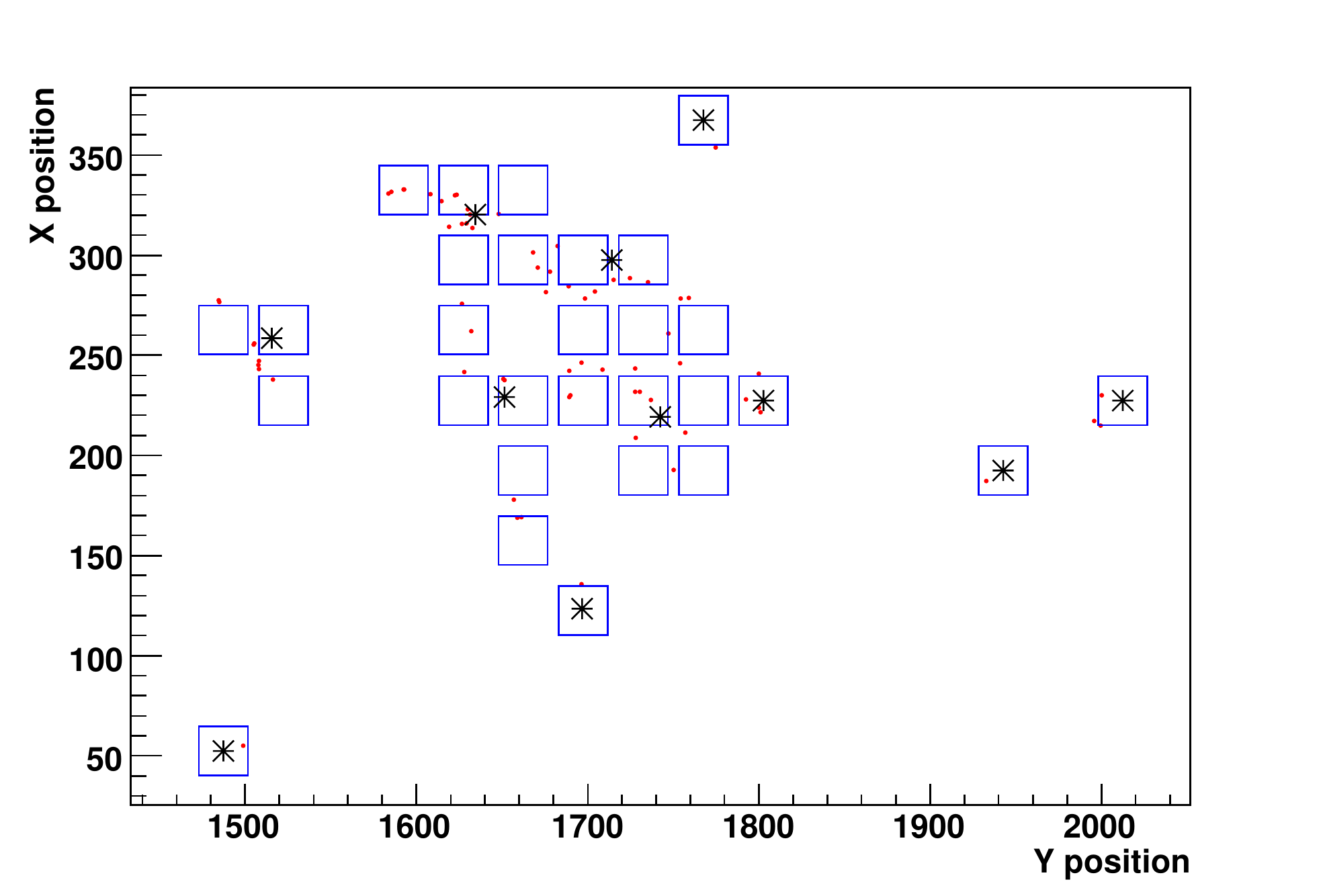}
  \end{center}
  \caption{The digitisation and voxel clustering of an example event:
    (top left) bending plane view, (top right) non bending plane,
    (bottom) an individual scintillator plane. The individual hits are
    small dots (in red), the blue squares are the voxels and the black
    asterisks represent the centroid positions of the clusters.}
  \label{fig:voxclust}
\end{figure}

A simplified digitisation model was considered for this
simulation. Two-dimensional boxes -- termed voxels -- represent
view-matched $x$ and $y$ readout positions. Any deposit which falls
within a voxel has its energy deposit added to the voxel total raw
energy deposit. The thickness of two centimetres of scintillator per
plane assumes 1~cm per view.  Voxels with edge lengths of 3.5~cm were
chosen to match the required point resolution of 1~cm
($3.5/\sqrt{12}$), assuming a uniform hit distribution along the width
of the scintillator bar. The response of the scintillator bars is
derived from the raw energy deposit in each voxel, read out using
wavelength shifting (WLS) fibres with attenuation length $\lambda =
5$~m, as reported by the MINERvA
collaboration~\cite{PlaDalmau:2005dp}. Assuming that approximately
half of the energy will come from each view, the deposit is halved and
the remaining energy at each edge in $x$ and $y$ is calculated. This
energy is then smeared according to a Gaussian with $\sigma/E = 6\%$
to represent the response of the electronics and then recombined into
$x$, $y$ and total = $x+y$ energy deposit per voxel. An output
wavelength of 525~nm, a photo-detector quantum efficiency of
$\sim$30\% and a threshold of 4.7 photo electrons (pe) per view (as in
MINOS~\cite{Michael:2008bc}) were assumed. Any voxel in which the two
views do not make this threshold is cut. If only one view is above
threshold, then only the view below the cut is excluded (see
section~\ref{Sec:RecG4}). The digitisation of an example event is
shown in figure \ref{fig:voxclust}.

\subsection{Event reconstruction}
\label{Sec:RecG4}
The reconstruction package was described in detail in
\cite{Cervera:2010rz}. We present here an update of the reconstruction
based on the MIND simulation generated using GENIE and GEANT4.

Many traversing particles, particularly hadrons, deposit energy in
more than one voxel. Forming clusters of adjacent voxels reduces event
complexity and can improve pattern recognition in the region of the
hadron shower. The clustering algorithm is invoked at the start of
each event. The voxels of every plane in which energy has been
deposited are considered in sequence. Where an active voxel is in
contact with no other active voxel, this voxel becomes a cluster. If
there are adjacent voxels, the voxel with the largest total deposit
(at scintillator edge) is sought and all active voxels in the
surrounding 3$\times$3 area are considered part of the
cluster. Adjacent deposits that do not fall into this area are
considered separate.  The cluster position is calculated independently
in the $x$ and $y$ views as the energy-weighted sum of the individual
voxels. One voxel, two voxel and three voxel clusters were found to
have position resolutions of 9.4~mm, 8.0~mm and 7.2~mm,
respectively. The improved resolution due to clusters with multiple
voxels is due to the charge sharing between voxels. The clusters
formed from the hit voxels of an event are then passed to the
reconstruction algorithm.

The separation of candidate muons from hadronic activity is achieved
using two methods: a Kalman filter algorithm provided by
RecPack~\cite{CerveraVillanueva:2004kt} and a cellular automaton
method (based on \cite{Emeliyanov_otr/itr-cats:tracking}), both
algorithms are described in detail in~\cite{Cervera:2010rz}. The
Kalman filter method requires a section of at least five planes where
only one cluster is present in the highest $z$ region of the event
that is associated with particle tracks. Between 85\% and 95\% of
$\nu_\mu~(\overline{\nu}_\mu)$ CC interactions and $\sim$2.5\% of NC
interactions fall into this category. This section is used to form a
seed, which is projected back through the high occupancy planes using
a helix model. Events which do not have such a section (generally high
$Q^2$ or low neutrino energy events) are subject to the cellular
automaton which tests a number of possible tracks to find a 
potential muon candidate. Between 5\% and 13\% of
$\nu_\mu~(\overline{\nu}_\mu)$ CC interactions and $\sim$83\% of NC
are presented to the cellular automaton for consideration. NC events
produce a candidate muon which is successfully fitted as such in
$\sim$60\% of cases sent to the Kalman filter. Of the $\sim$28\%
$\nu_\mu~(\overline{\nu}_\mu)$ CC events sent to the cellular
automaton method 99\% of $\nu_{\mu}$ and 45\% of $\bar{\nu}_{\mu}$ are
successfully fitted.

Compared to the method used and described in detail
in~\cite{Cervera:2010rz}, an additional step has been added to the
reconstruction method to take into account that fully-contained muons
(particularly $\mu^-$) can have additional deposits at their endpoint
due to captures on nuclei or due to decays. Long, well defined tracks
can be rejected if there is added energy deposited at the muon end
point, since this can be interpreted as hadronic activity and rejected
by the Kalman filter method, thereby confusing the track finding
algorithm of the cellular automaton. Therefore, after sorting clusters
into increasing \emph{z} position, an additional algorithm is used to
identify such activity and extract the track section for seeding and
projection. The details of this algorithm can be found in
\cite{andrewsthesis}, but it relies on identifying isolated muon-like
hits at the end of a track and removing the high activity region in
the choice of seeds to perform the track fit.

\begin{figure}
  \begin{center}$
    \begin{array}{cc}
      \includegraphics[width=8cm, height=6cm]{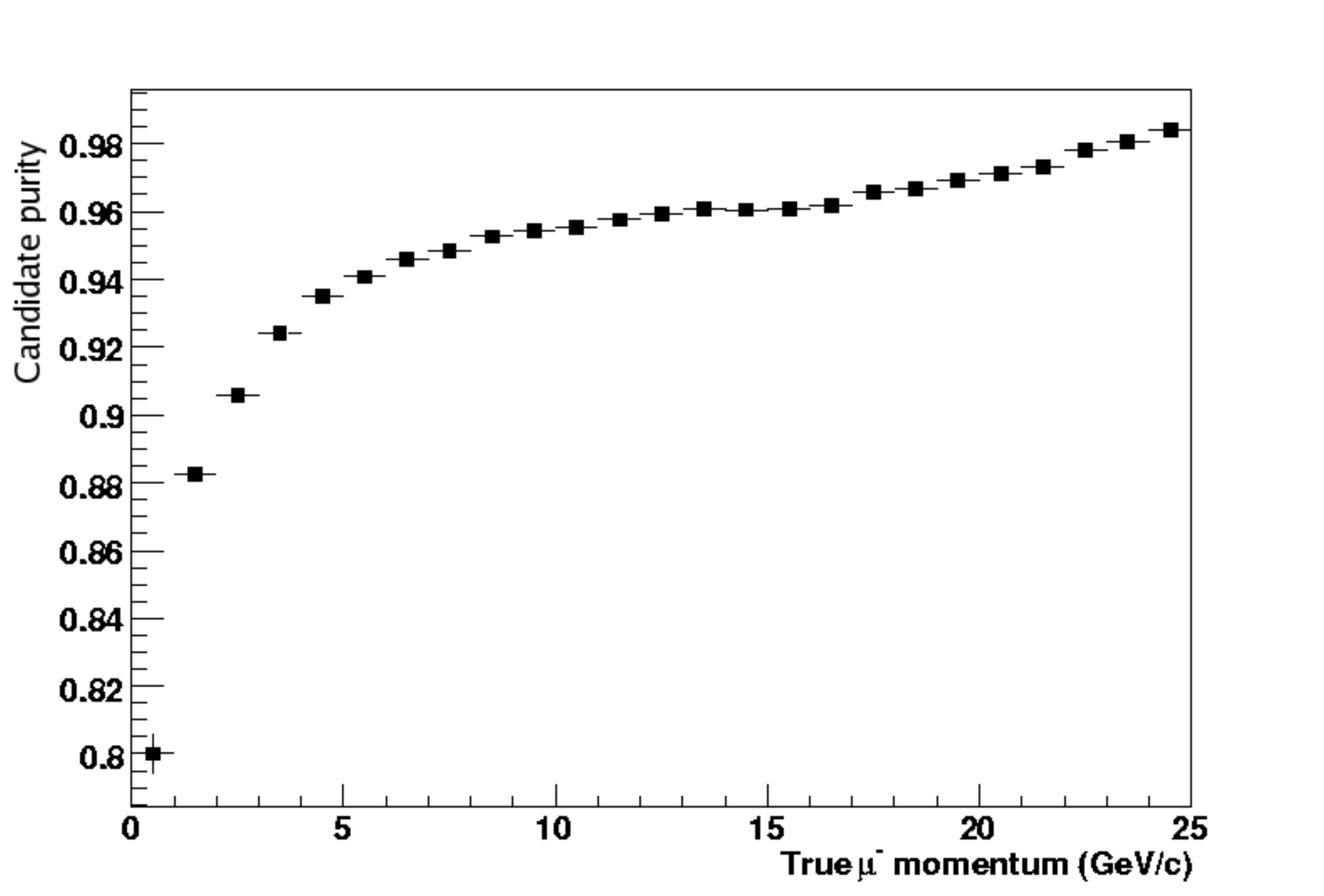} &
      \includegraphics[width=8cm, height=6cm]{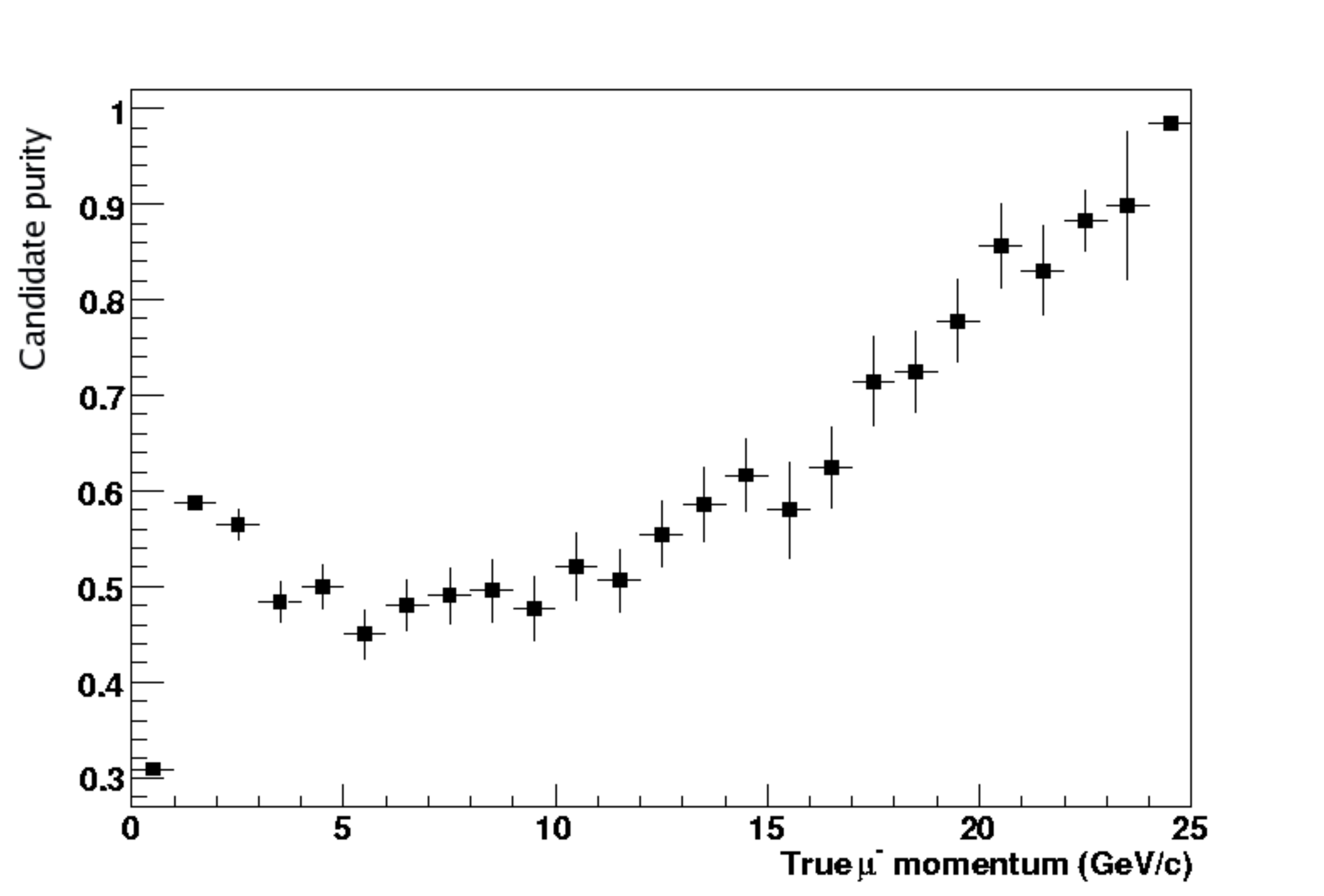}\\
      \includegraphics[width=8cm, height=6cm]{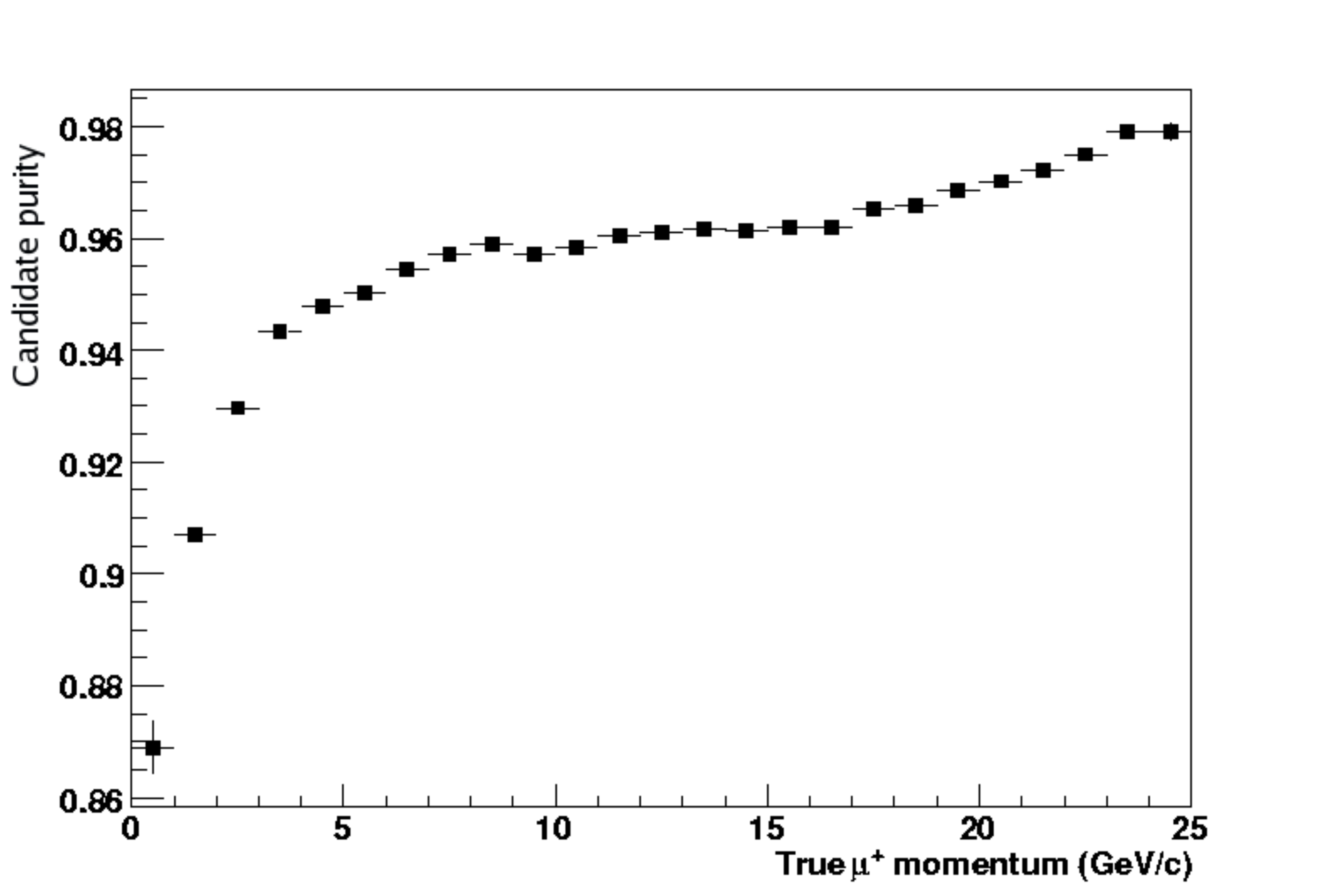} &
      \includegraphics[width=8cm, height=6cm]{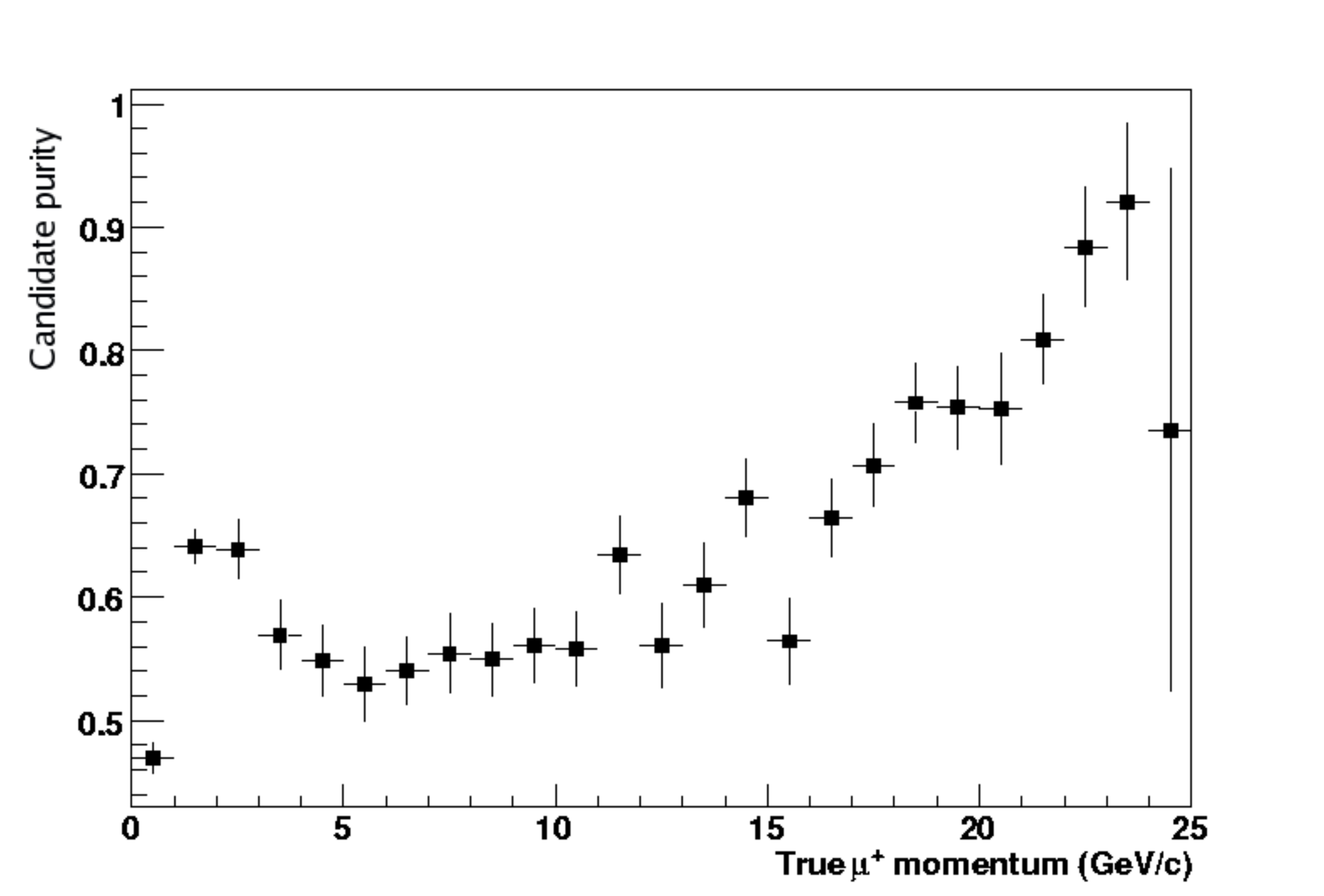}
    \end{array}$
  \end{center}
  \caption{Muon candidate hit purity for $\nu_\mu$ CC (top) and
    $\overline{\nu}_\mu$ CC (bottom) interactions extracted using
    (left) Kalman filter method and (right) cellular automaton
    method.}
  \label{fig:G4purity}
\end{figure}

The complete pattern-recognition chain using these algorithms leads to
candidate purity (fraction of candidate hits of true muon origin) for
$\nu_\mu~(\overline{\nu}_\mu)$ CC events as shown in figure
\ref{fig:G4purity}. A cluster is considered to be of muon origin if
greater than 80\% of the raw deposits contained within the cluster
were recorded as muon deposits.

\begin{figure}
  \begin{center}$
    \begin{array}{cc}
      \includegraphics[width=8cm, height=6cm]{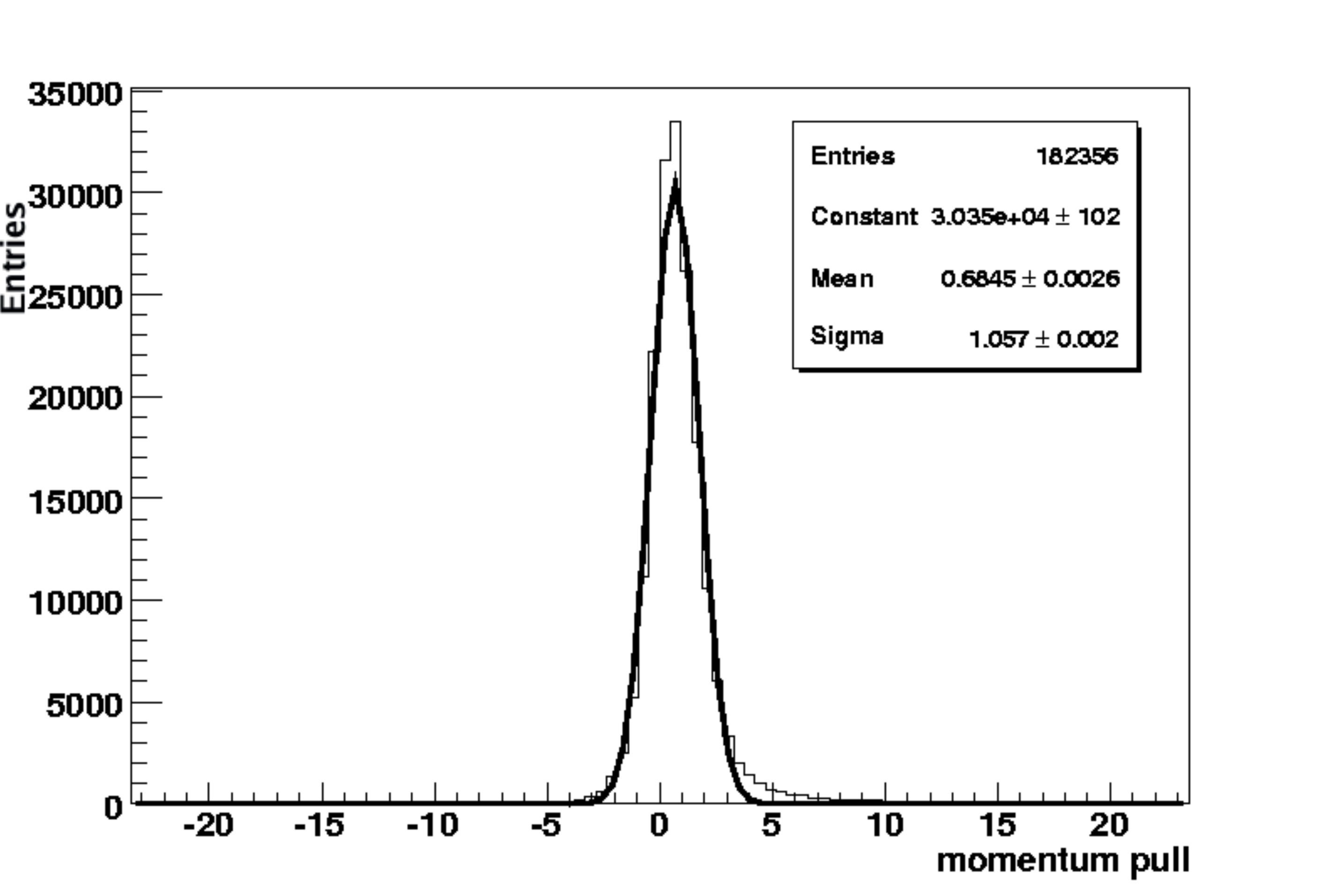} &
      \includegraphics[width=8cm, height=6cm]{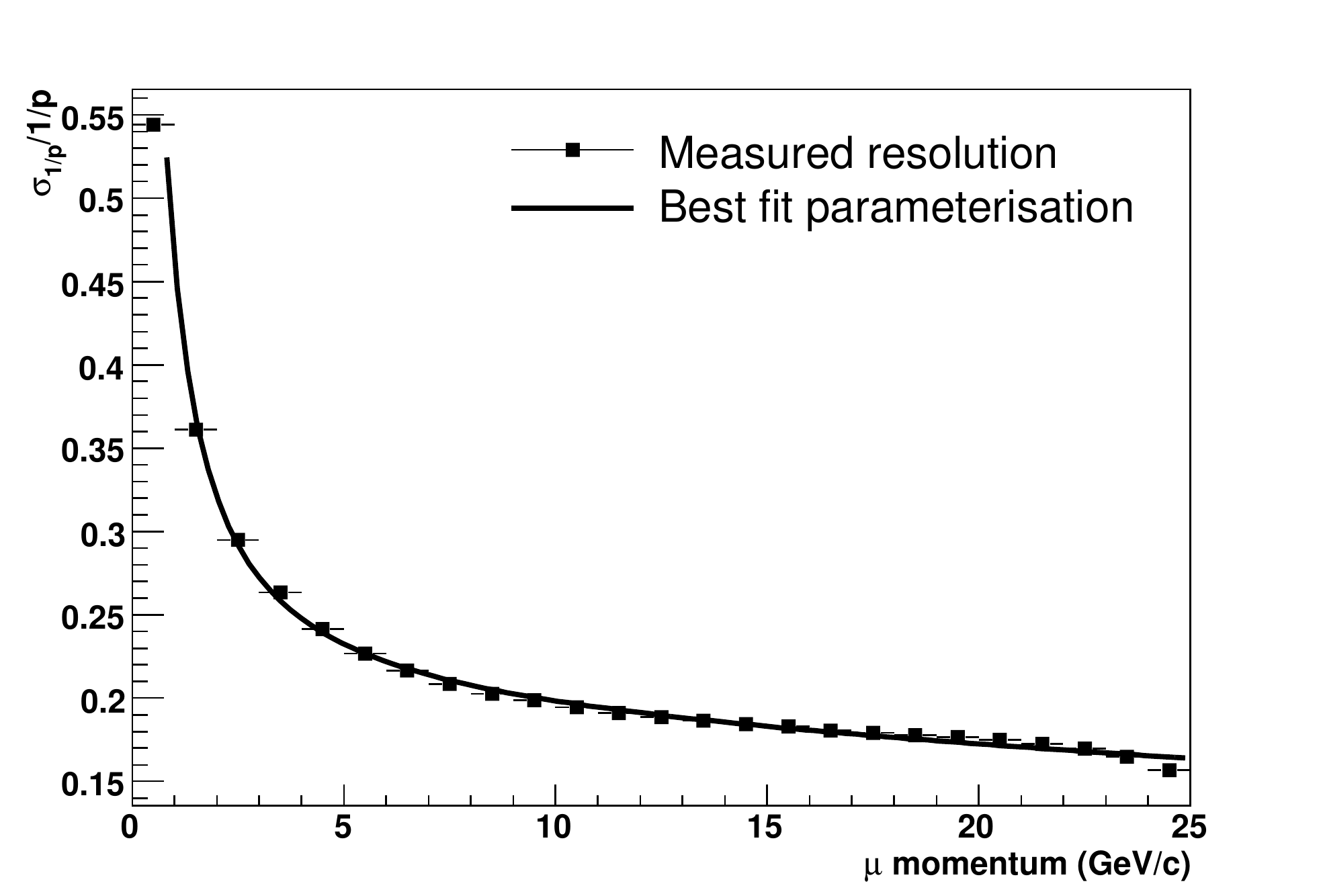}
    \end{array}$
  \end{center}
  \caption{Pull on the reconstructed momentum (the difference between
    the true and reconstructed momentum divided by the measured error)
    (left) and momentum resolution (right).}
  \label{fig:G4pulls}
\end{figure}

Fitting of the candidates proceeds using a Kalman filter to fit a
helix to the candidate, using an initial seed estimated by a quartic
fit, and then refitting any successes. Projecting successful
trajectories back to the true vertex \emph{z} position, the quality of
the fitter can be estimated by comparing to the pull distribution of
the reconstructed momentum, defined as the difference in true and
reconstructed momentum, divided by the measured error in the momentum
from the fit (see figure \ref{fig:G4pulls}). The error in the momentum pull is 
as expected, but the mean pull has a bias (+0.68), due to an incomplete 
energy-loss model in the Kalman filter.  This small bias is taken into account 
in the migration matrices derived for this analysis (see Appendix). 
Further improvements to the energy-loss model within Recpack are being 
carried out and should reduce any residual bias.
An empirical parametrisation of the momentum resolution is also shown in figure \ref{fig:G4pulls}, which can be written as follows:
\begin{equation}
  \label{eq:G4resol}
  \displaystyle\frac{\sigma_{1/p}}{1/p} = 0.18 + \frac{0.28}{p (GeV)} - 1.17\times 10^{-3}p (GeV).
\end{equation}

Neutrino energy is generally reconstructed as the sum of the muon and
hadronic energies, with hadronic reconstruction currently performed
using a smear on the true quantities as described in
ref. \cite{Cervera:2010rz}.  The reconstruction of the hadronic energy
$E_{had}$ assumes a resolution $\delta E_{had}$ from the MINOS CalDet
testbeam~\cite{Michael:2008bc,Adamson:2006xv}:
\begin{equation}
  \label{CalDet1}
  \frac{\delta E_{had}}{E_{had}} = \frac{0.55}{\sqrt{E_{had}}} \oplus 0.03.
\end{equation}
The hadronic shower direction vector was also smeared according to the
angular resolution found by the Monolith test-beam~\cite{Bari:2003bt}:
\begin{equation}
  \label{eq:MONang}
  \delta \theta_{had} = \frac{10.4}{\sqrt{E_{had}}} \oplus \frac{10.1}{E_{had}}.
\end{equation}

In the case of QE interactions, where there is no hadronic jet, the
neutrino energy reconstruction was carried out using the formula:
\begin{equation}
  \label{eq:quasiEng}
  E_\nu = \displaystyle\frac{m_NE_\mu + \frac{1}{2}\left(m_{N'}^2 - m_\mu^2 - m_N^2\right)}{m_N - E_\mu + |p_\mu|\cos\vartheta} \, ;
\end{equation}
where $\vartheta$ is the angle between the muon momentum vector and
the beam direction, $m_N$ is the mass of the initial state nucleon,
and $m_{N'}$ is the mass of the outgoing nucleon for the interactions
$\nu_\mu + n \rightarrow \mu^- + p$ and $\overline{\nu}_\mu + p
\rightarrow \mu^+ + n$ (see for example \cite{Blondel:2004cx}). The
current algorithm only uses this formula for the case of events
consisting of a single unaccompanied track, however, its use could be
extended by selecting QE interactions using their distribution in
$\vartheta$ and their event-plane occupancy among other
parameters. Should the use of equation~\ref{eq:quasiEng} result in a
negative value for the energy, it is recalculated as the total energy
of a muon with its reconstructed momentum.

\section{Analysis of potential signal and background}
\label{sec:analysis}
There are four principal sources of background to the wrong sign muon
search: charge mis-identification of the primary muon in $\nu_\mu$
charged current (CC) interactions, wrong sign muons from hadron decay
in $\overline{\nu}_\mu$ CC events, neutral current (NC) from all
species and $\nu_e$ CC events wrongly identified as $\nu_\mu$
CC. Typically, a $\nu_\mu$ charged current event has greater length in
the beam direction than a NC or $\nu_e$ CC event, due to the
penetrating muon. Any muons produced from the decay of primary
interaction hadrons will tend to be less isolated from other hadronic
activity. Additionally, the $\nu_e$ spectrum at a Neutrino Factory has
a lower average energy than the $\nu_\mu$ spectrum which results in
reduced probability of producing a high energy particle in the
interaction.

The previous general principles are used to define a series of offline
cuts that reject the dominant background while maintaining good signal
efficiency. These can be organised in three categories: 1) track
quality cuts; 2) charged current selection cut; and 3) kinematic
cuts. We will describe these cuts in detail in
sub-sections~\ref{subsec:qP}, \ref{subsec:CCsel} and
\ref{subsec:G4kinCut}. A summary of the performance of each of the
cuts on signal and background will be presented in
sub-section~\ref{subsec:Csumm} and table~\ref{tab:G4cutSum}.

\subsection{Track quality cuts}
\label{subsec:qP}
The quality of the reconstruction and the error on the momentum
parameter of the Kalman filter are powerful handles in the rejection
of backgrounds. We commence by imposing the reconstruction criteria
from the previous section to guarantee fully reconstructed neutrino
events. We then proceed to impose a fiducial cut requiring that $z1$, which is the 
cluster with the lowest \emph{z} in the candidate, be at least 2~m from the end of the
detector ($z1-z_{end}\geq 2000$~mm), to reduce the mis-identification of
candidates originating at high \emph{z}. Additionally, a maximum value
for the reconstructed muon momentum is imposed at 16~GeV to improve
energy resolution and remove backgrounds caused by very straight
particles, which confuse the fitter.

\begin{figure}
  \begin{center}$
    \begin{array}{cc}
      \includegraphics[width=0.45\linewidth]{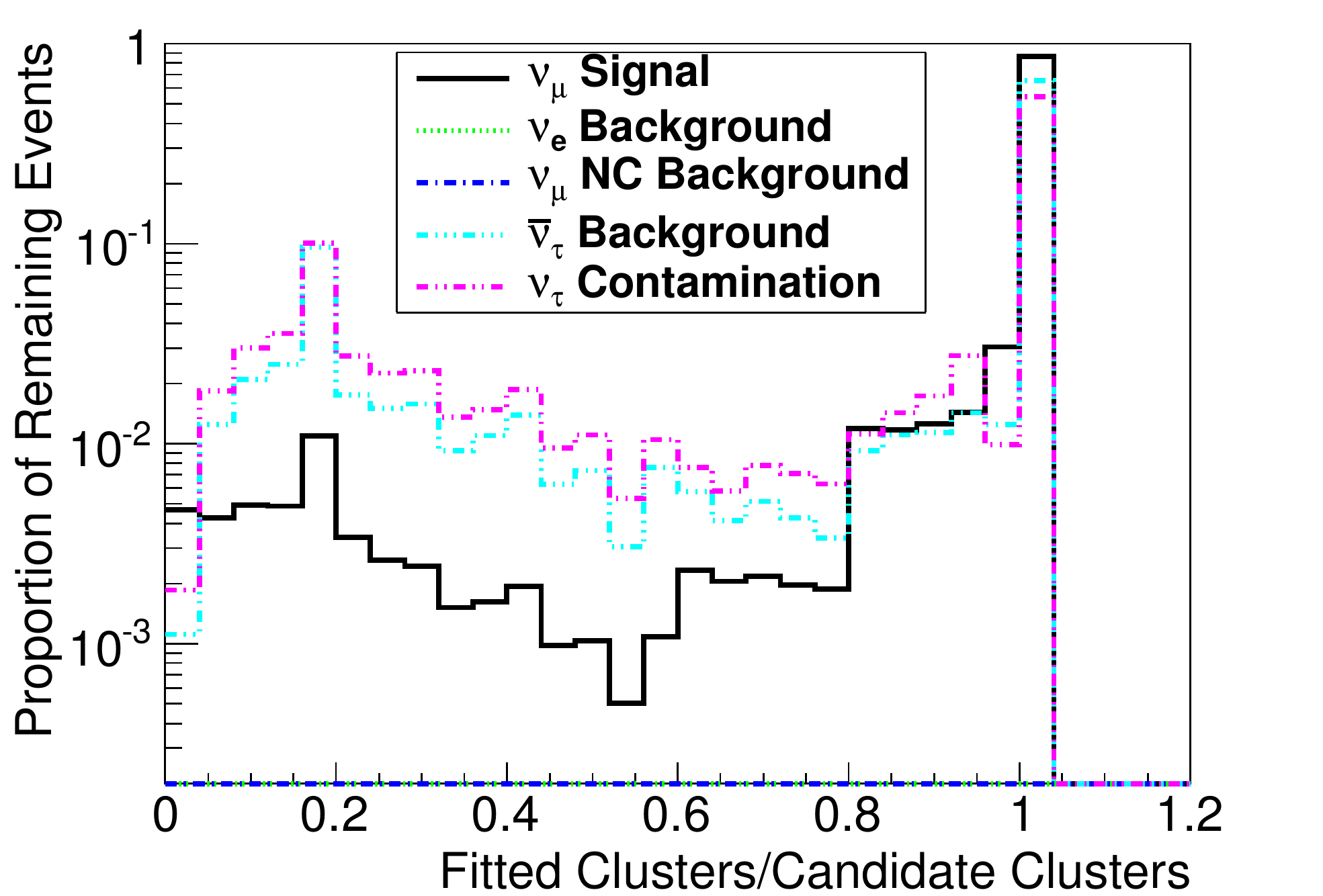} &
      \includegraphics[width=0.45\linewidth]{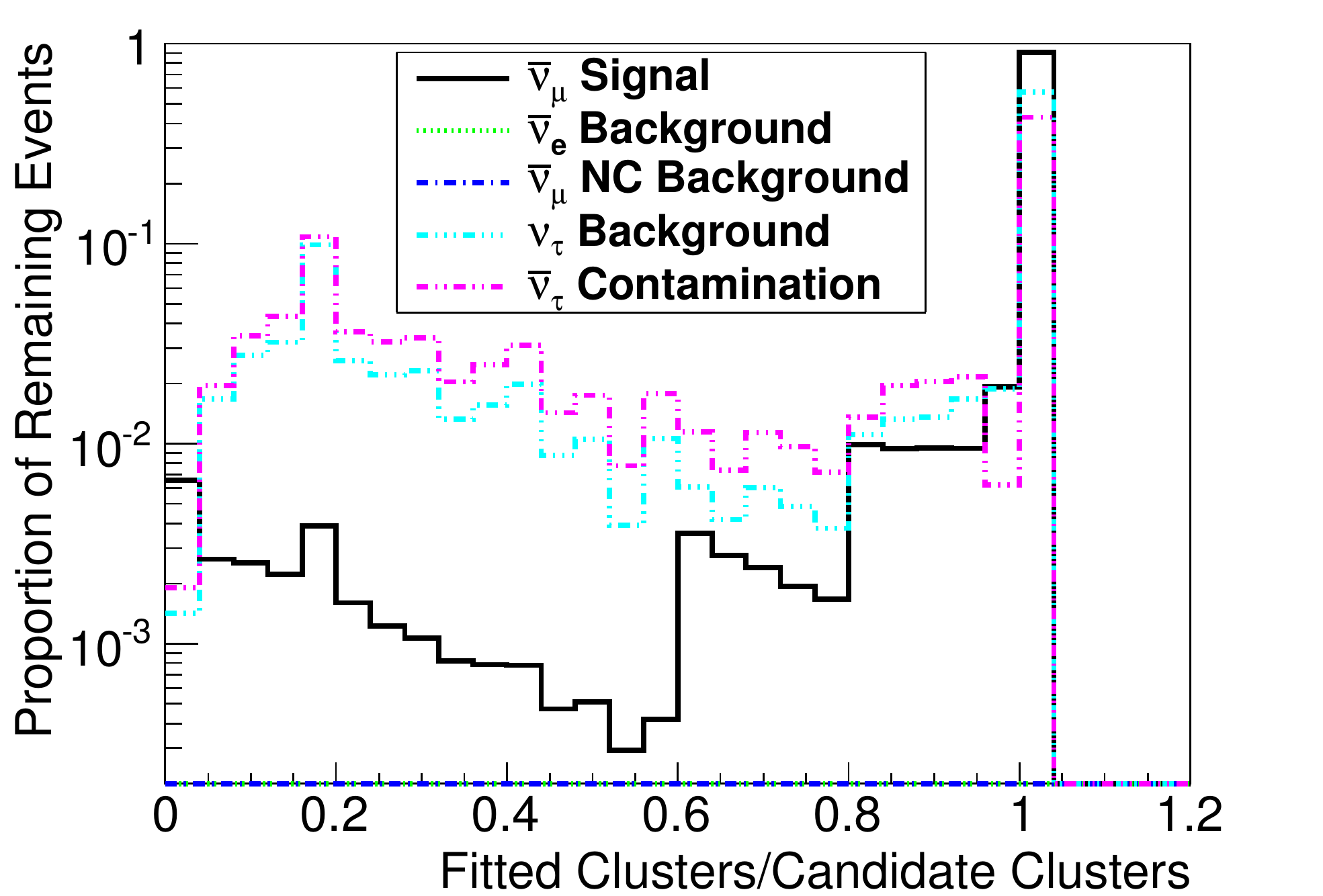}
      \end{array}$
  \end{center}
  \caption{Distribution of the proportion of clusters fitted in the
    trajectory for $\nu_\mu$ appearance (left) and
    $\overline{\nu}_\mu$ appearance (right), normalised to total
    remaining events individually for each interaction type.}
  \label{fig:fitnode}
\end{figure}

Tracks dominated by multiple scattering or incorporating deposits made
by particles not left by a muon can contribute significantly to
backgrounds. However, these tracks will tend to be fitted only
partially or with a larger error on the momentum variables. As such,
cuts on these variables can be used to reduce the effect of these
backgrounds. The distribution of the ratio of the candidate clusters
which are fitted with respect to the total number of candidate
clusters for signal and background is shown in figure
\ref{fig:fitnode}. Accepting only those events in which a candidate has more than
60\% of its clusters fitted reduces the background levels.

\begin{figure}
  \begin{center}$
    \begin{array}{cc}
      \includegraphics[width=0.45\linewidth]{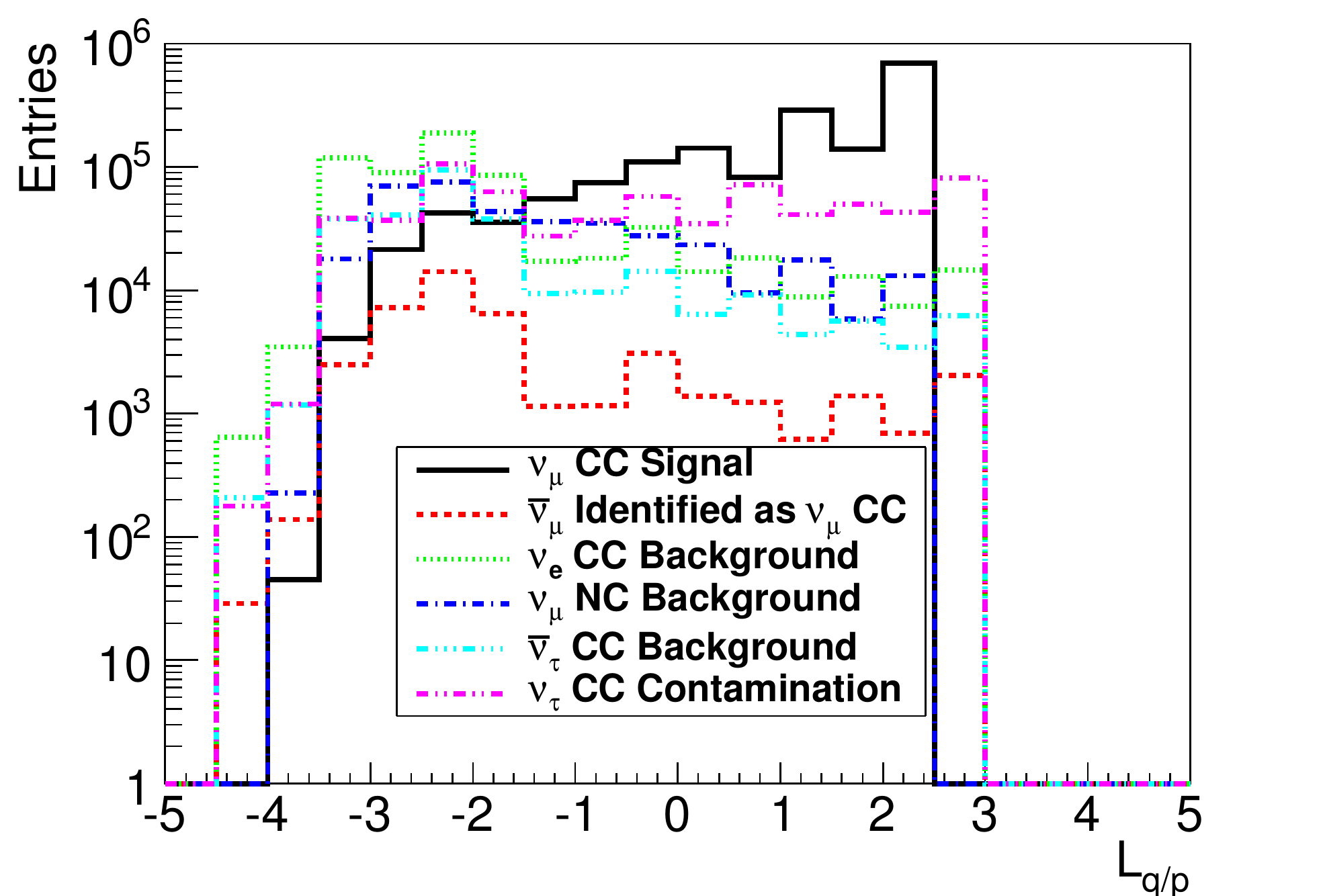} & 
      \includegraphics[width=0.45\linewidth]{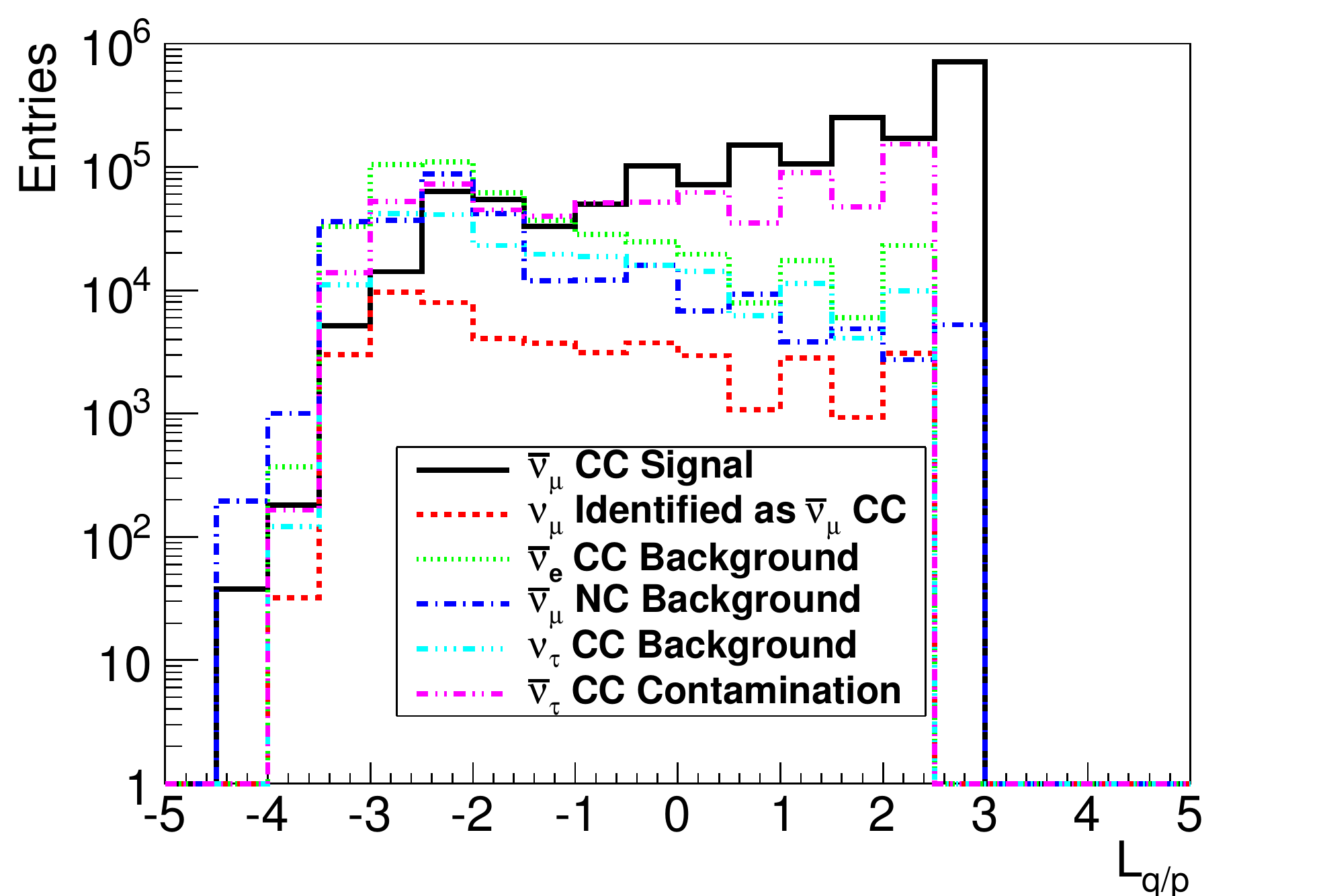} 
    \end{array}$
  \end{center}
  \caption{Log likelihood distribution ($\mathcal{L}_{q/p}$) to
    separate wrong-sign muons for signal and background for
    $\nu_{\mu}$ (left) and $\bar{\nu}_{\mu}$ (right) appearance experiments.} 
  \label{fig:qPlike}
\end{figure}

Further reduction is achieved by performing a cut related to the
relative error in the momentum of the candidate muon
$\frac{\sigma_{q/p}}{q/p}$, where $q$ is the charge of the muon and
$p$ its momentum. A log-likelihood distribution $\mathcal{L}_{q/p}$
based on the ratio of $\frac{\sigma_{q/p}}{q/p}$ for both signal and
background is shown in figure \ref{fig:qPlike}. The signal events are
selected as those with a log-likelihood parameter
$\mathcal{L}_{\sigma/p} > -0.5$.

\begin{figure}
  \begin{center}$
    \begin{array}{cc}
      \includegraphics[width=0.45\linewidth]{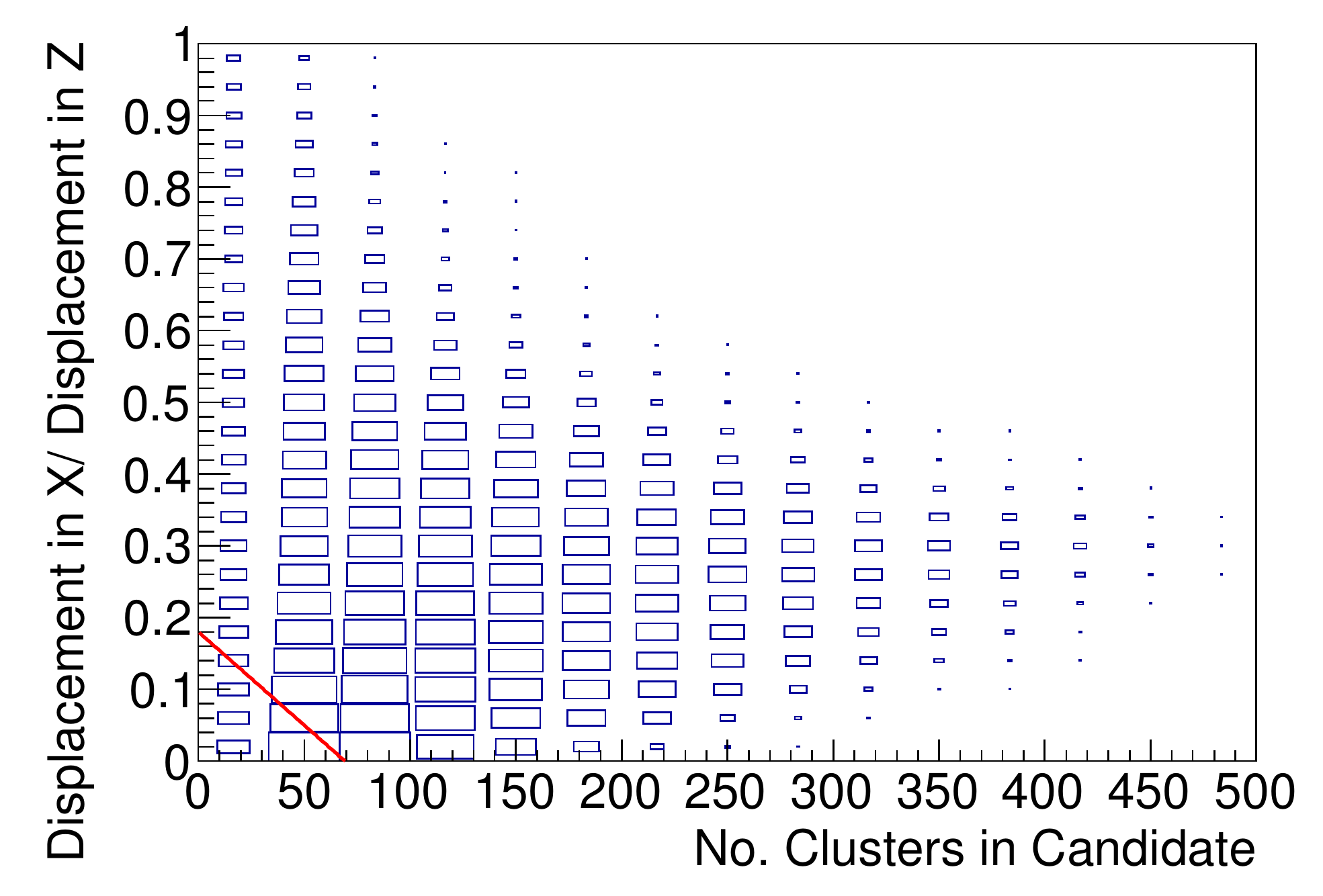} &
      \includegraphics[width=0.45\linewidth]{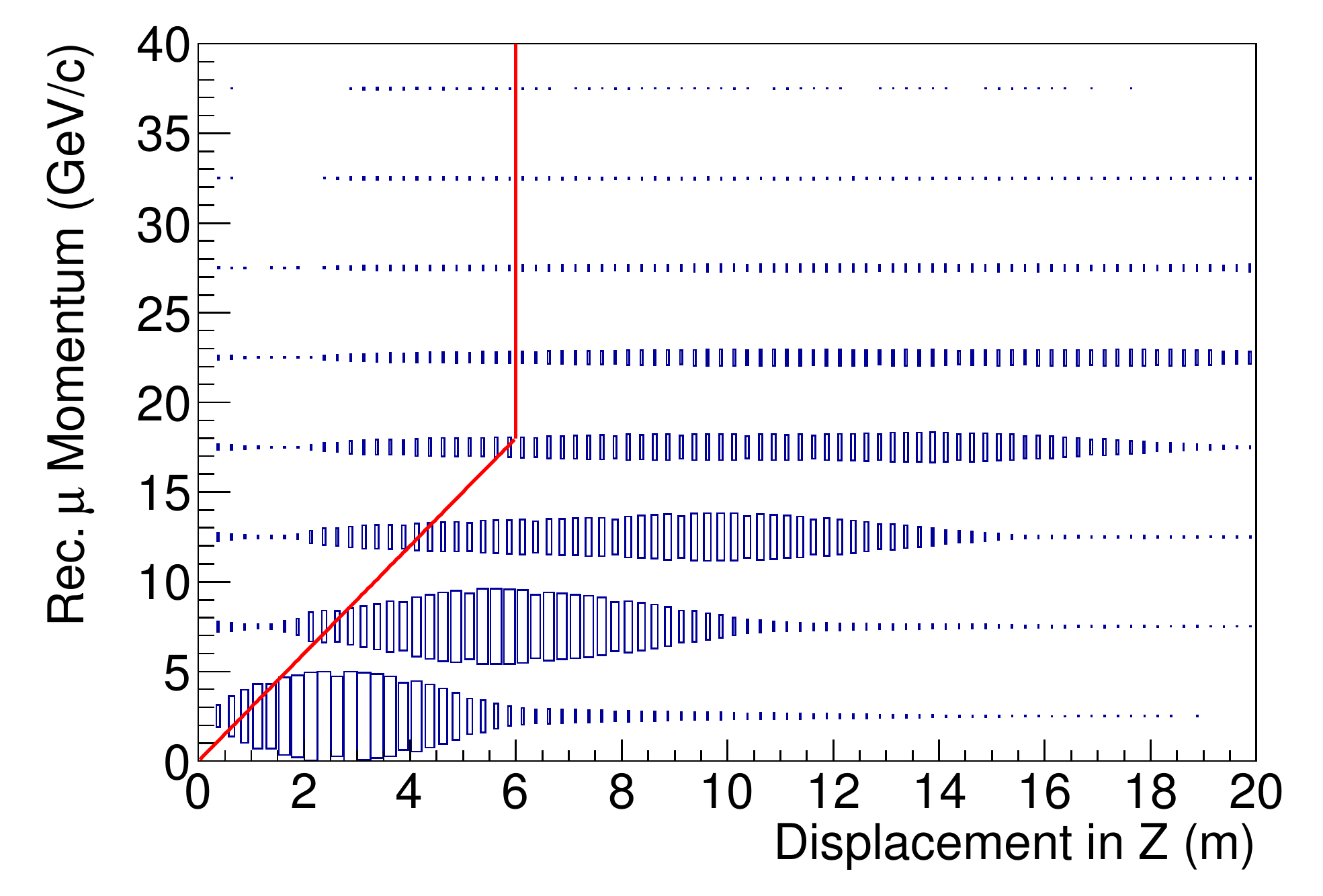} \\
      \includegraphics[width=0.45\linewidth]{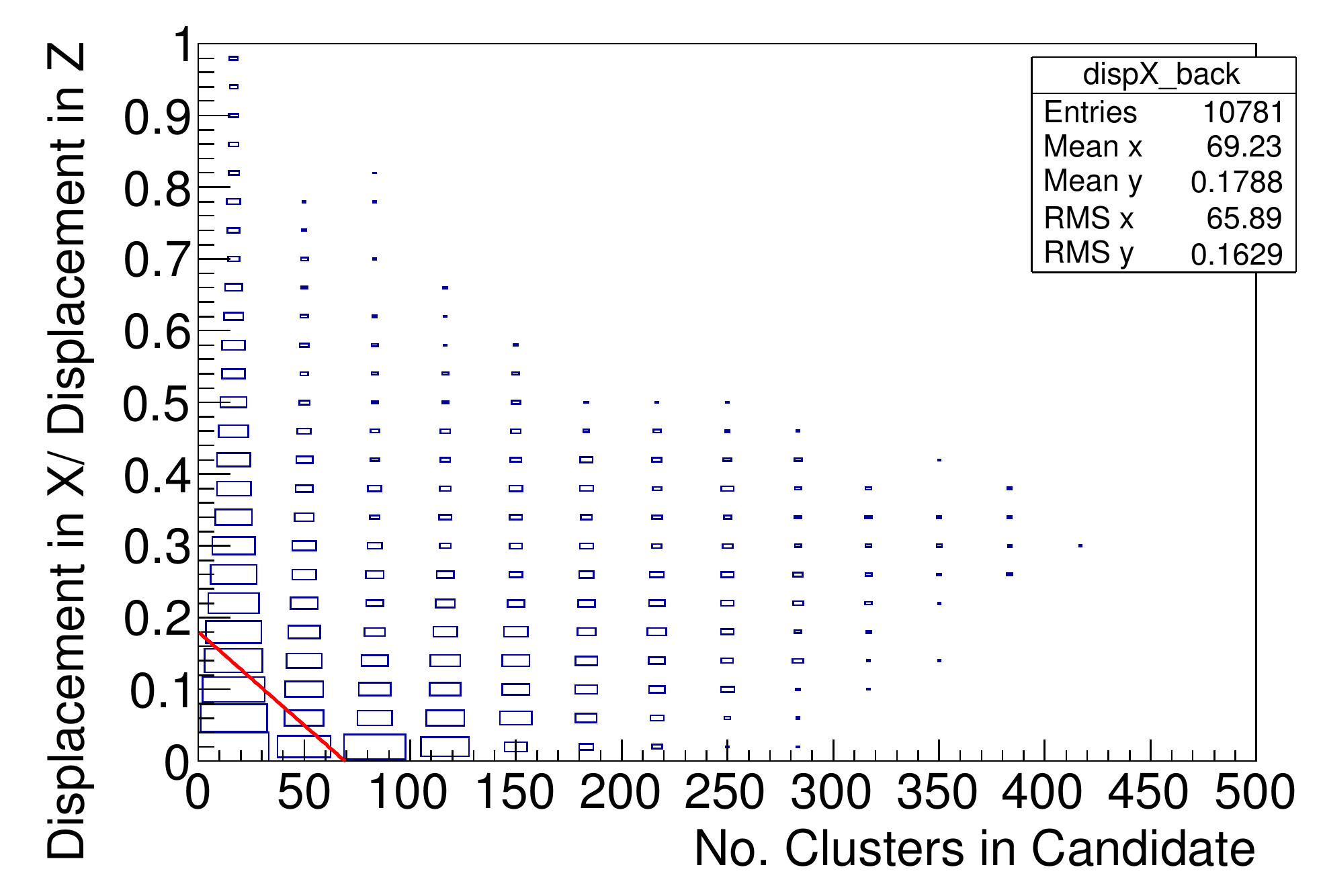} &
      \includegraphics[width=0.45\linewidth]{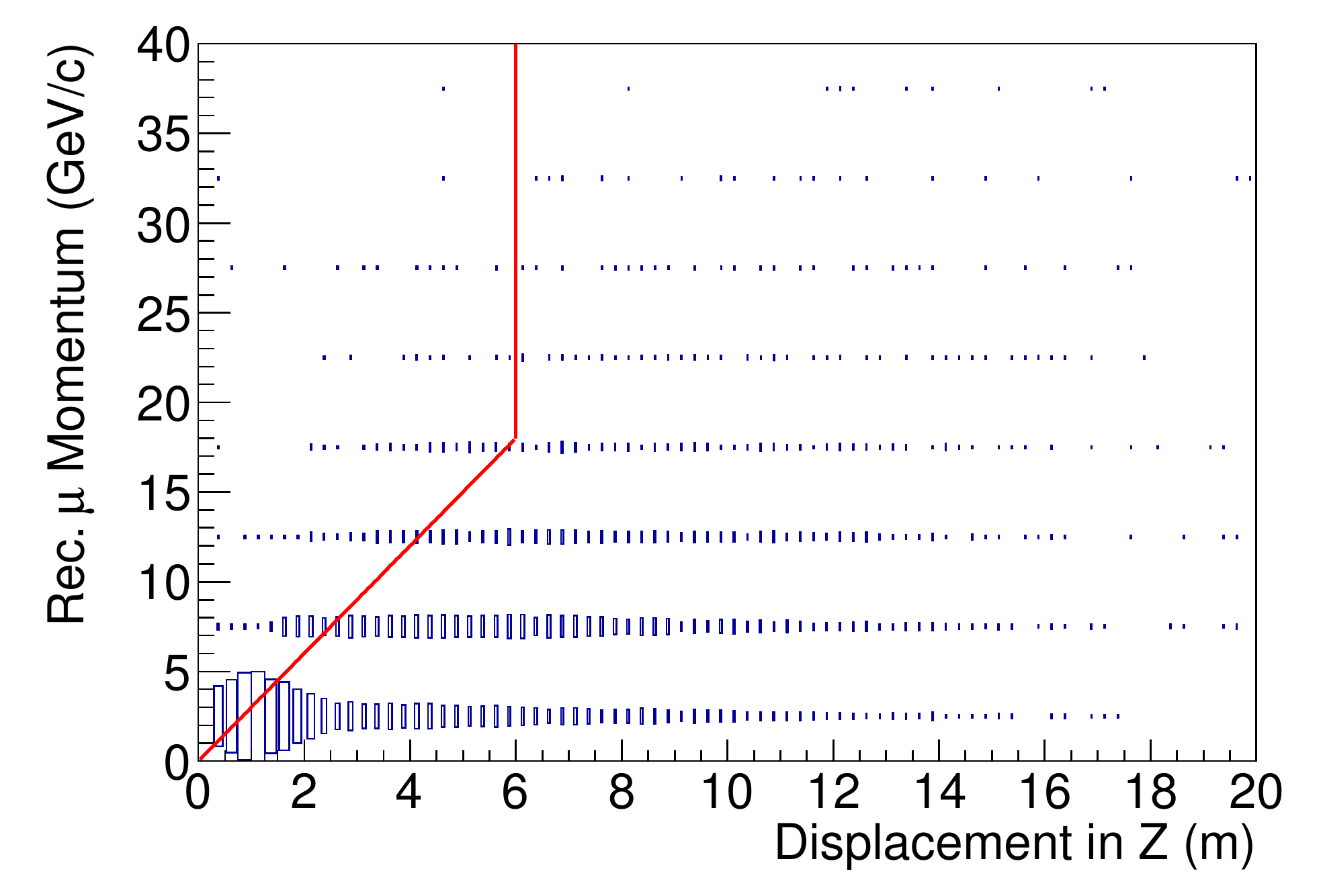} \\
    \end{array}$
  \end{center}
  \caption{Distributions of displacement and momentum with cut levels:
    (top left) relative displacement in the bending plane to the $z$
    direction against candidate hits for signal events, (top right)
    reconstructed momentum against displacement in $z$ for signal events
    and (bottom) as top for $\nu_\mu~(\overline{\nu}_\mu)$ CC
    backgrounds. The red lines represent the cuts from equations
    \ref{eq:dispMom} and \ref{eq:dispMom2}. }
  \label{fig:dispMom}
\end{figure}

After the preceding cuts there remain some background events which
exhibit little bending due to the magnetic field or are reconstructed
with relatively high momentum despite being relatively short tracks
(see figure \ref{fig:dispMom}, right) as a result of high levels of
multiple scattering. As can be seen in figure \ref{fig:dispMom}, left,
removing short events in which the end point is displaced in the
bending plane by an amount that is relatively small compared to the
displacement in the lateral view ($dispX/dispZ$) effectively reduces
background. Events are accepted if they meet the conditions described
in equations~\ref{eq:dispMom} and \ref{eq:dispMom2}, illustrated by the red lines in
figure \ref{fig:dispMom}:
\begin{eqnarray}
  \label{eq:dispMom}
  \displaystyle \frac{dispX}{dispZ} &>& 0.18 - 0.0026\cdot N_h \, ; {\rm and} \\
  \label{eq:dispMom2}
  dispZ~&>&~6000\mbox{ mm} \mbox{~~~or~~~} p_\mu~\leq~3\cdot~dispZ \, ;
\end{eqnarray}
where $N_h$ is the number of clusters in the candidate, $dispZ$ is in
units of mm, and $p_\mu$ in units of MeV/c.

\begin{figure}
  \begin{center}$
    \begin{array}{cc}
      \includegraphics[width=0.45\linewidth]{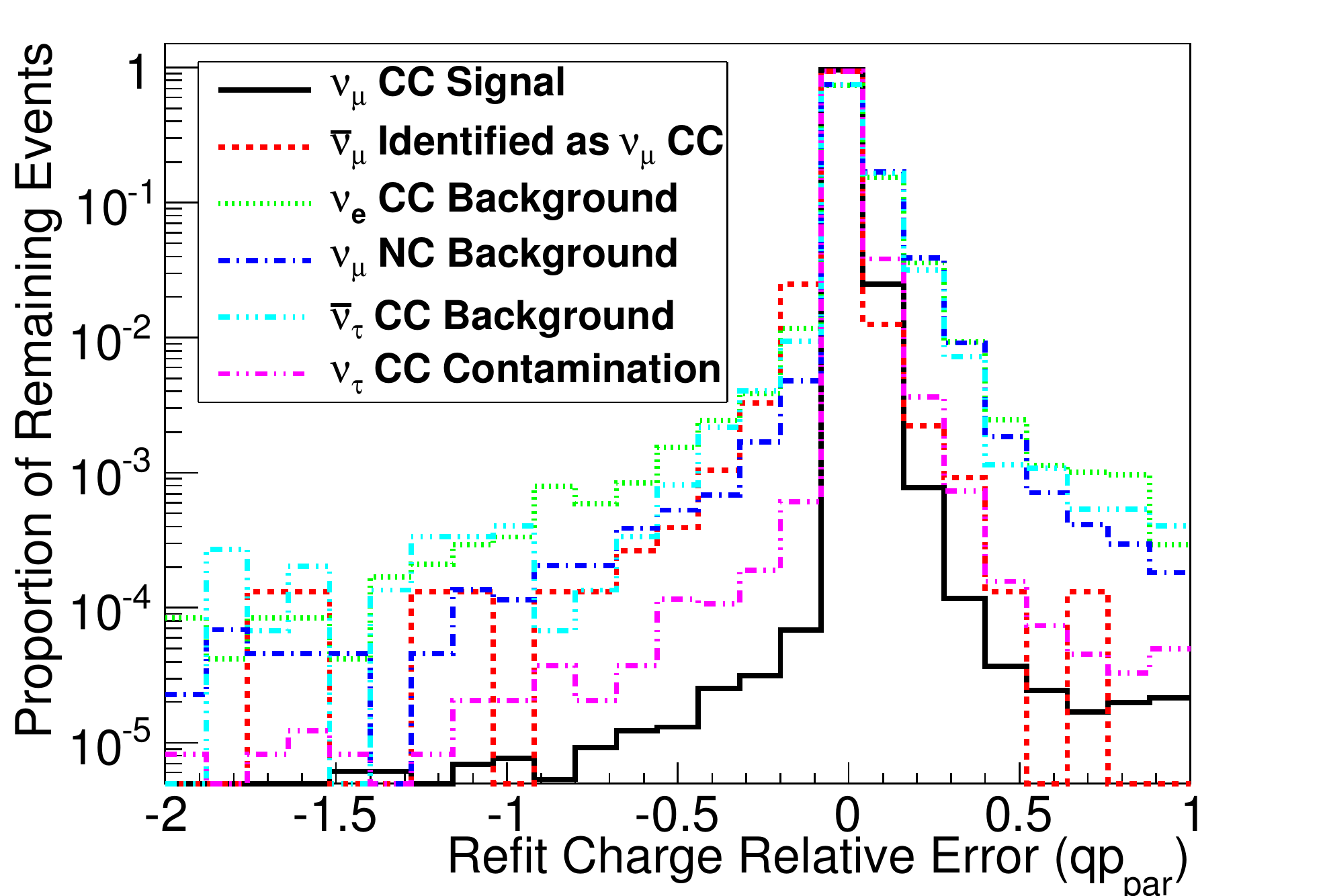}
      &
      \includegraphics[width=0.45\linewidth]{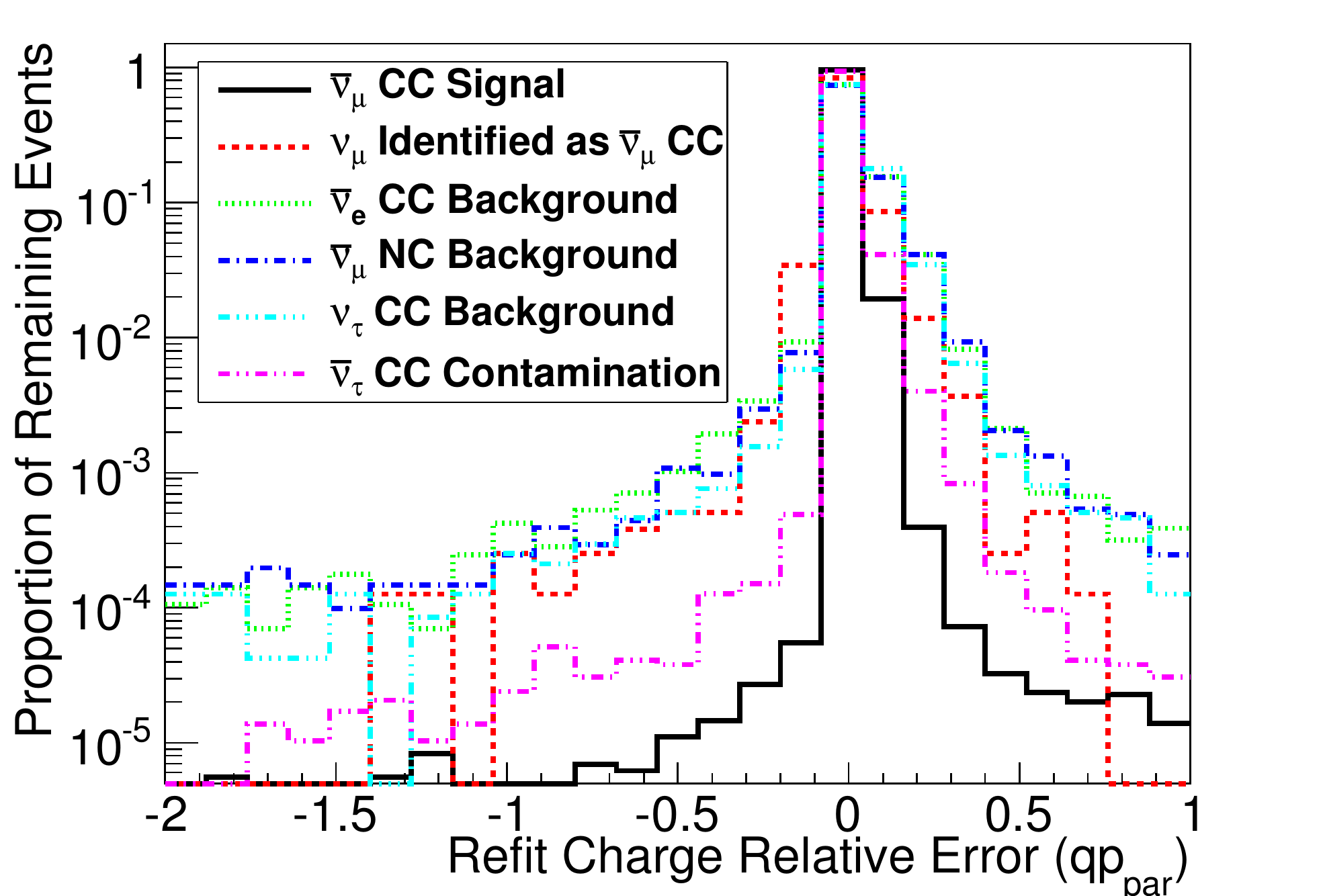}
    \end{array}$
  \end{center}
  \caption{Distribution of the $qp_{par}$ variable, with the region
    where the parameter is $<$ 0 representing those candidates fitted
    with charge opposite to the initial Kalman filter for $\nu_{\mu}$
    (left) and $\bar{\nu}_{\mu}$ appearance experiments. The
    distributions are normalised to the total remaining events,
    individually, for each interaction type.}
  \label{fig:qcharge}
\end{figure}

The final quality cut involves fitting to a parabola the candidate's
projection onto the bending plane. In the current simulation a
negatively-charged muon bends upwards, so that for a parabola defined
as $a + bz + cz^2$ the parameter $c$ would be positive and the charge
of the muon is $Q_{par} = -sign(c)$. If the charge fitted is opposite
to that found by the Kalman filter, the quality of the fit is assessed
using the variable:
\begin{equation}
  \label{eq:qCharge}
  \begin{array}{ll}
  qp_{par} &= \left\{ \begin{array}{c} \left|\displaystyle\frac{\sigma_c}{c}\right|\mbox{, if } Q_{par} = Q_{kal} \, ;\\ \\
      -\left|\displaystyle\frac{\sigma_c}{c}\right|\mbox{, if } Q_{par} = -Q_{kal} \, ;\end{array}\right.
  \end{array}
\end{equation}
where $Q_{kal}$ is the charge fitted by the Kalman filter
fit. Defining the parameter in this way ensures that the cut is
independent of the initial fitted charge. Events with no charge change
($qp_{par} > 0.0$) are accepted as signal. Additionally, those fitted
badly with a charge change ($qp_{par} < -1.0$) are also accepted. In
this way, background events which have remained in the sample due to
local variations affecting the Kalman fitter can be removed without
rejecting viable events in which the Kalman fitter ignored a section
after a high angle scatter. The distribution of $qp_{par}$ is shown
in figure~\ref{fig:qcharge}.

\subsection{Charged current selection}
\label{subsec:CCsel}
\begin{figure}
  \includegraphics[width=0.45\linewidth]{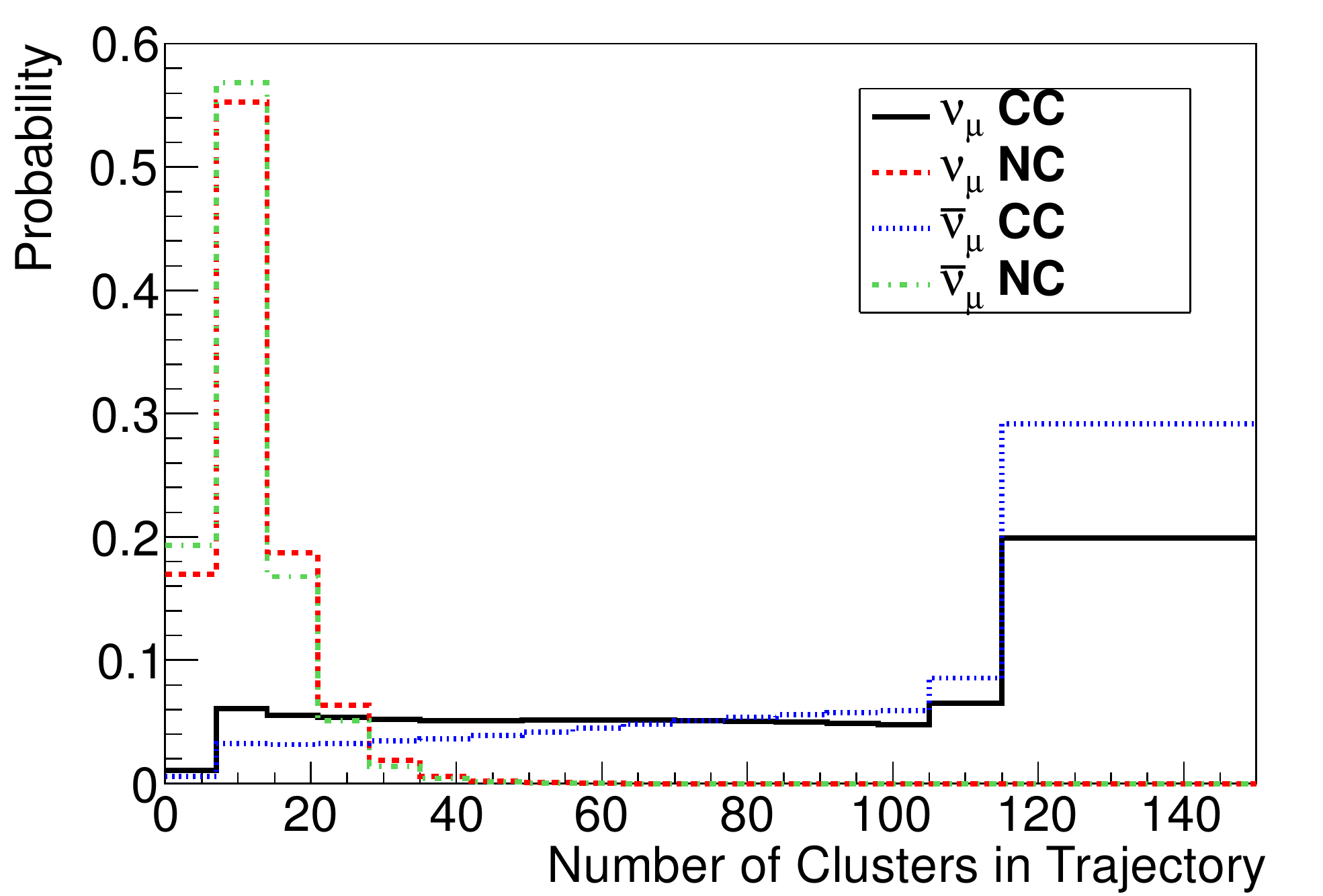}  
  \caption{Distribution of the number of fit clusters in candidate
    track used to calculate the log likelihood based charged current
    selection.}
  \label{fig:NCpdfs}
\end{figure}

\begin{figure}
  \subfloat[Distribution for $\nu_{\mu}$ detection]{
    \includegraphics[width=0.45\linewidth]{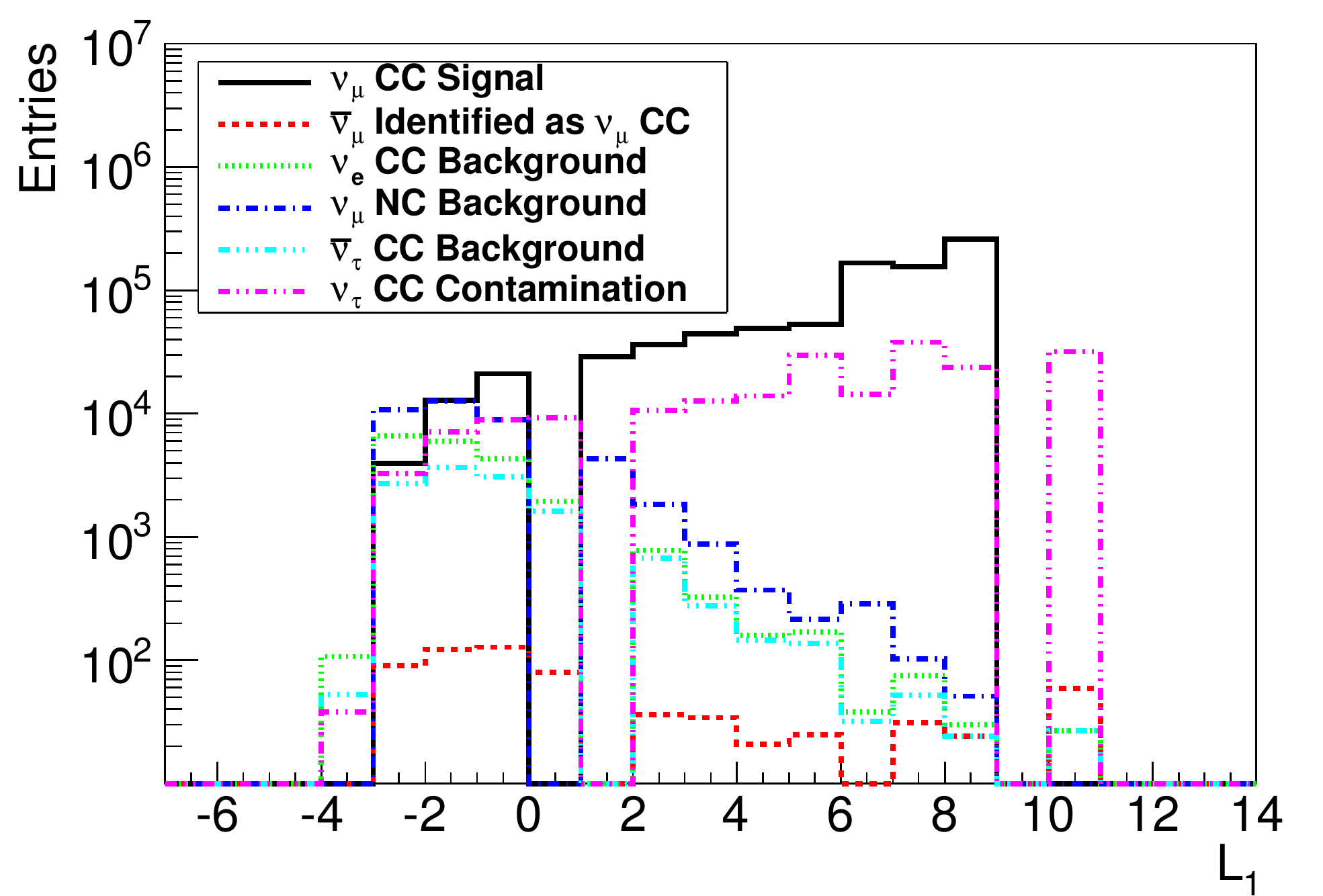}
  }
  \subfloat[Distribution for $\bar{\nu}_{\mu}$ detection]{
    \includegraphics[width=0.45\linewidth]{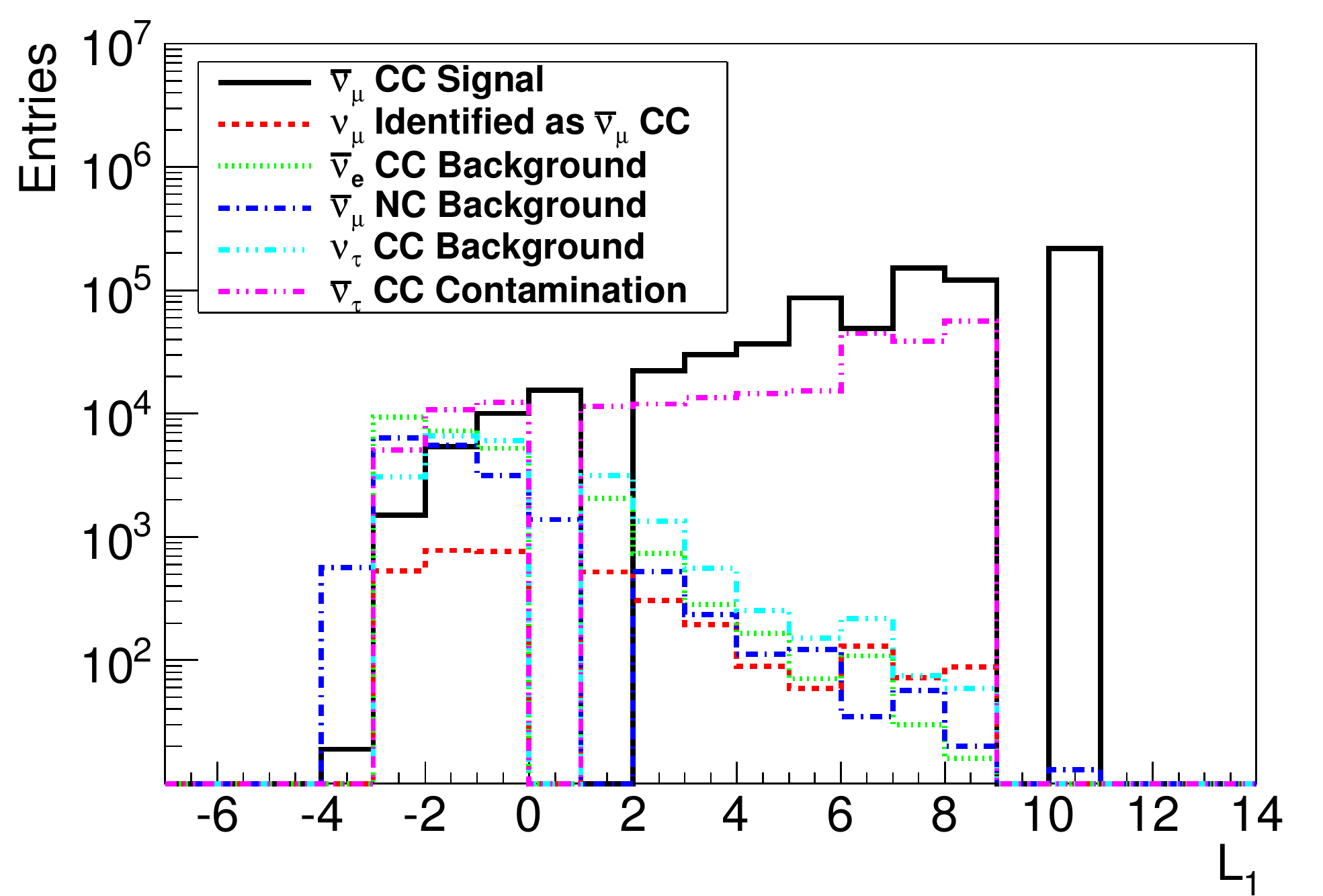}
  }
  \caption{Distribution of $\mathcal{L}_{1}$ likelihood ratios used to
    reject NC and other background signals.}
  \label{fig:L4dist}
\end{figure}

Selection of charged currents and rejection of neutral current events
is most efficiently performed by exploiting the propertiy that
$\nu_\mu$ CC events tend to have greater length in \emph{z} than NC
events, since a true muon only interacts electromagnetically where a
pion or kaon of similar momentum can interact via the strong force
and will tend to stop after a shorter distance. Hence, the
number of hits, $l_{hit}$, was used to generate Probability Density
Functions (PDF) for charged and neutral current events (see figure
\ref{fig:NCpdfs}). One can see that the NC events have fewer
reconstructed clusters than the equivalent $\nu_\mu$ CC events. For
the event selection, candidates with greater than 150 clusters are
considered signal, otherwise, the log likelihood rejection parameter:
\begin{equation}
  \mathcal{L}_1 = \log \left( \frac{l_{hit}^{CC}}{l_{hit}^{NC}}
  \right) \, ;
\end{equation}
is used, which is shown in figure \ref{fig:L4dist}. Allowing
only those candidates where the log parameter is $\mathcal{L}_{1} >
1.0$ to remain in the sample ensures that the sample is pure. This
analysis is similar, but simpler, than that employed by MINOS
\cite{Adamson:2007gu}. The effect of the CC event selection is to
reduce the background by one order of magnitude (see
table~\ref{tab:G4cutSum}) while having minimal effect on the signal
efficiency.

\subsection{Kinematic cuts}
\label{subsec:G4kinCut}
\begin{figure}
  \begin{center}$
    \begin{array}{cc}
      \includegraphics[width=7.5cm]{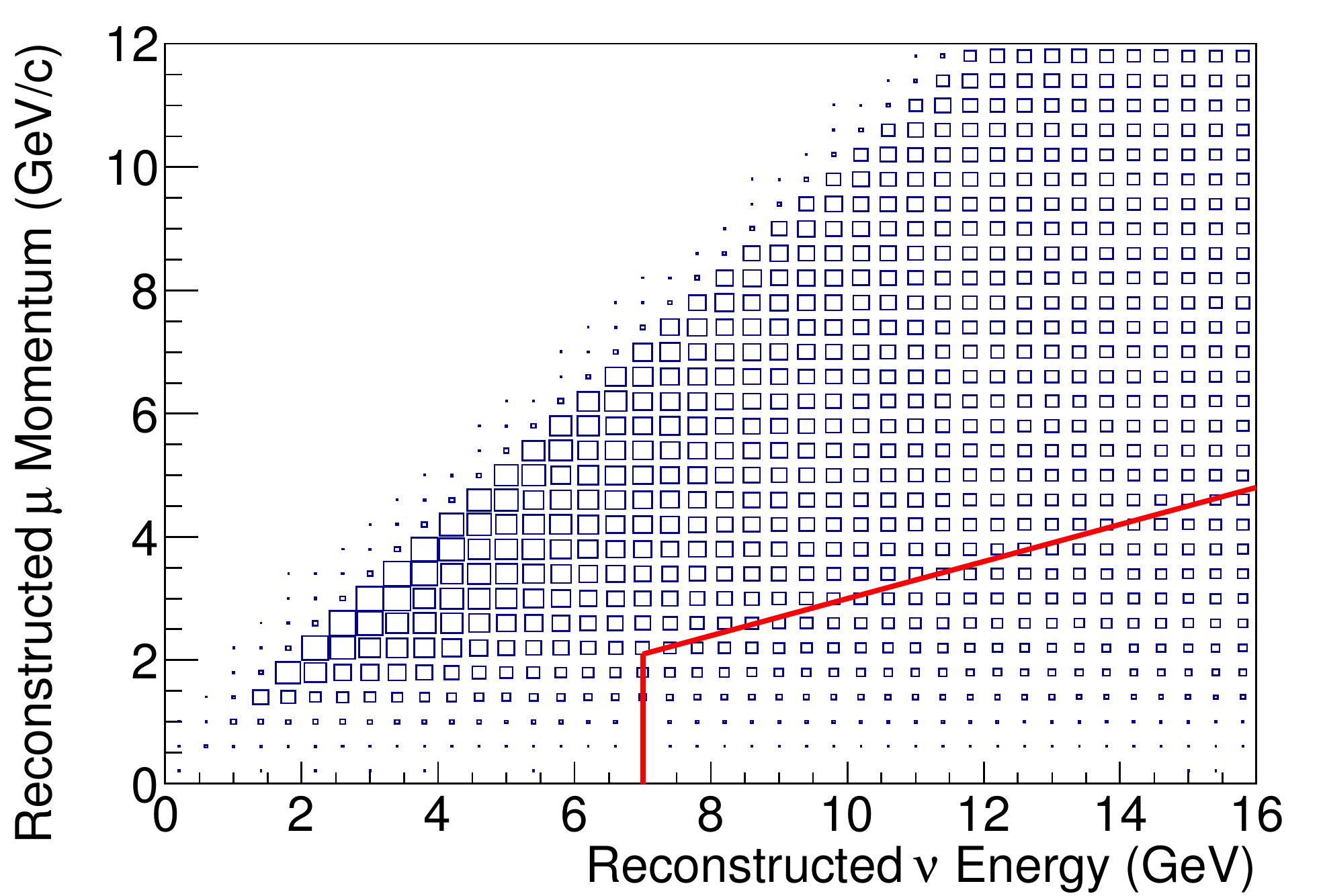} &
      \includegraphics[width=7.5cm]{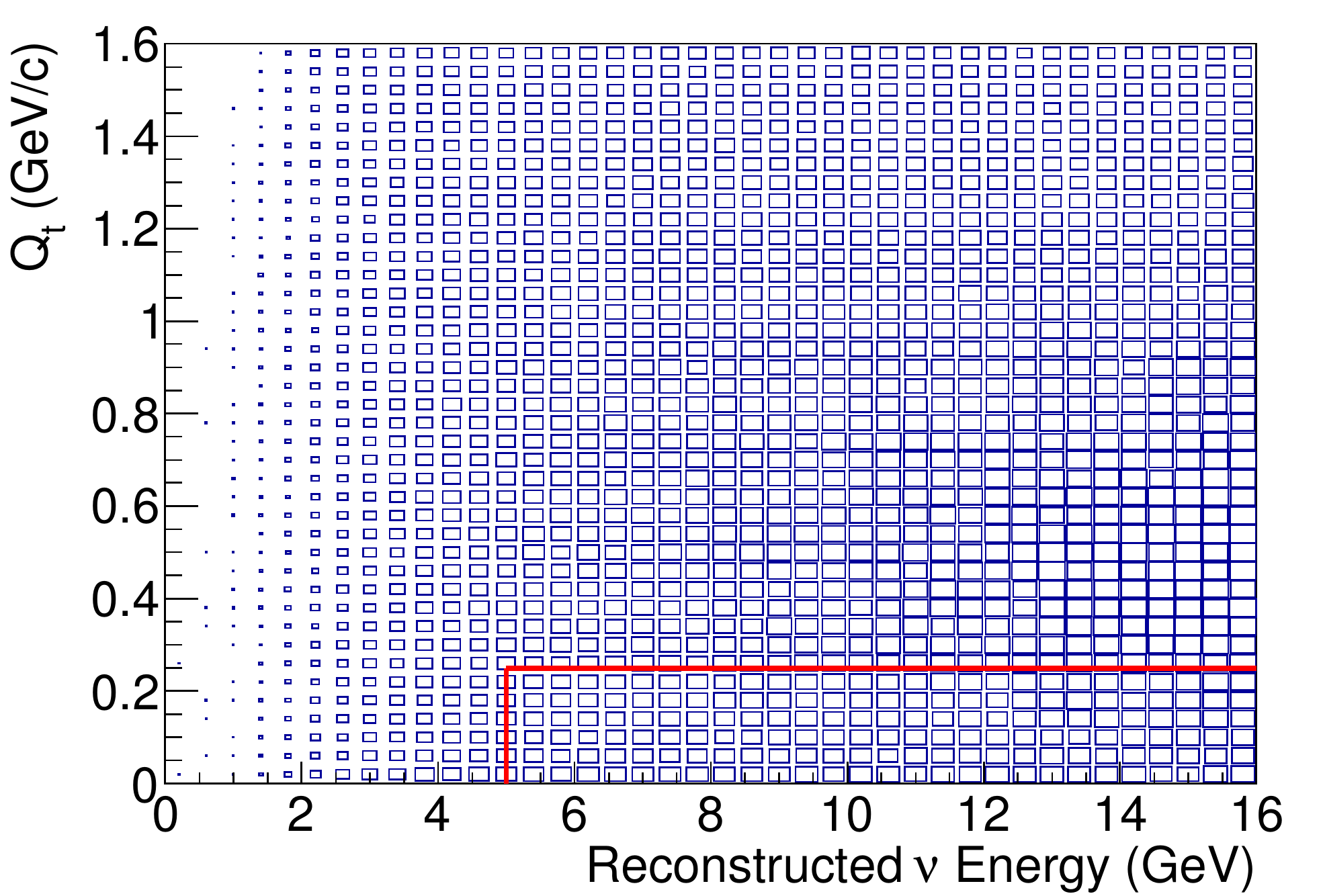} \\
      \includegraphics[width=7.5cm]{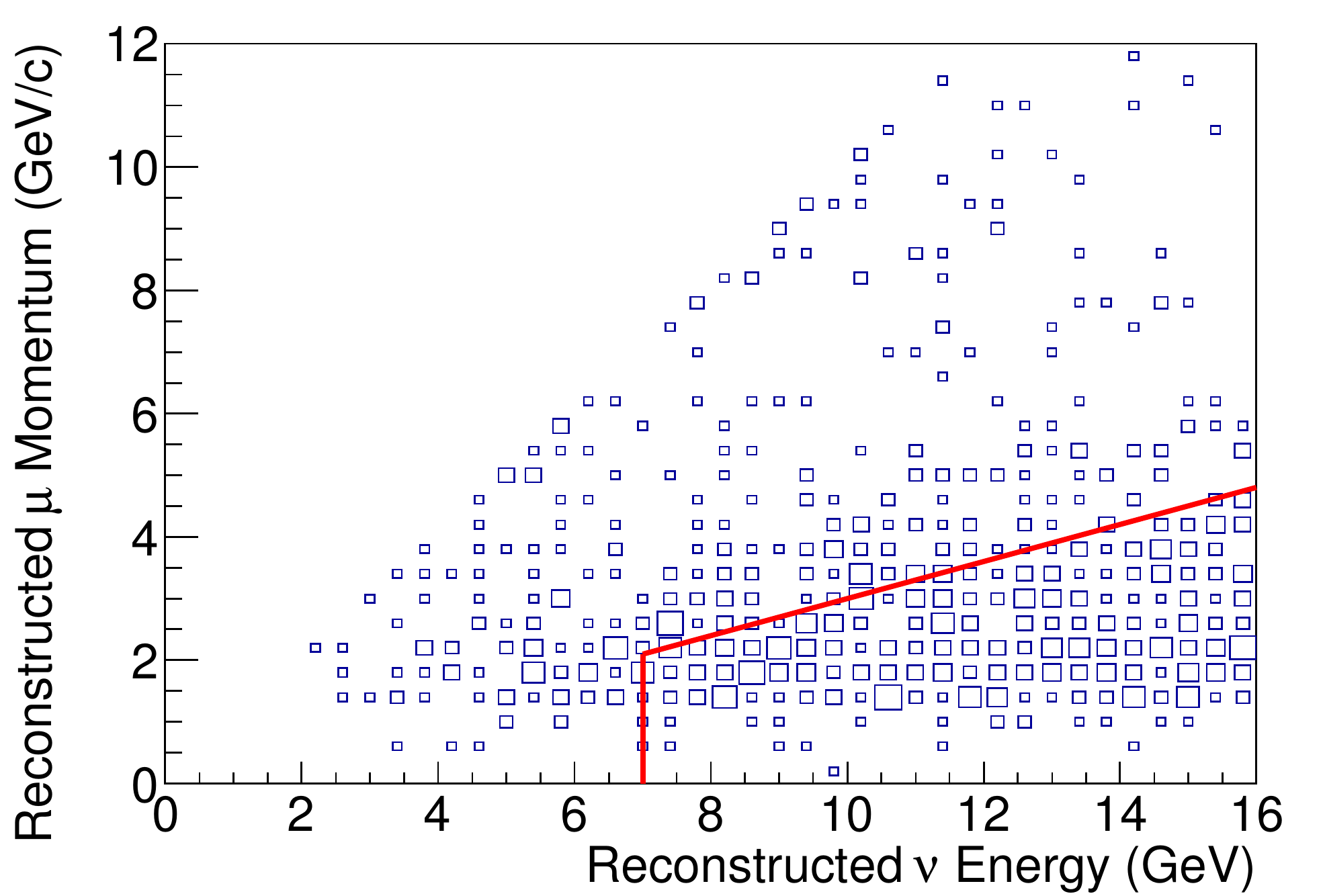} &
      \includegraphics[width=7.5cm]{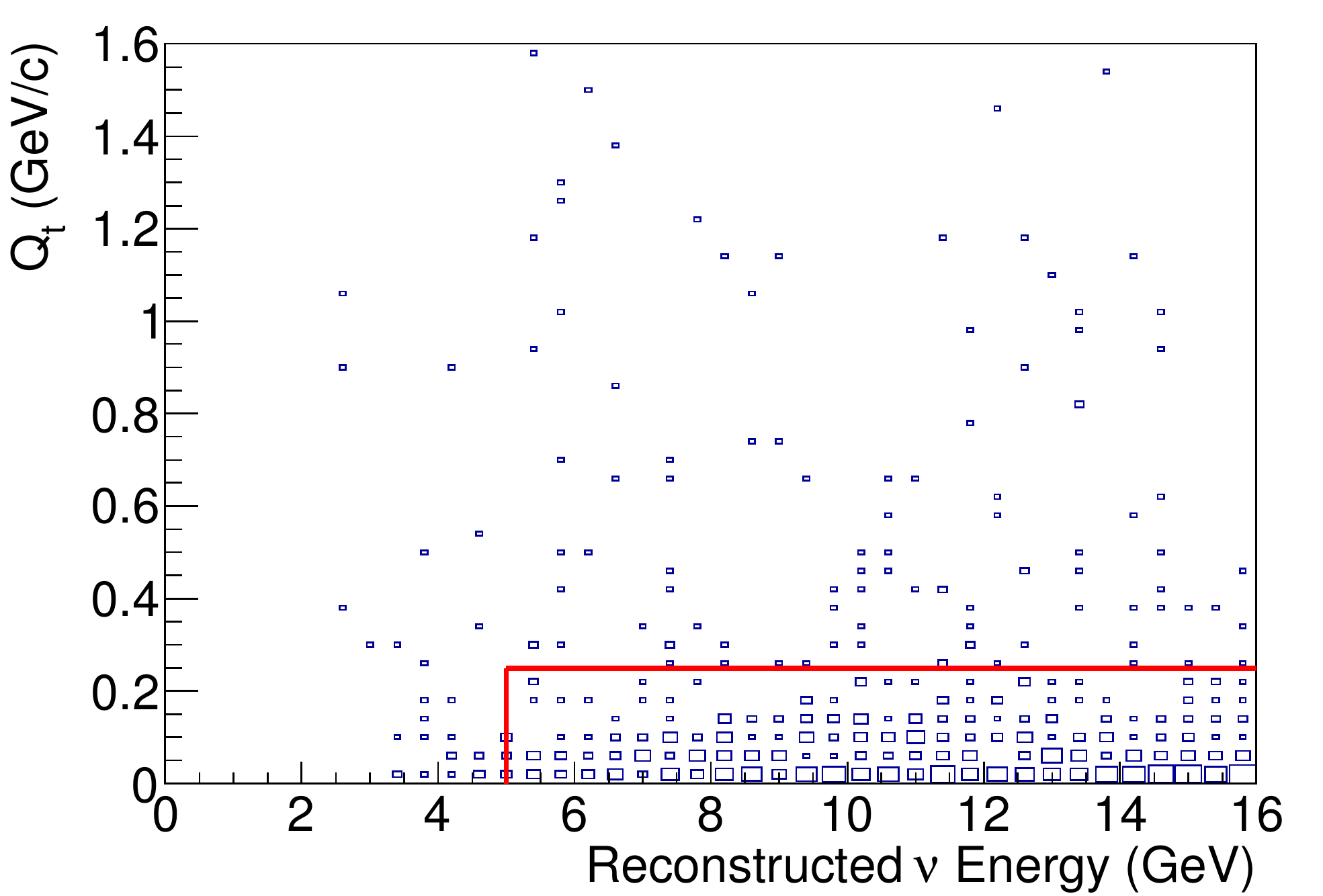} \\
      \includegraphics[width=7.5cm]{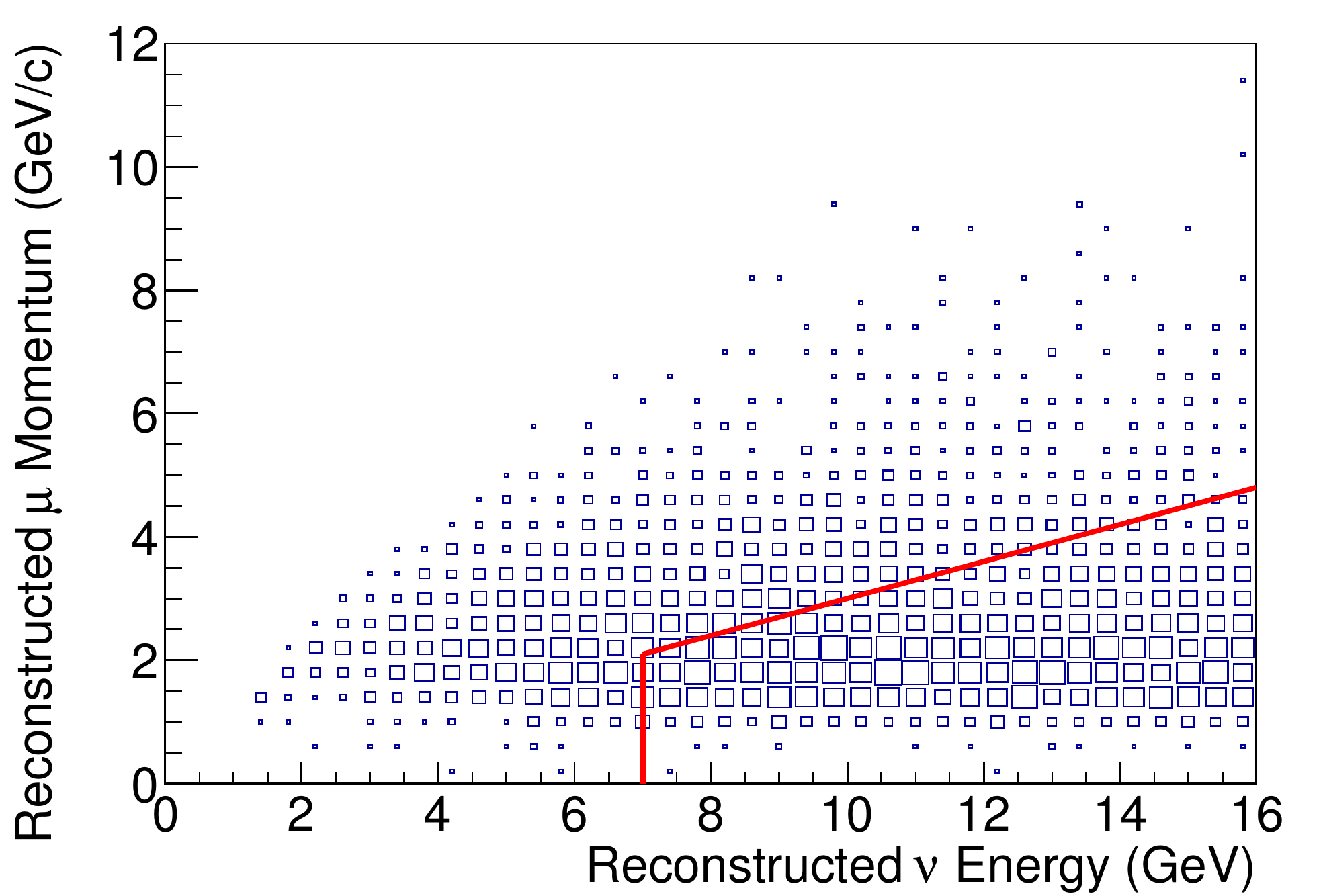} &
      \includegraphics[width=7.5cm]{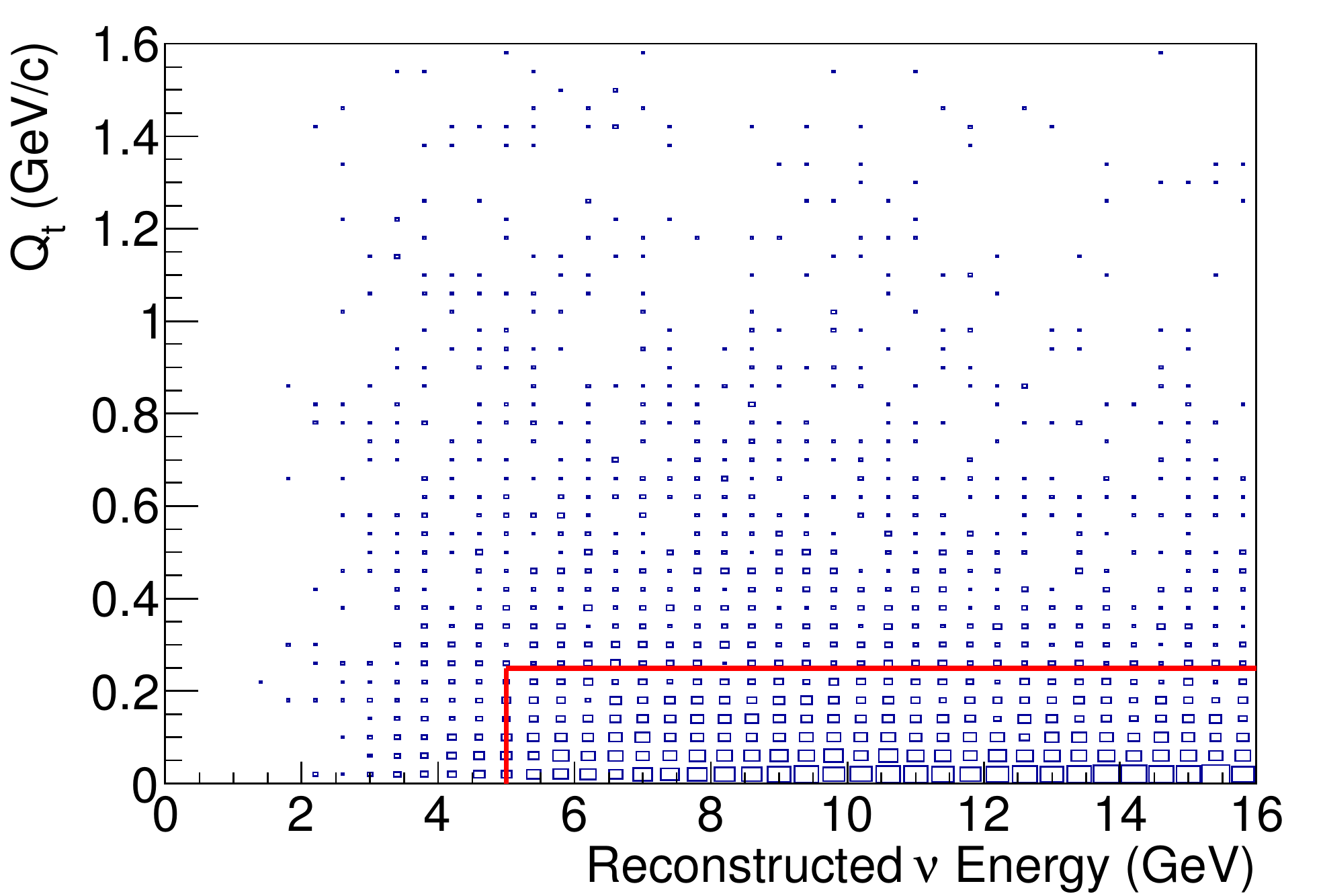} \\
      \includegraphics[width=7.5cm]{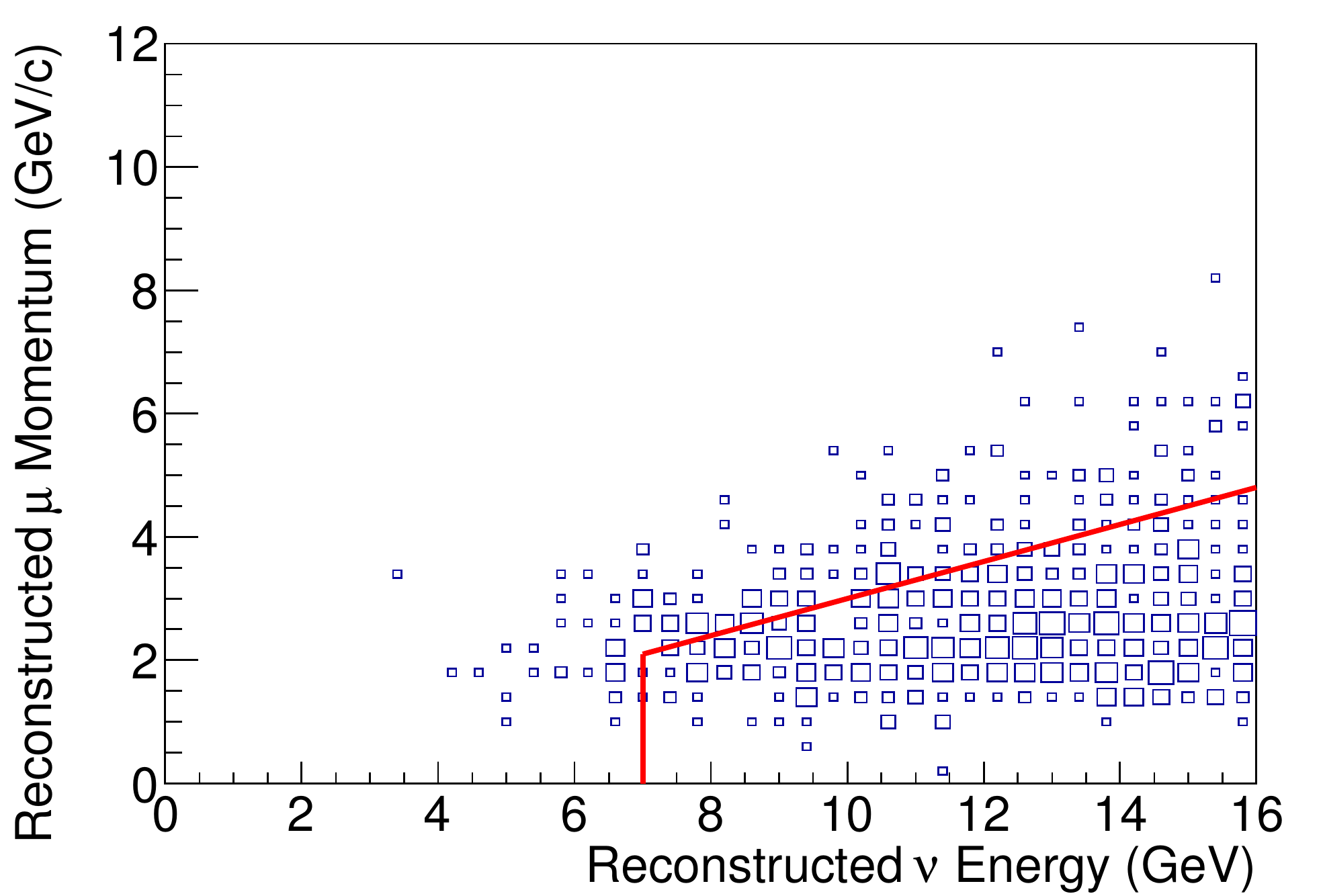} &
      \includegraphics[width=7.5cm]{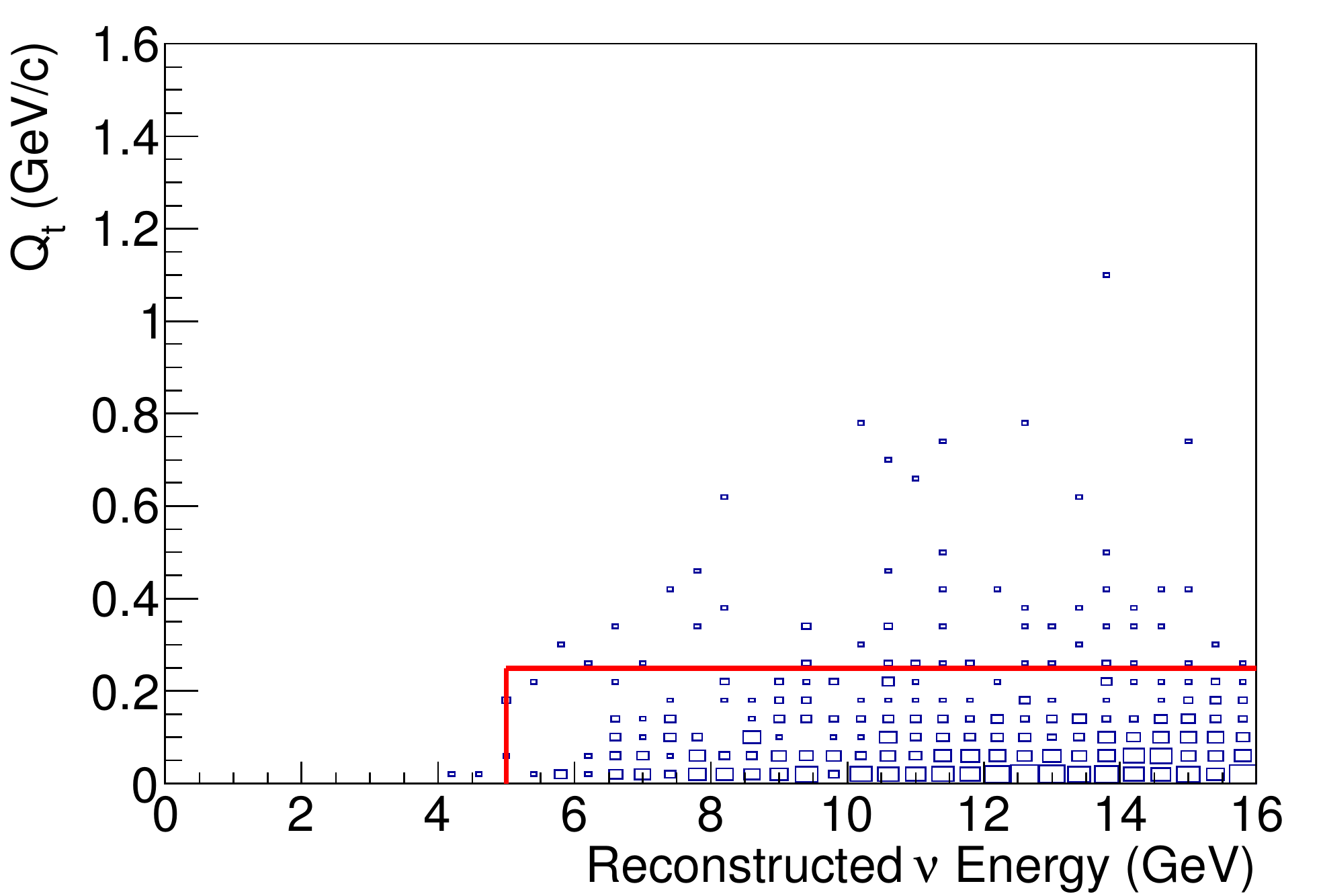} \\
    \end{array}$
  \end{center}
  \caption{Distributions of kinematic variables: (left) Reconstructed
    muon momentum with reconstructed neutrino energy for
    (top$\rightarrow$bottom) $\nu_\mu~(\overline{\nu}_\mu)$ signal,
    $\nu_\mu~(\overline{\nu}_\mu)$ CC background, NC background,
    $\nu_e~(\overline{\nu}_e)$ CC background and (right) $Q_t$
    variable (in the same order). The red lines represent the cuts
    from equations \ref{eq:G4kin1} and \ref{eq:G4kin2}. }
  \label{fig:G4kin}
\end{figure}

Kinematic cuts based on the momentum and isolation of the candidate,
in relation to the reconstructed energy of the event $E_{rec}$, can be
used to reduce backgrounds from hadron decays. The isolation of the
candidate muon is described by the variable $Q_t=p_\mu \sin^2\theta$,
where $\theta$ is the angle between the muon candidate and the
hadronic-jet vector. The muon from a true CC event is generally
isolated from the hadronic jet so, on average, the $Q_t$ is larger for
CC events than for NC events, in which a hadron associated with the
hadronic jet decays to a muon. Cuts based on this variable and on the
reconstructed momentum compared to the reconstructed energy are an
effective way to reduce all of the relevant beam related
backgrounds. The distributions after the application of the preceding
cuts are shown in figure \ref{fig:G4kin}, where the red lines
illustrate the acceptance conditions defined in
equations~\ref{eq:G4kin1} and~\ref{eq:G4kin2}:
\begin{eqnarray}
  \label{eq:G4kin1}
  E_{rec}~\leq~5\mbox{~GeV~~~or~~~} &Q_t~>~0.25\mbox{~GeV/c} \,\mbox{~~and~};\\
  \label{eq:G4kin2}
  E_{rec}~\leq~7\mbox{~GeV~~~or~~~} &p_\mu~\geq~0.3\cdot E_{rec} \, .
\end{eqnarray}
QE like events (see section~\ref{Sec:RecG4}) and those events passing
the conditions of equation~\ref{eq:G4kin1} must also pass the
conditions of equation~\ref{eq:G4kin2} to remain in the data-set for
the next series of cuts. The effect of these cuts is to reduce the
background by a further order of magnitude, while only having a modest
effect on the signal efficiency, as can be seen in
table~\ref{tab:G4cutSum}.

\subsection{Cut summary}
\label{subsec:Csumm}
In  summary, after tuning the cuts described in the previous sub-sections to
a test statistic, these were applied to independent simulated data
leading to an absolute efficiency of 51\% for $\nu_\mu$ selection and
62\% for $\overline{\nu}_\mu$ selection, while reducing the background
to a level below 10$^{-3}$. A summary of all the cuts, with their
effect on the signal and absolute background, can be found in
Table~\ref{tab:G4cutSum}. The species which would be
expected to contribute the greatest amount of background interactions
for an example oscillation parameter set is also identified at each level.

\begin{table}[ht]
\caption{Summary of cuts applied to select the golden channel
  appearance signals. The level of absolute efficiency and, for a
  100~ktonne MIND 2000~km from the NF and $\theta_{13} = 9.0^{\circ}$
  and $\delta_{CP} = 45^{\circ}$, the
  proportion of the total non-golden channel interactions remaining in
  the sample after each cut are also shown, along with the species
  contributing the greatest number of interactions.}
  \label{tab:G4cutSum}
  \begin{center}
    \begin{tabular}{|c|c|c|c|c|c|}
      \hline
      {\bf Cut} & {\bf Acceptance level} & \multicolumn{2}{c|}{{\bf
          Eff. after cut}}& \multicolumn{2}{c|}{{\bf background
          {\small ($\times 10^{-3}$)}}} \\
      \cline{3-6}
                &                        & $\nu_\mu$& $\overline{\nu}_\mu$& $\nu_\mu$ & $\overline{\nu}_\mu$ \\
      \hline
      \hline
      & successful pattern rec. and fit & 0.91 & 0.93 & 419 ($\nu_e$)
      & 153 ($\overline{\nu}_{\mu} NC)$\\
      \hline
      Fiducial & $z1-z_{end} \leq$ 2000~mm & 0.88 & 0.90 &
      400 ($\nu_e$) & 147 ($\overline{\nu}_{\mu} NC$)\\
      \hline
      Max. momentum & $P_\mu \leq 16$~GeV & 0.85 & 0.89 & 158
      ($\nu_e$) & 108 ($\overline{\nu}_e$)\\
      \hline
      Fitted proportion & $N_{fit}/N_h \geq 0.6$& 0.81 & 0.87 & 74.4
      ($\nu_e$) & 71.3 ($\bar{\nu}_e$)\\
      \hline
      Track quality & $\mathcal{L}_{q/p} > -0.5$ & 0.70 & 0.76 & 13.6
      ($\nu_e$) & 20.3 ($\nu_\tau$)\\
      \hline
      Displacement & $dispX/dispZ > 0.18 - 0.0026N_h$ & 0.65 & 0.72 &
      13.6 ($\nu_\mu$ NC) & 10.9 ($\nu_\tau$) \\
      & $dispZ > 6000~mm \mbox{ or } P_\mu \leq 3dispZ$ &  &  &  & \\     
      \hline
      Quadratic fit & $qp_{par} < -1.0$ or $qp_{par} > 0.0$ & 0.65 &
      0.72 & 10.3 ($\nu_\mu$ NC) & 10.9 ($\nu_\tau$) \\
      \hline
      CC selection & $\mathcal{L}_1 > 1.0$ & 0.63 & 0.70 & 2.1
      ($\nu_\mu$ NC) & 3.0 ($\nu_\tau$)\\
      \hline
      Kinematic & $E_{rec} \leq 5~GeV \mbox{ or } Q_t > 0.25$ & 0.51 &
      0.62 & 0.3 ($\bar{\nu}_\mu$ NC) & 0.9 ($\nu_\tau$)\\
      & $E_{rec} \leq 7~GeV \mbox{ or } P_\mu \geq 0.3E_{rec}$&  &  &  &  \\
      \hline
    \end{tabular}
  \end{center}
\end{table}

\section{MIND response to the Golden Channel}
\label{sec:response}
Using a data-set of $3\times 10^6$ events each of $\nu_\mu$ CC,
$\overline{\nu}_\mu$ CC, $\nu_e$ CC, $\overline{\nu}_e$ CC and
$7\times 10^6$ NC interactions from neutrinos and anti-neutrinos
generated using GENIE and tracked through the GEANT4 representation
of MIND, the expected efficiency and background suppression for the
reconstruction and analysis of the golden channel in MIND have been
evaluated for both $\nu_\mu$ and $\overline{\nu}_\mu$
appearance. Additionally, the expected level of contamination of the
signal from other appearance oscillation channels is considered.

\subsection{Signal efficiency and beam neutrino background suppression}
\label{subsec:efficiency}
The resultant efficiencies for both polarities and the corresponding
background levels expected for the appearance channels are summarised
in Figs.~\ref{fig:CCback}~--~\ref{fig:G4Eff}. Numeric response
matrices for each of the channels may be found in
the Appendix.  As can be seen in figure
\ref{fig:CCback} the expected level of background from CC
mis-identification is around $10^{-4}$, which is significantly below
$10^{-3}$ at all energies for the new simulation and re-optimised
analysis. This is also below the background levels achieved in
\cite{Cervera:2010rz}, mainly due to the additional quality cuts.
\begin{figure}
  \begin{center}$
    \begin{array}{cc}
      \includegraphics[width=8cm, height=6cm]{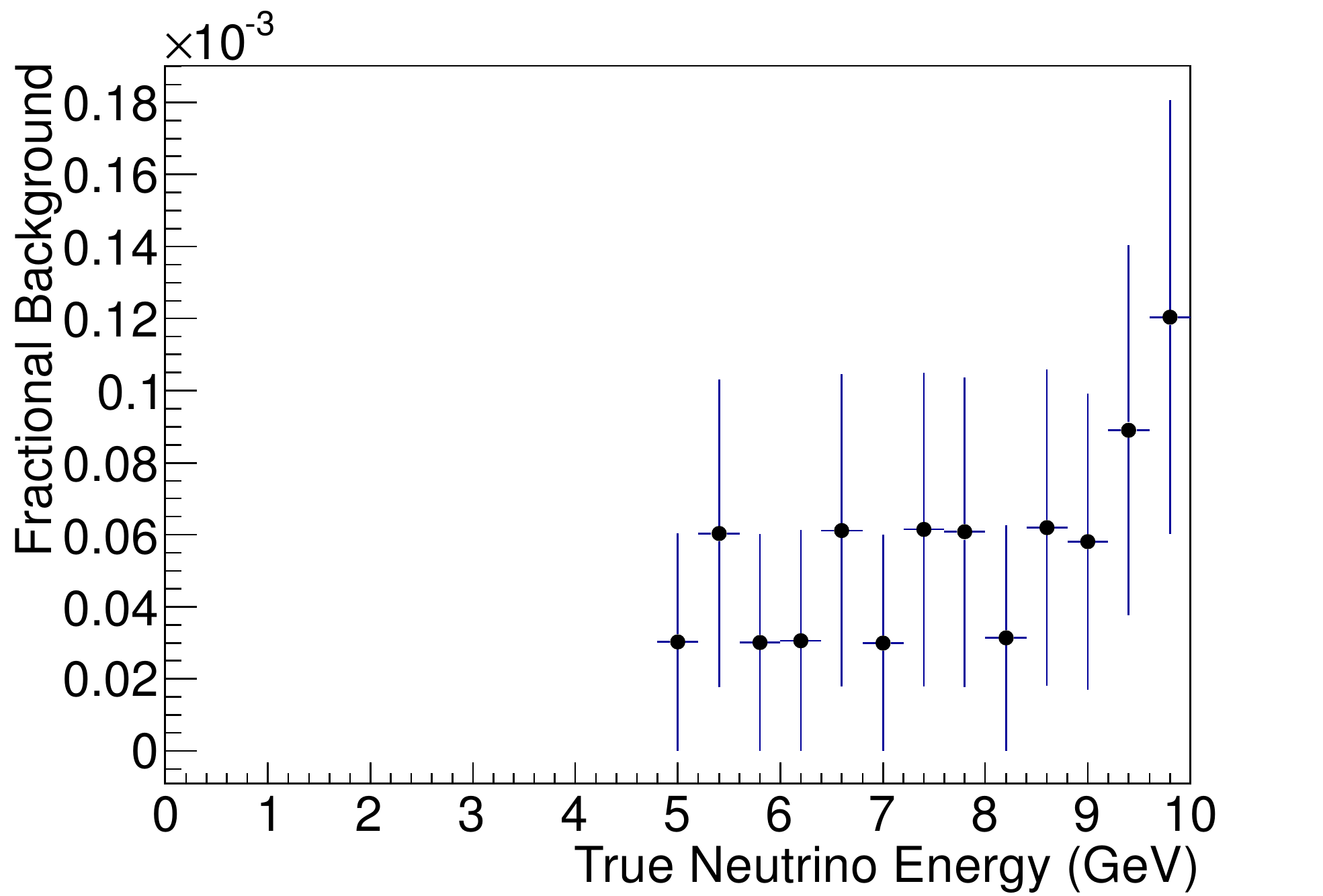} &
      \includegraphics[width=8cm, height=6cm]{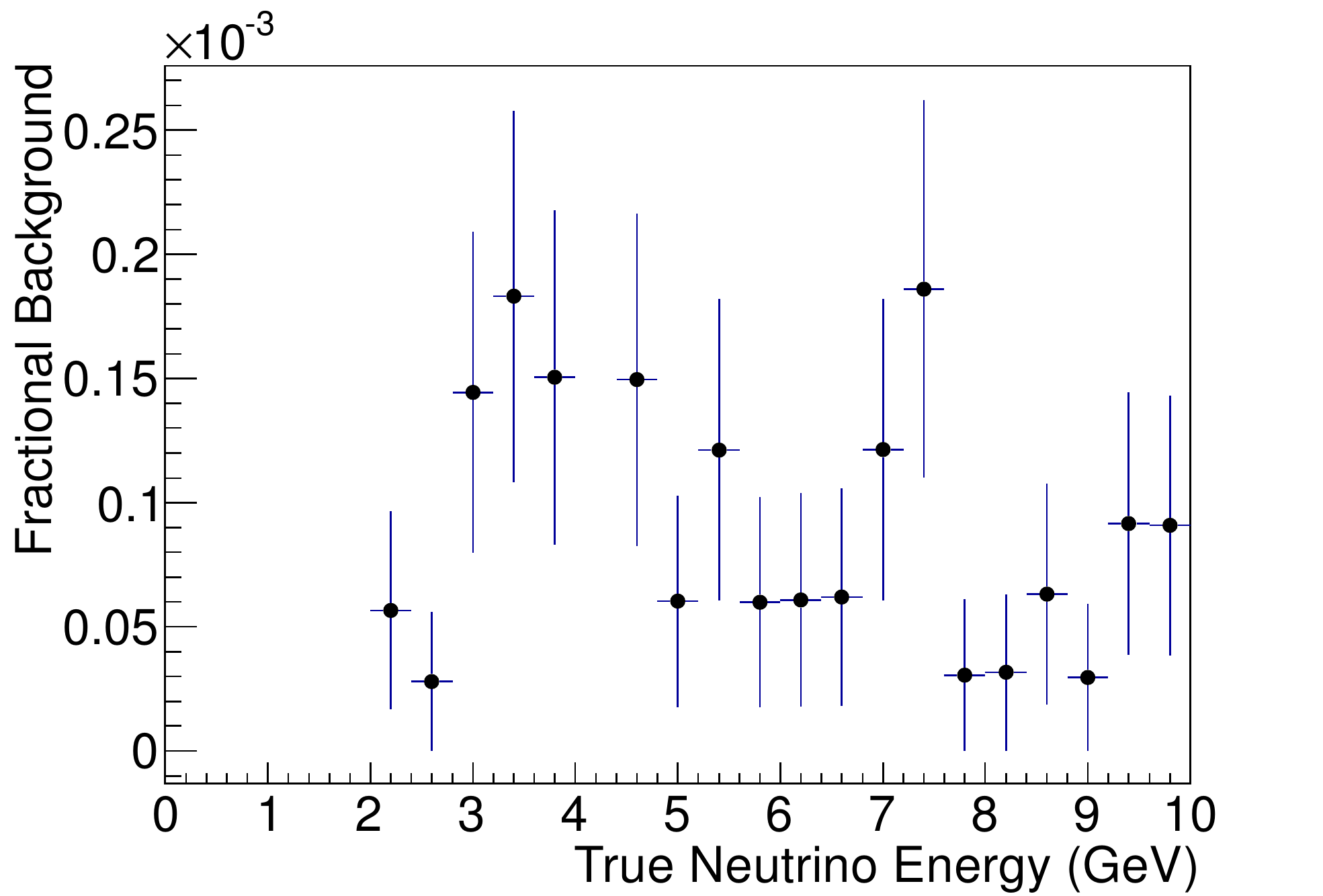} \\
    \end{array}$
  \end{center}
  \caption{Background from mis-identification of
    $\nu_\mu~(\overline{\nu}_\mu)$ CC interactions as the opposite
    polarity. (left) $\overline{\nu}_\mu$ CC reconstructed as
    $\nu_\mu$ CC, (right) $\nu_\mu$ CC reconstructed as
    $\overline{\nu}_\mu$ CC as a function of true energy.}
  \label{fig:CCback}
\end{figure}

The background from neutral current interactions lies at or below
the $10^{-3}$ level, with the high energy region exhibiting a higher
level than the low-energy region due to the dominance of DIS
interactions. The increased particle multiplicity and greater
likelihood of producing a penetrating pion that can mimic a primary
muon are the primary reasons for this increase. As expected, the NC
background tends to be reconstructed at low energy due to the missing
energy (see appendix~\ref{app:response}).
\begin{figure}
  \begin{center}$
    \begin{array}{cc}
      \includegraphics[width=7.5cm, height=5.5cm]{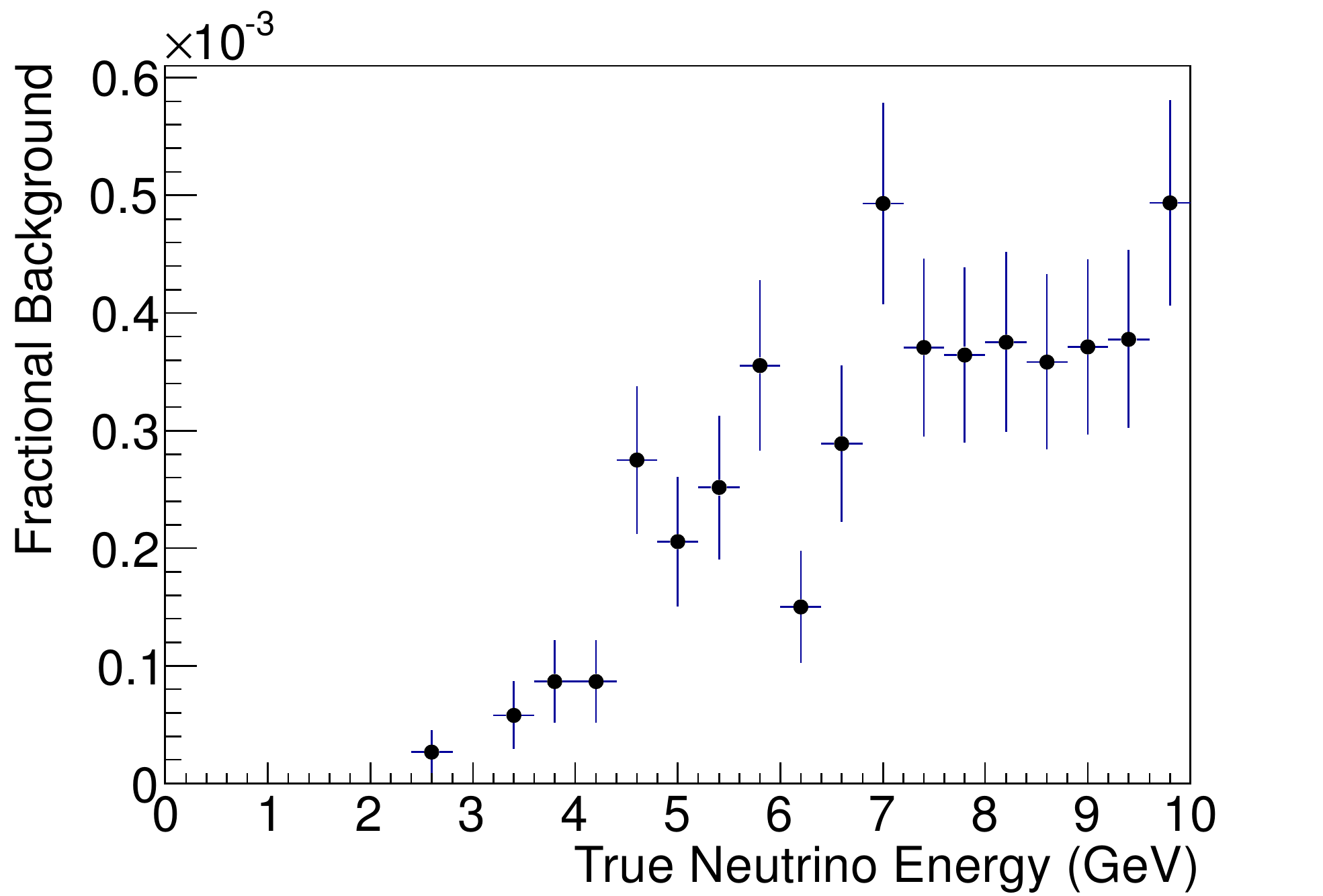} &
      \includegraphics[width=7.5cm, height=5.5cm]{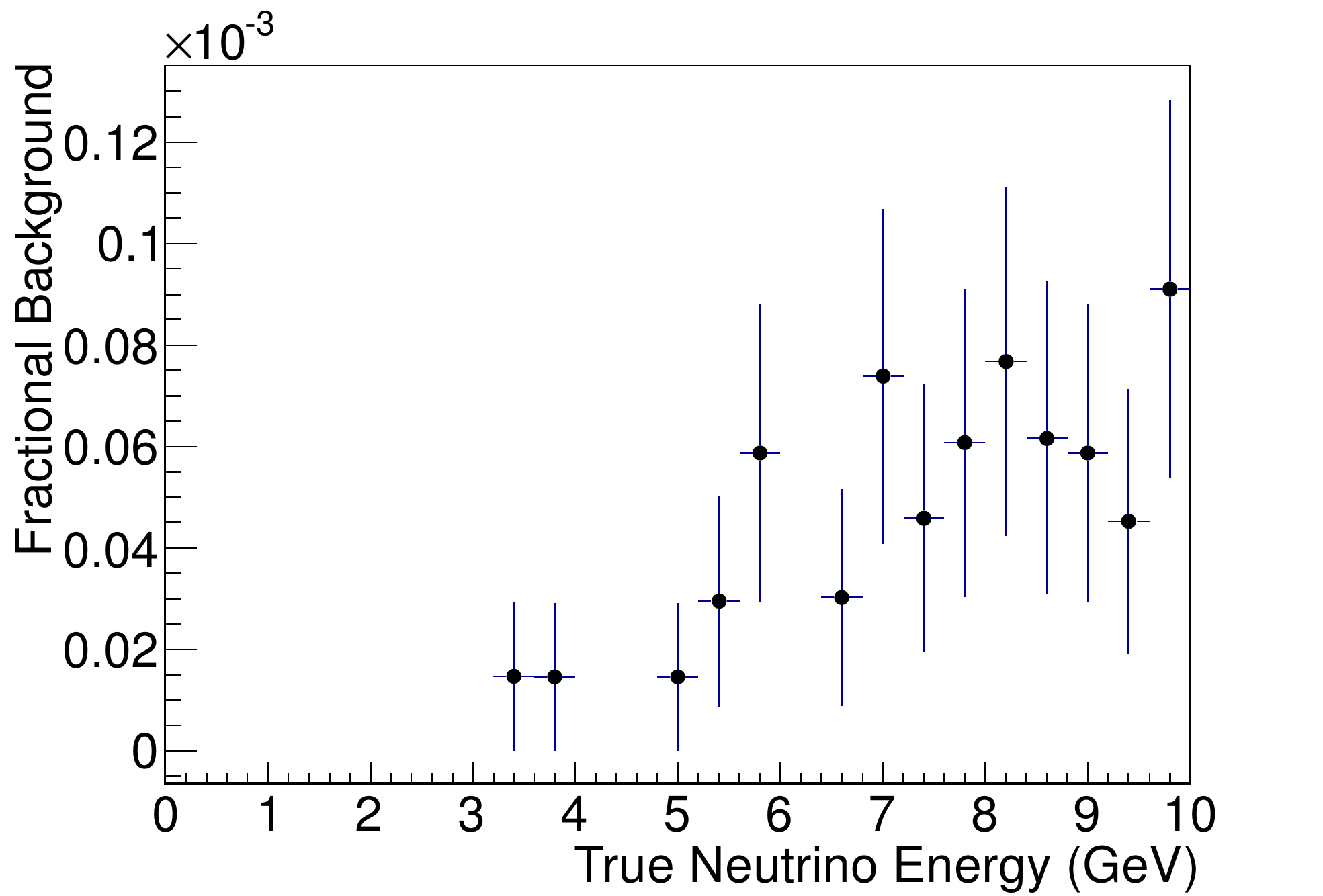} \\
    \end{array}$
  \end{center}
  \caption{Background from mis-identification of NC interactions as
    $\nu_\mu~(\overline{\nu}_\mu)$ CC interactions. (left) NC
    reconstructed as $\nu_\mu$ CC, (right) NC reconstructed as
    $\overline{\nu}_\mu$ CC as a function of true energy.}
  \label{fig:NCback}
\end{figure}

The background from $\nu_e~(\overline{\nu}_e)$ CC interactions is once
again expected to constitute a very low level addition to the observed
signal. This background is particularly well suppressed due to the
electron shower overlapping with the hadron shower. If a particle from
the hadronic jet decays into a wrong-sign muon, it has a lower energy
and is less isolated than in the NC case, so the kinematic cuts
suppress this background more.
\begin{figure}
  \begin{center}$
    \begin{array}{cc}
      \includegraphics[width=7.5cm, height=5.5cm]{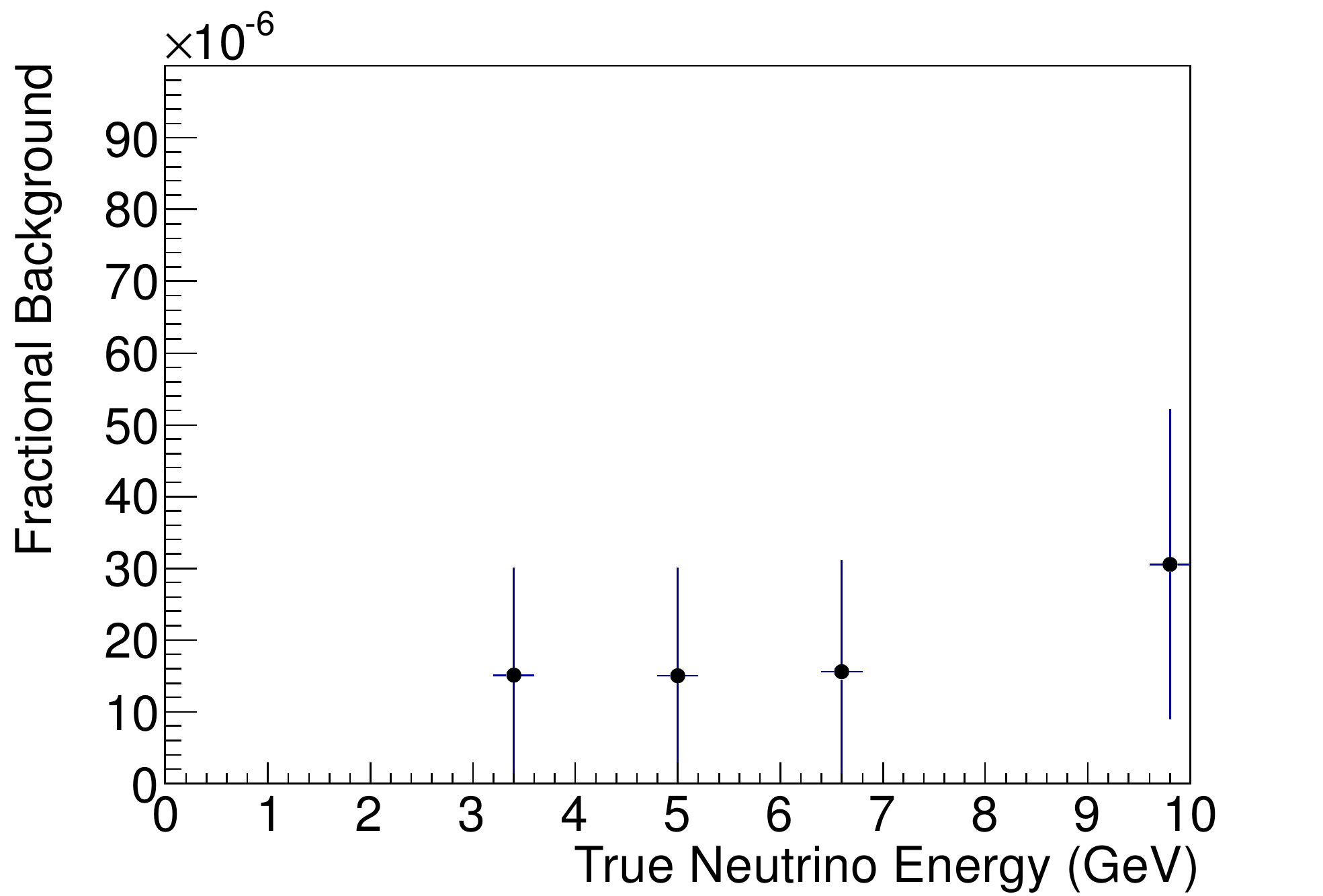} &
      \includegraphics[width=7.5cm, height=5.5cm]{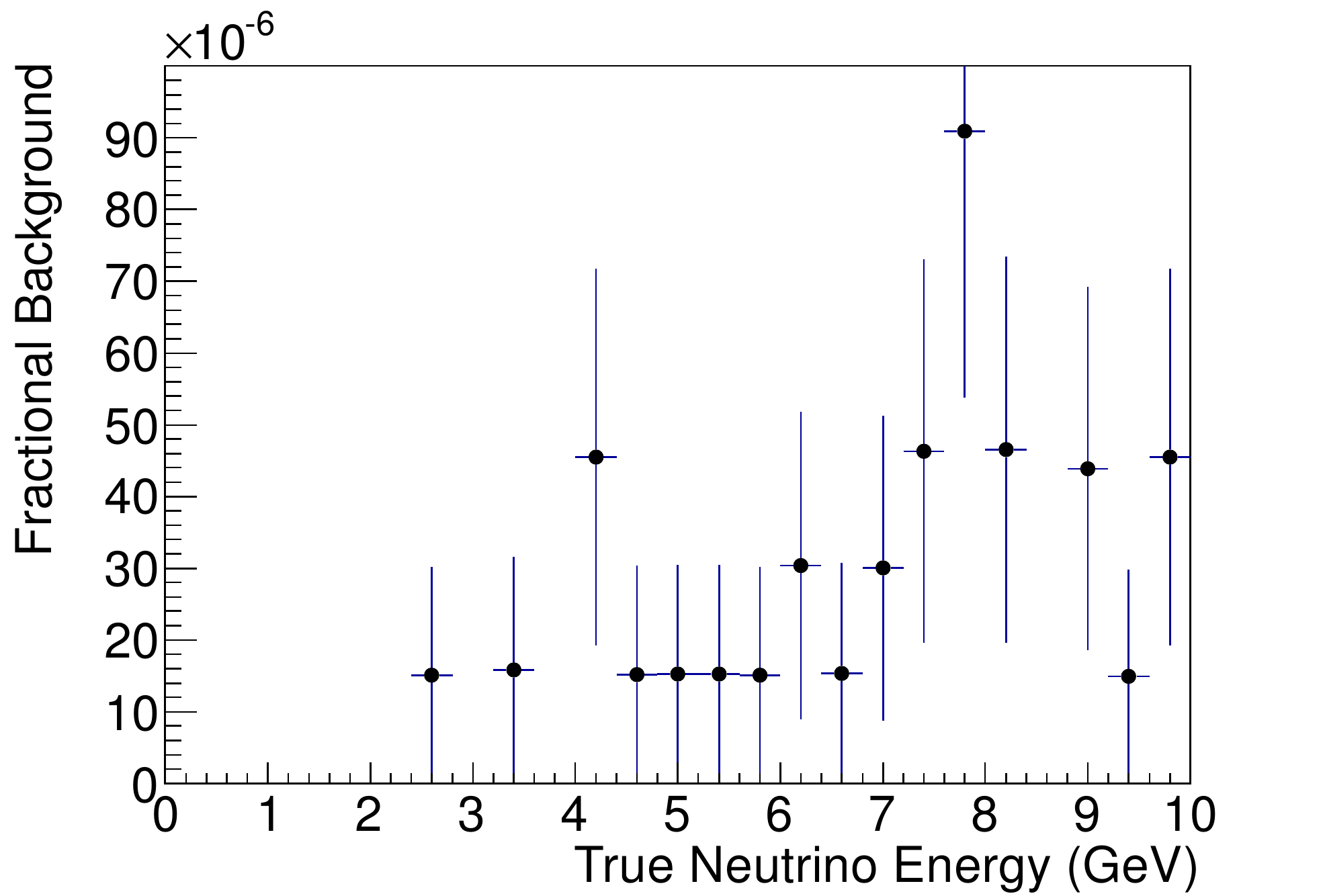} \\
    \end{array}$
  \end{center}
  \caption{Background from mis-identification of
    $\nu_e~(\overline{\nu}_e)$ CC interactions as
    $\nu_\mu~(\overline{\nu}_\mu)$ CC interactions. (left) $\nu_e$ CC
    reconstructed as $\nu_\mu$ CC, (right) $\overline{\nu}_e$ CC
    reconstructed as $\overline{\nu}_\mu$ CC as a function of true
    energy.}
  \label{fig:eCback}
\end{figure}

The efficiency of detection of the two $\nu_\mu$ polarities has
a threshold lower than that seen in previous studies
due to the presence of non-DIS interactions in the Monte Carlo
sample. The efficiencies expected for the current analysis are shown
in figure \ref{fig:G4Eff}. 
\begin{figure}
  \begin{center}$
    \begin{array}{cc}
      \includegraphics[width=7.5cm, height=5.5cm]{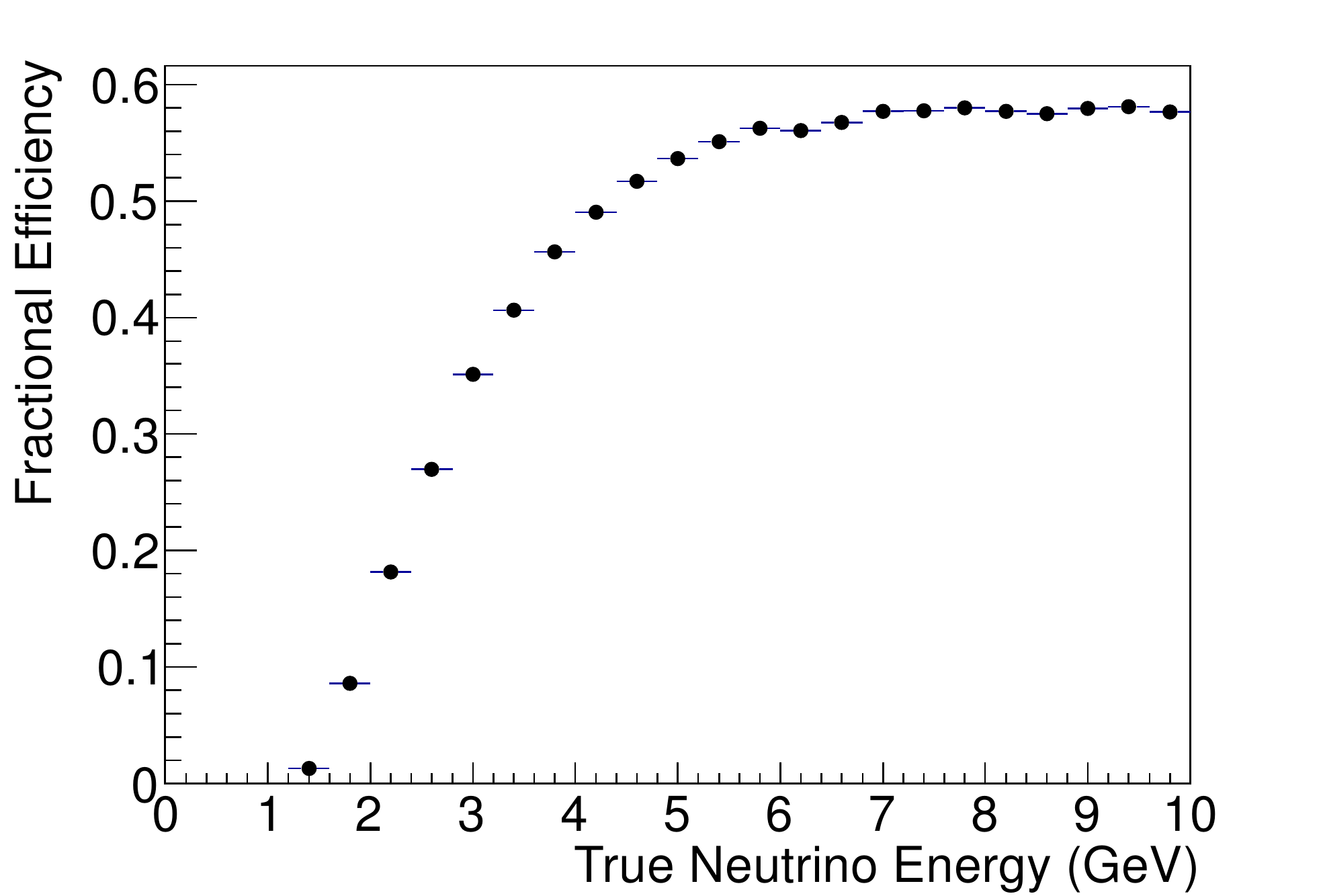} &
      \includegraphics[width=7.5cm, height=5.5cm]{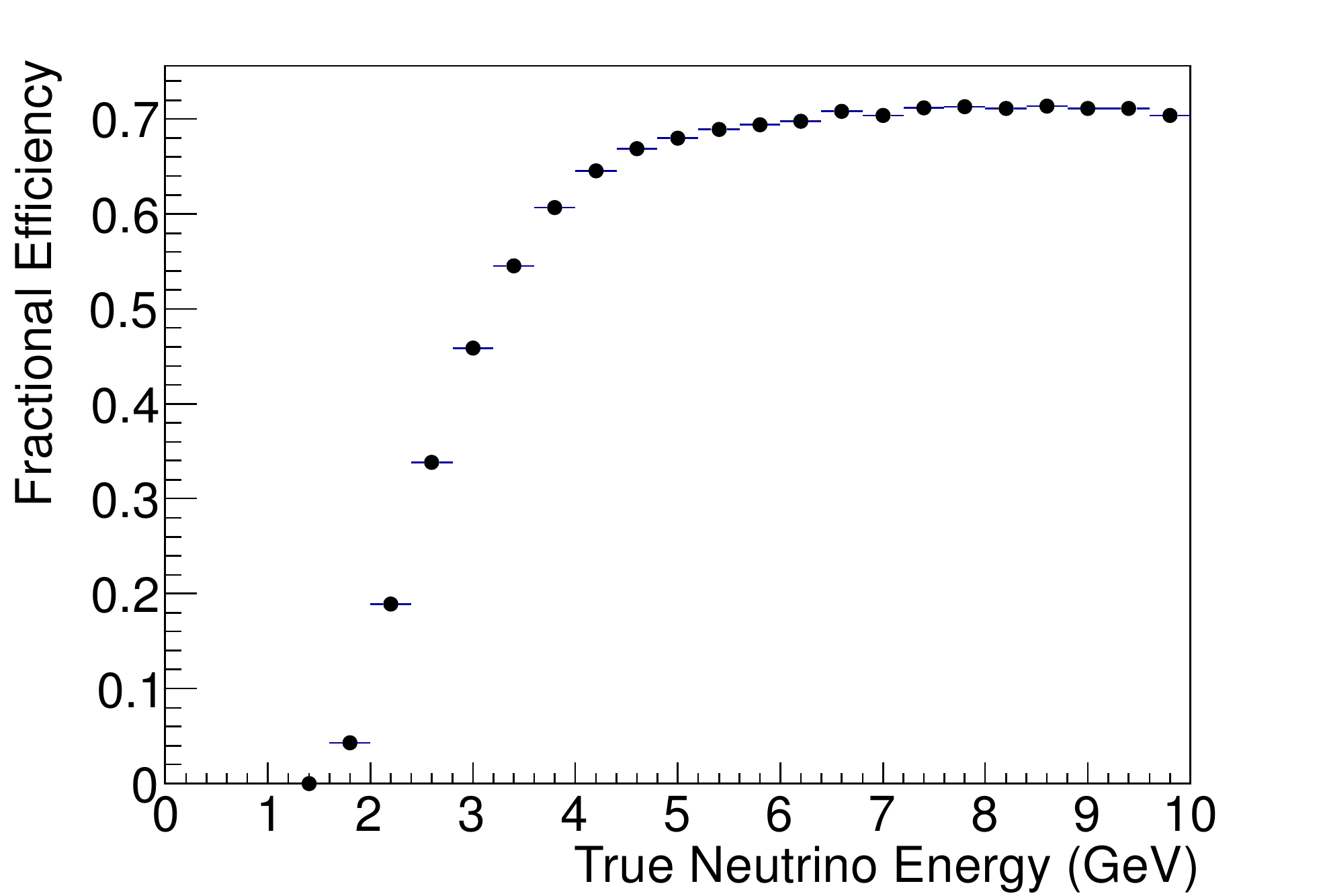} \\
    \end{array}$
  \end{center}
  \caption{Efficiency of reconstruction of $\nu_\mu~(\overline{\nu}_\mu)$ CC interactions. (left) $\nu_\mu$ CC efficiency, (right) $\overline{\nu}_\mu$ CC efficiency as a function of true energy.}
  \label{fig:G4Eff}
\end{figure}

A comparison of the resultant
$\nu_\mu$ and $\overline{\nu}_\mu$ efficiency can be made with that extracted in
previous studies.
Analyses performed in 2000 \cite{Cervera:2000kp} and 2005 \cite{CerveraVillanueva:2005ym}
assumed a 50~GeV Neutrino Factory, so were optimised
for high energy and low values of $\theta_{13}$. The background
rejection achieved was at the level of 10$^{-6}$, but at the expense
of signal efficiency, especially below 10~GeV. A more recent analysis
\cite{CerveraVillanueva:2008zz} was the first attempt at re-optimzing
for a 25~GeV Neutrino Factory, while that in 2010~\cite{Cervera:2010rz} 
was still based on a GEANT3 model, but included full event reconstruction and a
likelihood based analysis for the first time. There exists an improvement in
threshold for the current analysis, between 2--3~GeV, due to the
inclusion of QE and resonance events, since these events are easier to
reconstruct (see figure~\ref{fig:DISonly}).

\begin{figure}
  \begin{center}$
    \begin{array}{cc}
      \includegraphics[width=7.5cm, height=5.5cm]{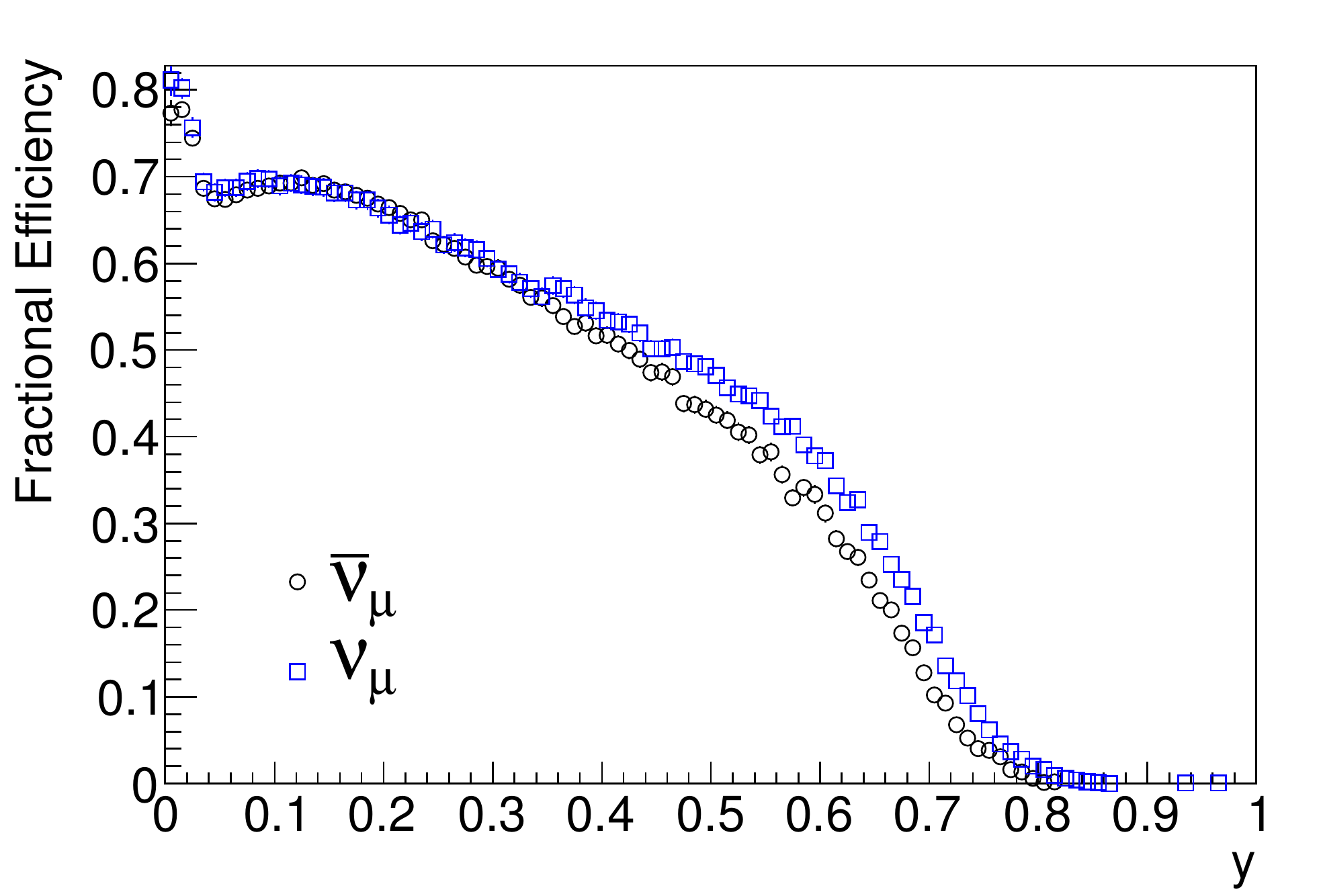} &
      \includegraphics[width=7.5cm, height=5.5cm]{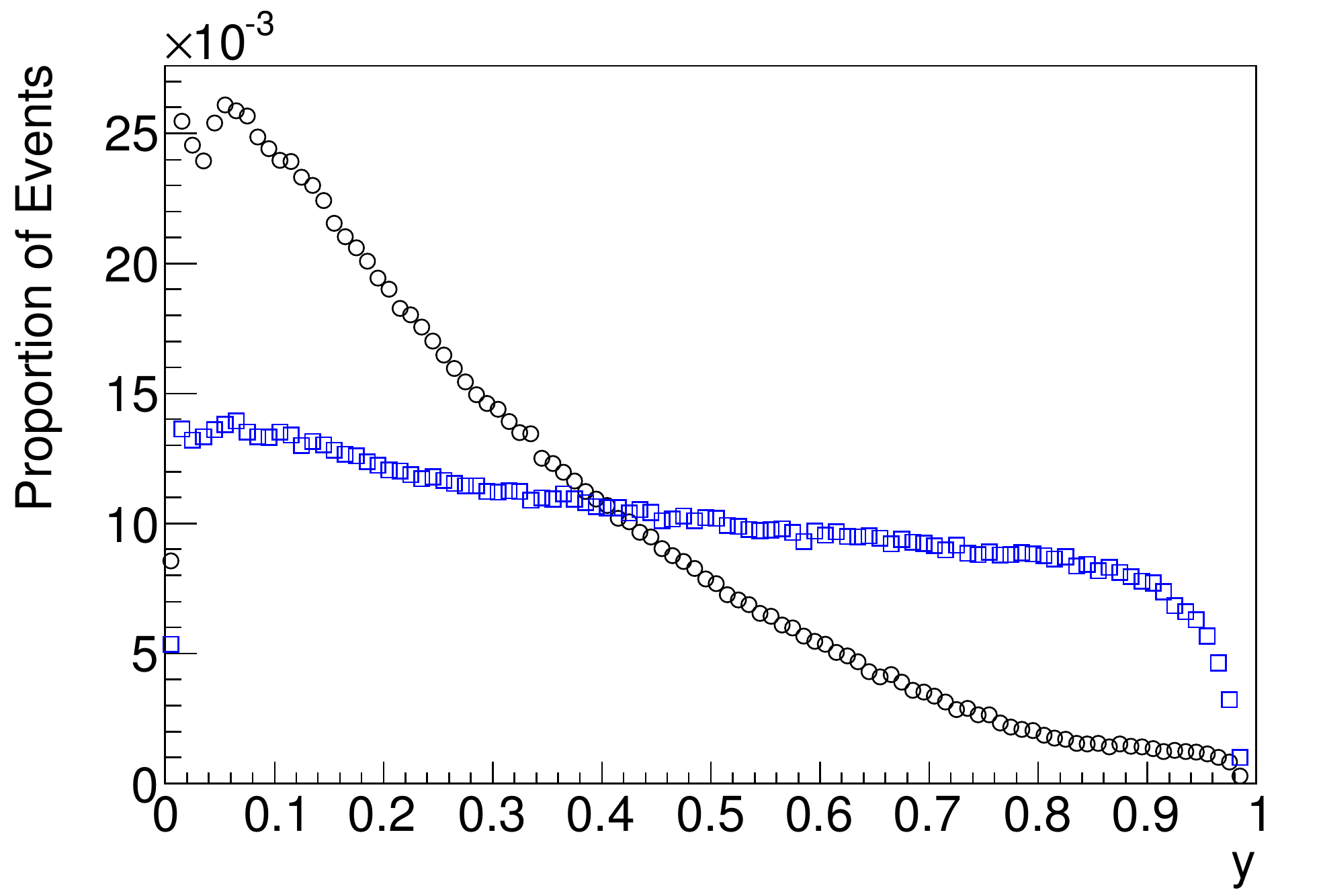}\\
    \end{array}$
  \end{center}
  \caption{$\nu_\mu$ CC and $\overline{\nu}_\mu$ CC signal detection
    efficiency as a function of $y$ (left) and the normalised
    distribution of all events considered in each polarity as a
    function of $y$ (right).}
  \label{fig:Effbjorken}
\end{figure}

The difference in efficiency between the two polarities is effectively
described by the difference in the inelasticity of neutrino and
anti-neutrino CC interactions. Neutrino DIS interactions with quarks
have a flat distribution in the Bjorken variable:
\begin{equation}
  y = \displaystyle\frac{E_\nu - E_l}{E_\nu} \, ,
\end{equation}
with $E_l$ the scattered-lepton energy. However, anti-neutrinos
interacting with quarks follow a $\propto~(1-y)^2$
distribution~\cite{Zuber:2004nz}. For this reason, neutrino
interactions generally involve a greater energy transfer to the
target. The efficiencies for the two species as a function of $y$ can
be seen from figure \ref{fig:Effbjorken}-(left). The shape of the efficiency
curves is a consequence of the ratio of DIS to non-DIS events in the
event samples. Neutrino and anti-neutrino efficiencies are very similar,
showing that the Bjorken $y$ of each event is the dominant contributor to the
efficiency. The difference in neutrino and anti-neutrino efficiencies, when translated 
into true neutrino-energy, can be explained by the greater abundance of neutrino
events at high $y$. However, since the cross section for the
interaction of neutrinos is approximately twice that for
anti-neutrinos, it is not expected that this reduced efficiency will
affect the fit to the observed spectrum significantly.
 
\subsection{Contamination from oscillation channels containing $\nu_\tau$ or $\overline{\nu}_\tau$}
\label{subsec:tau}
\begin{figure}[!ht]
  \begin{center}$
    \begin{array}{cc}
      \includegraphics[width=7.5cm, height=5.5cm]{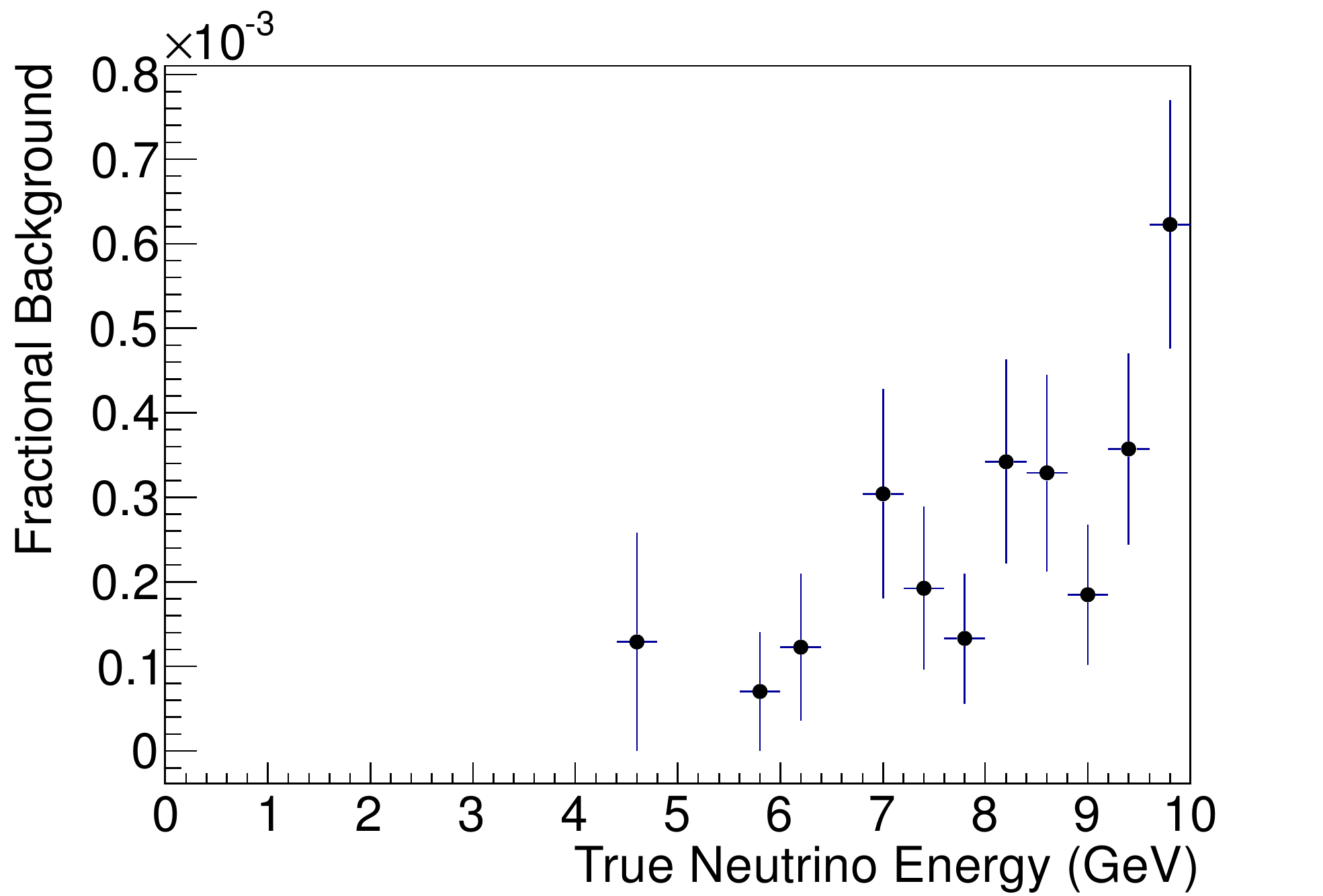}
      \includegraphics[width=7.5cm, height=5.5cm]{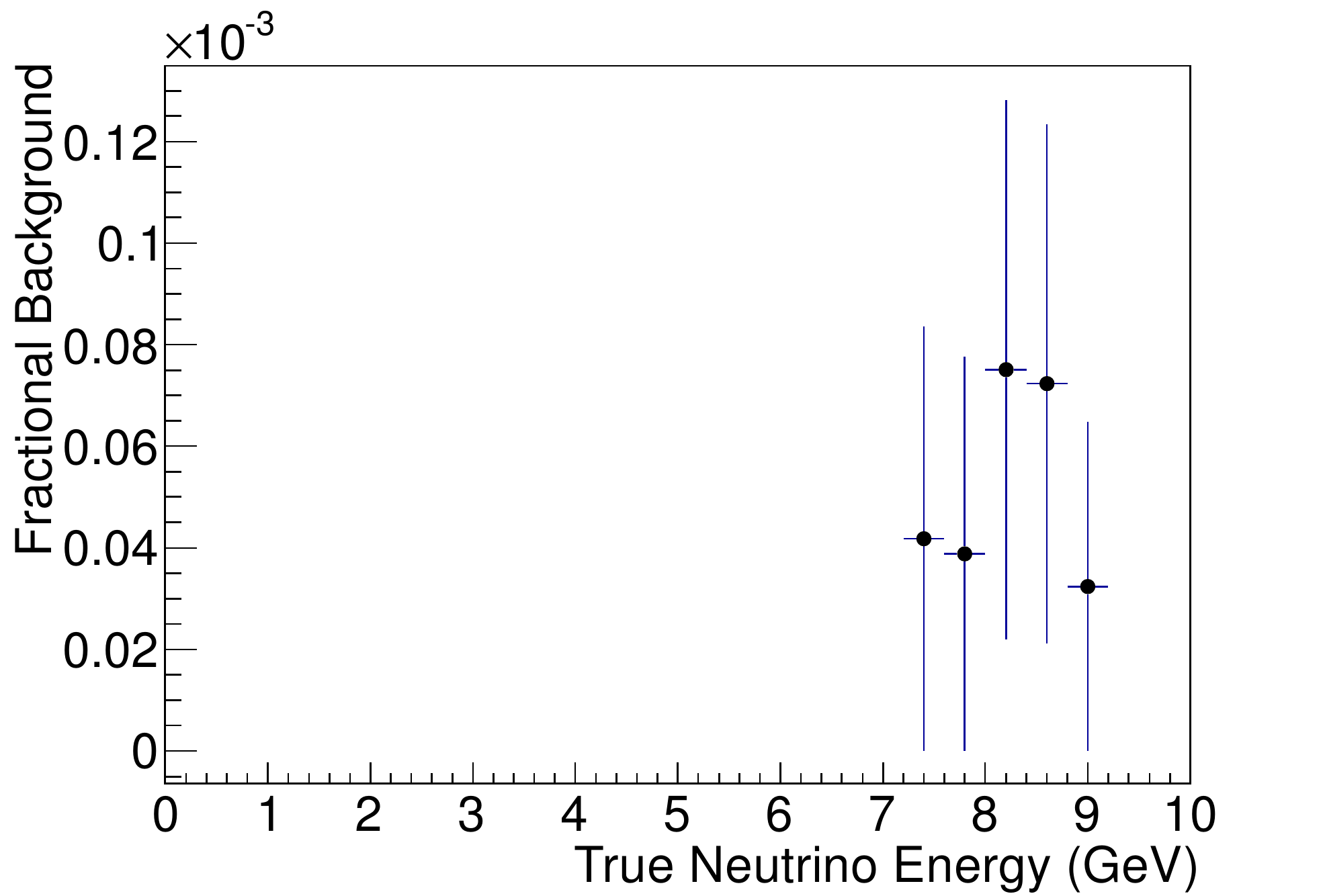}
    \end{array}$
  \end{center}
  \caption{Expected level of contamination from
      $\nu_\tau~(\overline{\nu}_\tau)$ CC interactions due to the
      dominant oscillation. (left) $\overline{\nu}_\tau$ CC
      reconstructed as $\nu_\mu$ CC, (right) $\nu_\tau$ CC
      reconstructed as $\overline{\nu}_\mu$ CC as a function of true
      energy.}
  \label{fig:tau2}
\end{figure}

Three million events of both $\nu_\tau$ and $\overline{\nu}_\tau$
interactions were generated using the GENIE
framework~\cite{Andreopoulos:2009rq} and passed through the GEANT4
simulation of MIND. These events were then subject to the same
digitization, reconstruction and analysis as the main beam
backgrounds. Matrices were extracted describing the expected level of
contamination in the golden channel data-set for the situation when a
viable muon candidate from a $\nu_\tau~(\overline{\nu}_\tau)$ interaction is 
reconstructed as a $\nu_\mu~(\overline{\nu}_\mu)$ candidate with the same 
and opposite charge to the true primary $\tau$.
\begin{figure}[!ht]
  \begin{center}$
    \begin{array}{cc}
      \includegraphics[width=7.5cm, height=5.5cm]{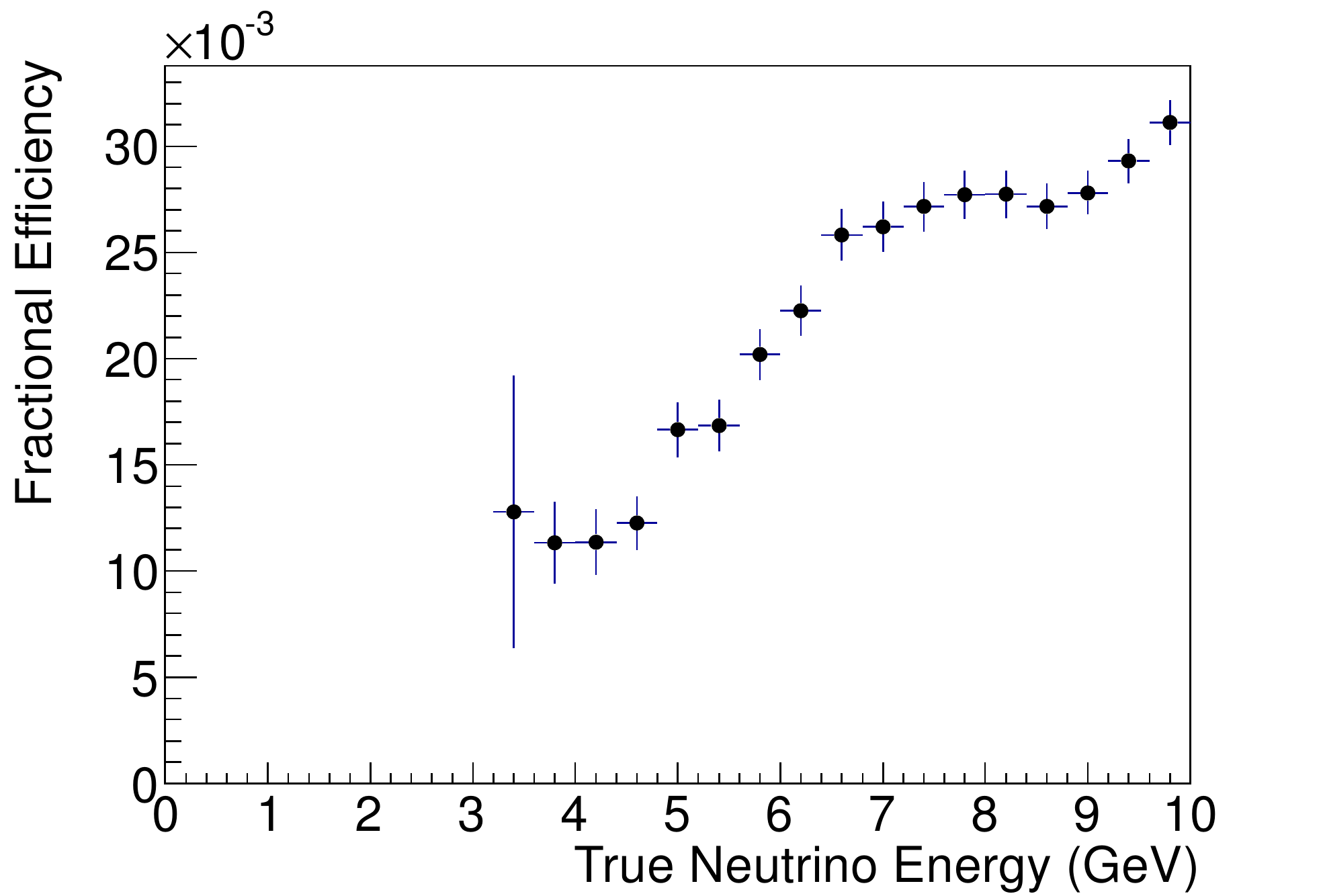}
      \includegraphics[width=7.5cm, height=5.5cm]{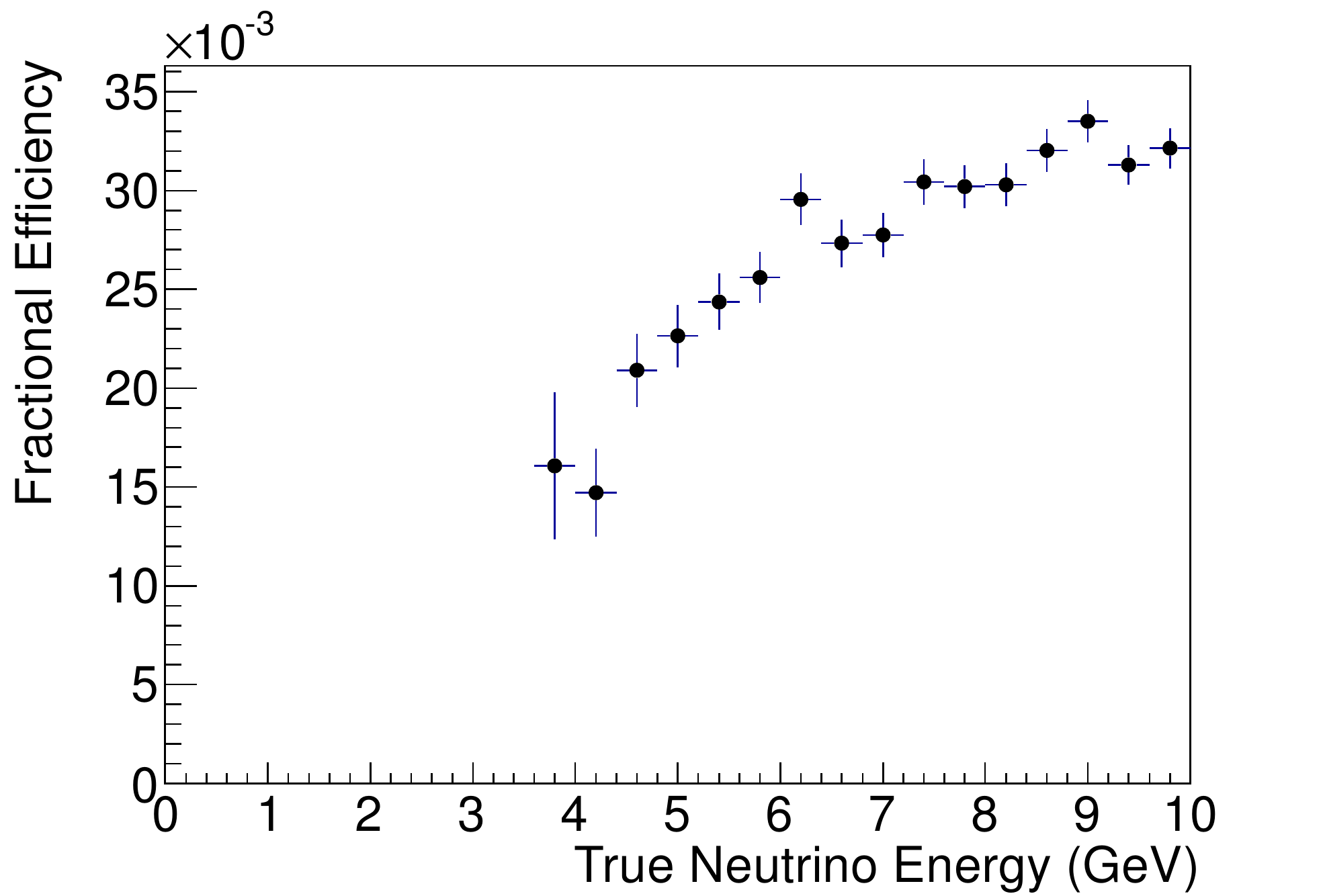}
    \end{array}$
  \end{center}
  \caption{Expected level of contamination from
      $\nu_\tau~(\overline{\nu}_\tau)$ CC interactions due to the
      platinum channel. (left) $\nu_\tau$ CC reconstructed as
      $\nu_\mu$ CC, (right) $\overline{\nu}_\tau$ CC reconstructed as
      $\overline{\nu}_\mu$ CC as a function of true energy.}
  \label{fig:tau1}
\end{figure}
As can be seen in figure~\ref{fig:tau1}, between 1\% and 3\% of the
$\nu_\tau~(\overline{\nu}_\tau)$ interactions are expected to be
identified as the golden $\nu_\mu~(\overline{\nu}_\mu)$
interactions. Considered properly, this contamination should not weaken
the extraction of the oscillation parameters
(see~\cite{Donini:2010xk}). Contamination from the dominant
oscillation (which requires reconstruction with the opposite primary
lepton charge) is expected to be below the $10^{-3}$ level (as shown
in figure~\ref{fig:tau2}). This contamination is taken into account, but does not deteriorate
the $\delta_{CP}$ fits, since the dominant oscillation is less sensitive to this parameter for 
large values of $\theta_{13}$.

\subsection{Interaction expectation for 10$^{21}$ muon decays\label{subsec:expect}}
\begin{figure}
  \begin{center}$
  \begin{array}{cc}
      \includegraphics[width=7.5cm, height=5.5cm]{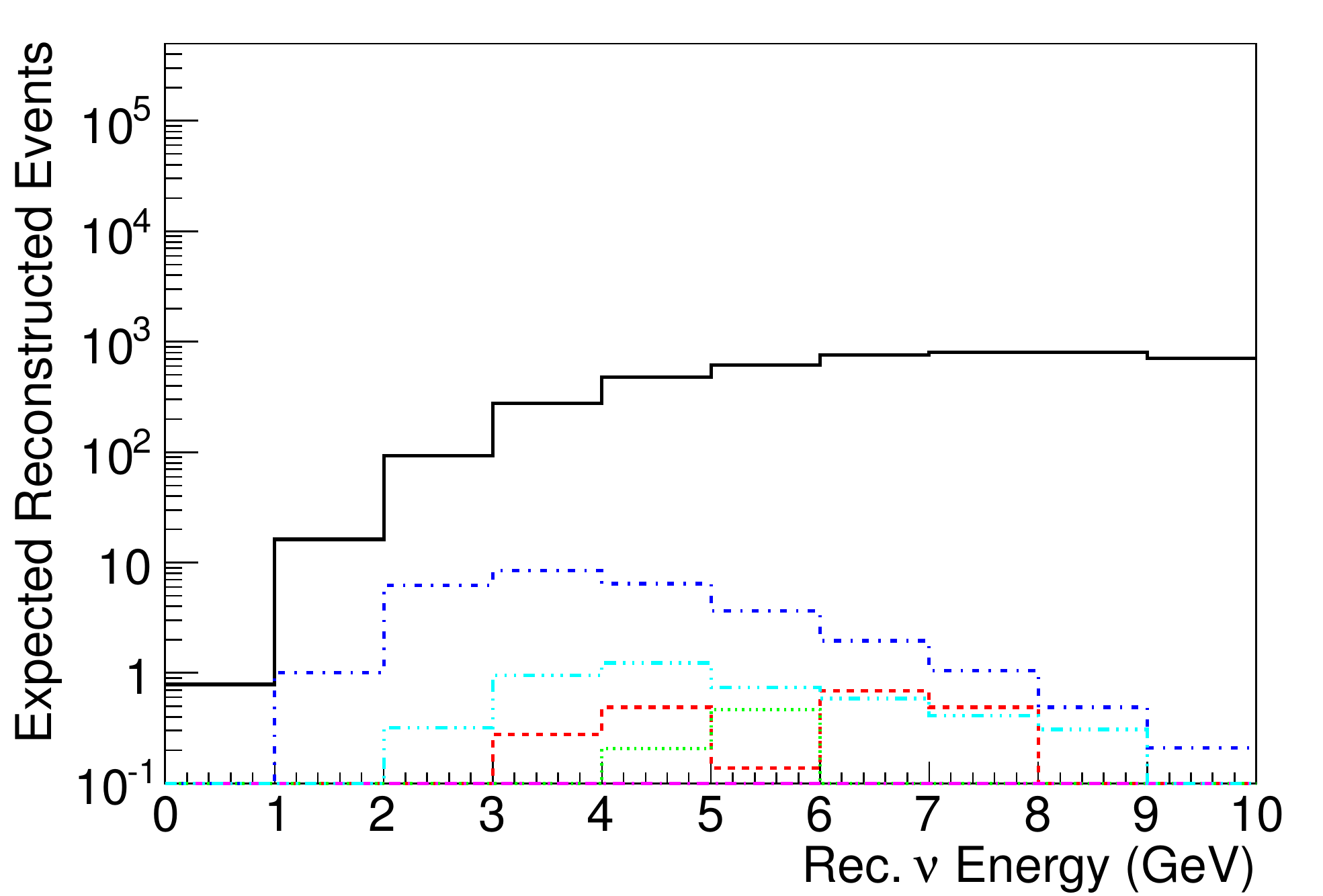} &
      \includegraphics[width=7.5cm, height=5.5cm]{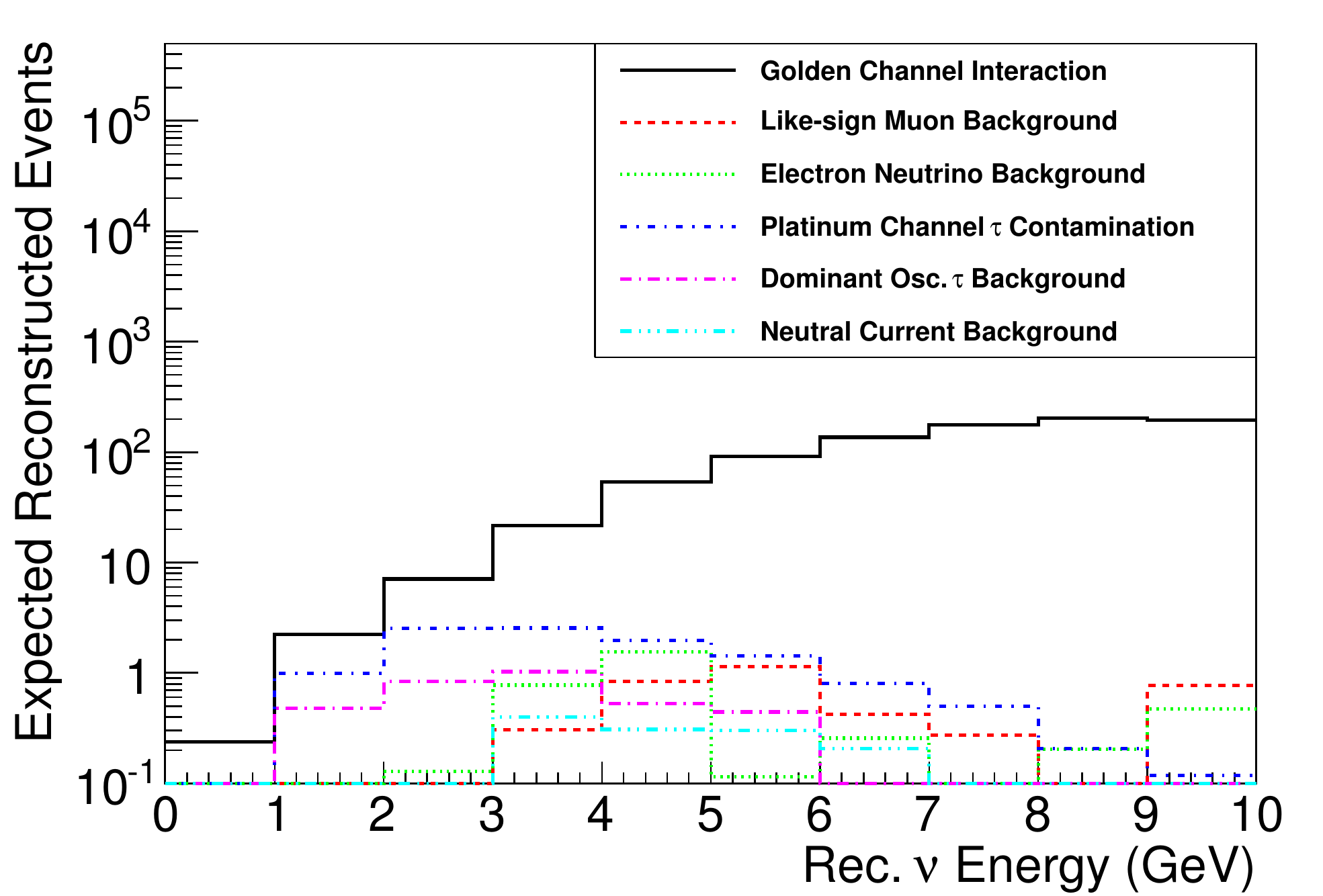}
    \end{array}$
  \end{center}
  \caption{Expected interactions in a 100~ktonne MIND 2000~km
      from the Neutrino Factory for $\theta_{13} = 9.0^{\circ}$.
      Left column for $\nu_\mu$ appearance and right for $\overline{\nu}_\mu$
      appearance.}
  \label{fig:expct}
\end{figure}

Using the response matrices extracted using the analysis described in
the preceding sections it is possible to make a prediction of the
expected contribution to the Monte Carlo sample from each of the
neutrino types in the beam. Figure~\ref{fig:expct} shows the expected
number of events for the best fit values of the currently measured
parameters taken from~\cite{Schwetz:2008er}: $\theta_{12} =
33.5^{\circ}\mbox{, } \theta_{23} = 45^{\circ}\mbox{, }\Delta m^2_{21}
= 7.65\times 10^{-5}\mbox{~eV$^2$, } \Delta m^2_{32} = 2.4\times
10^{-3}$~eV$^2$ for $\delta_{CP} = 45^{\circ}$ and calculating matter
effects using the PREM model~\cite{Dziewonski:1981xy}. The number of
interactions were calculated for a 100~ktonne MIND at a distance of
2000~km from the NF for a value of $\theta_{13}=9.0^{\circ}$, for an 
integrated flux due to 10$^{21}$ decays of each
polarity in the straight sections of the decay pipes.

\section{Study of systematic uncertainties}
\label{sec:syst}
The efficiencies and backgrounds described above will be affected by
several systematic effects. There will be many contributing factors
including uncertainty in the determination of the parameters used to
form the cuts in the analysis, uncertainty in the exclusive
cross-sections, uncertainty in the determination of the hadronic
shower energy and direction resolution, and any assumptions in the
representation of the detector and readout. While exact determination
of the overall systematic error in the efficiencies is complicated, an
estimate of the contribution of different factors can be obtained by
setting certain variables to the extremes of their errors.

\begin{figure}
  \begin{center}$
    \begin{array}{cc}
      \includegraphics[width=7.5cm, height=4.5cm]{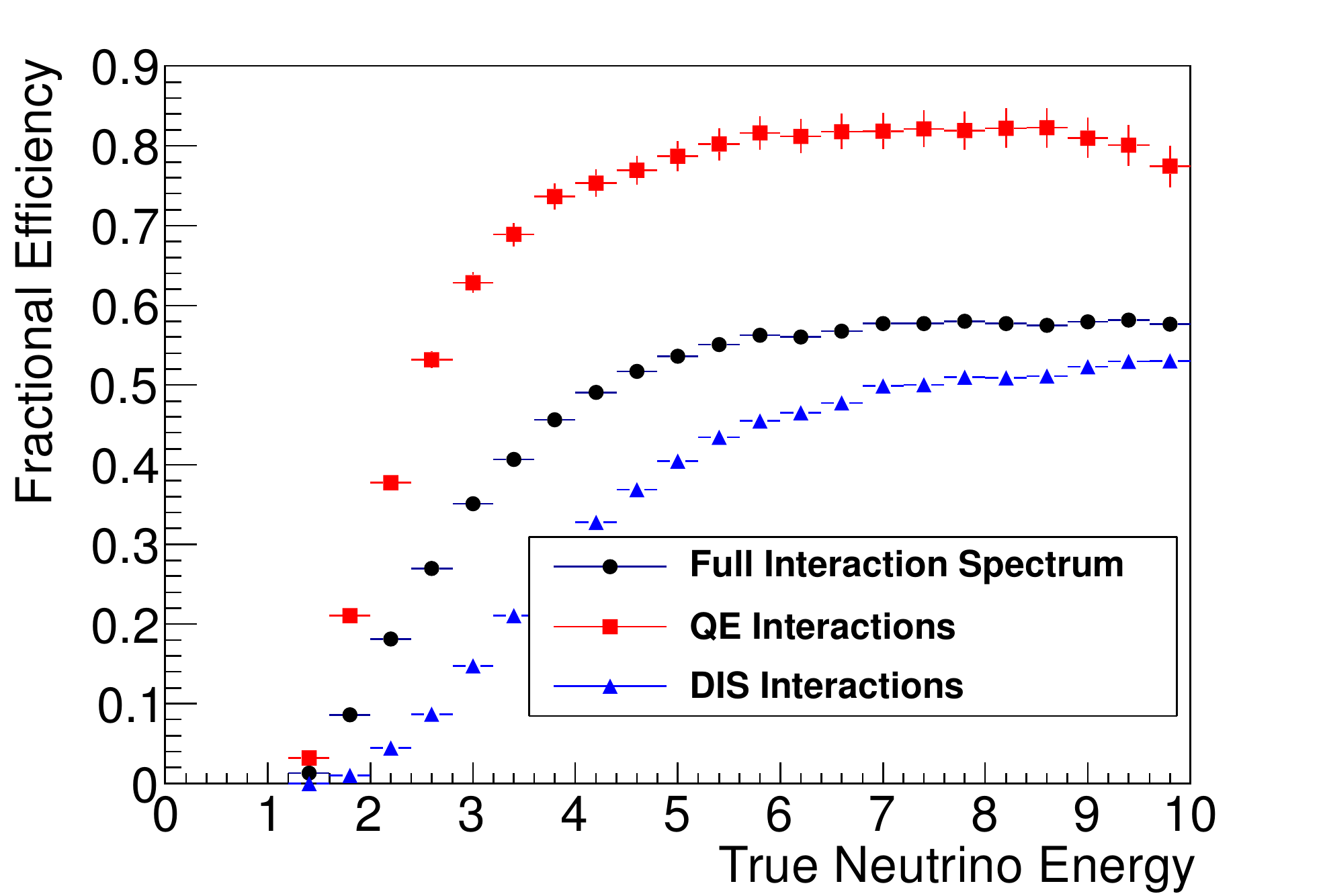} &
      \includegraphics[width=7.5cm, height=4.5cm]{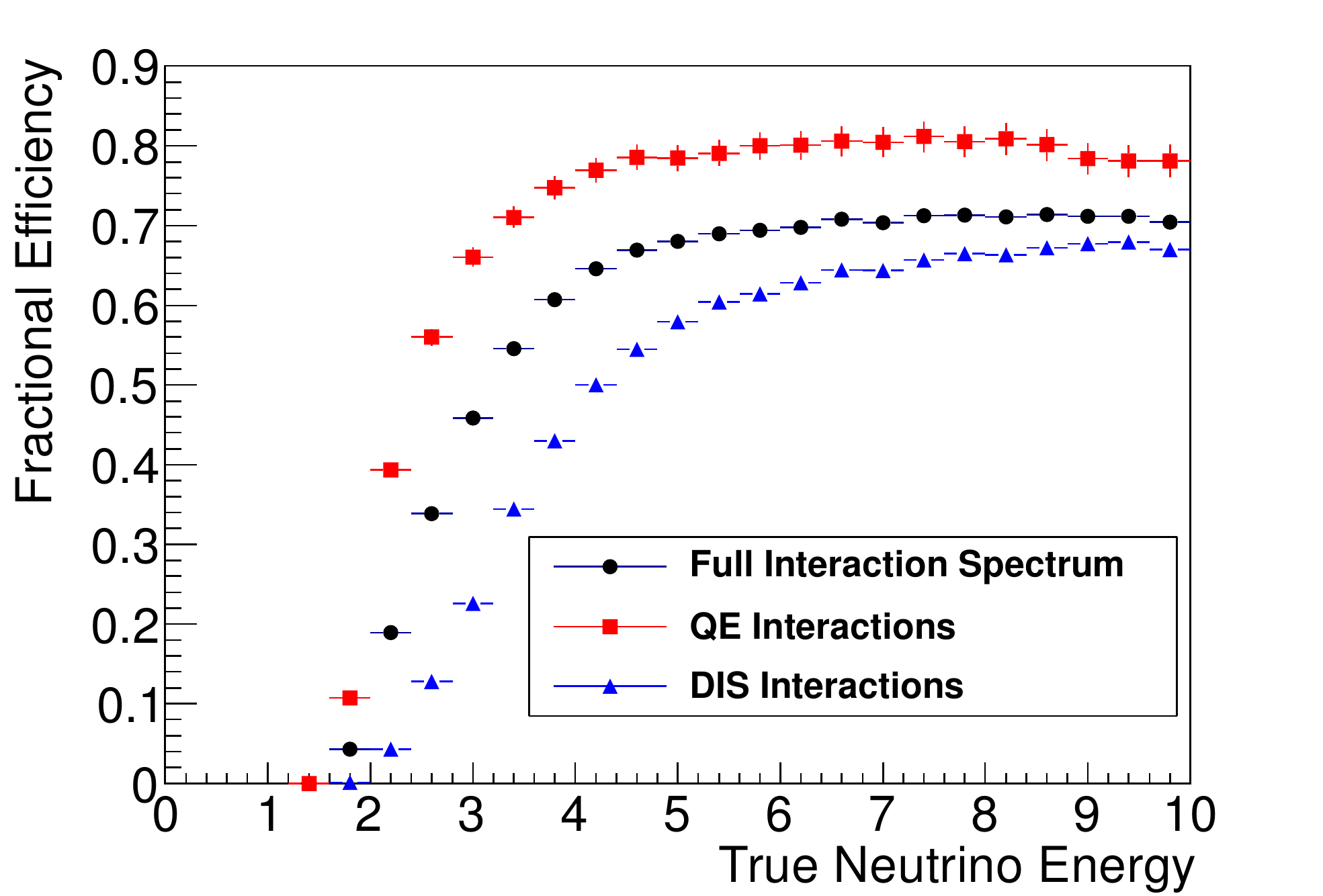}\\
      \includegraphics[width=7.5cm, height=4.5cm]{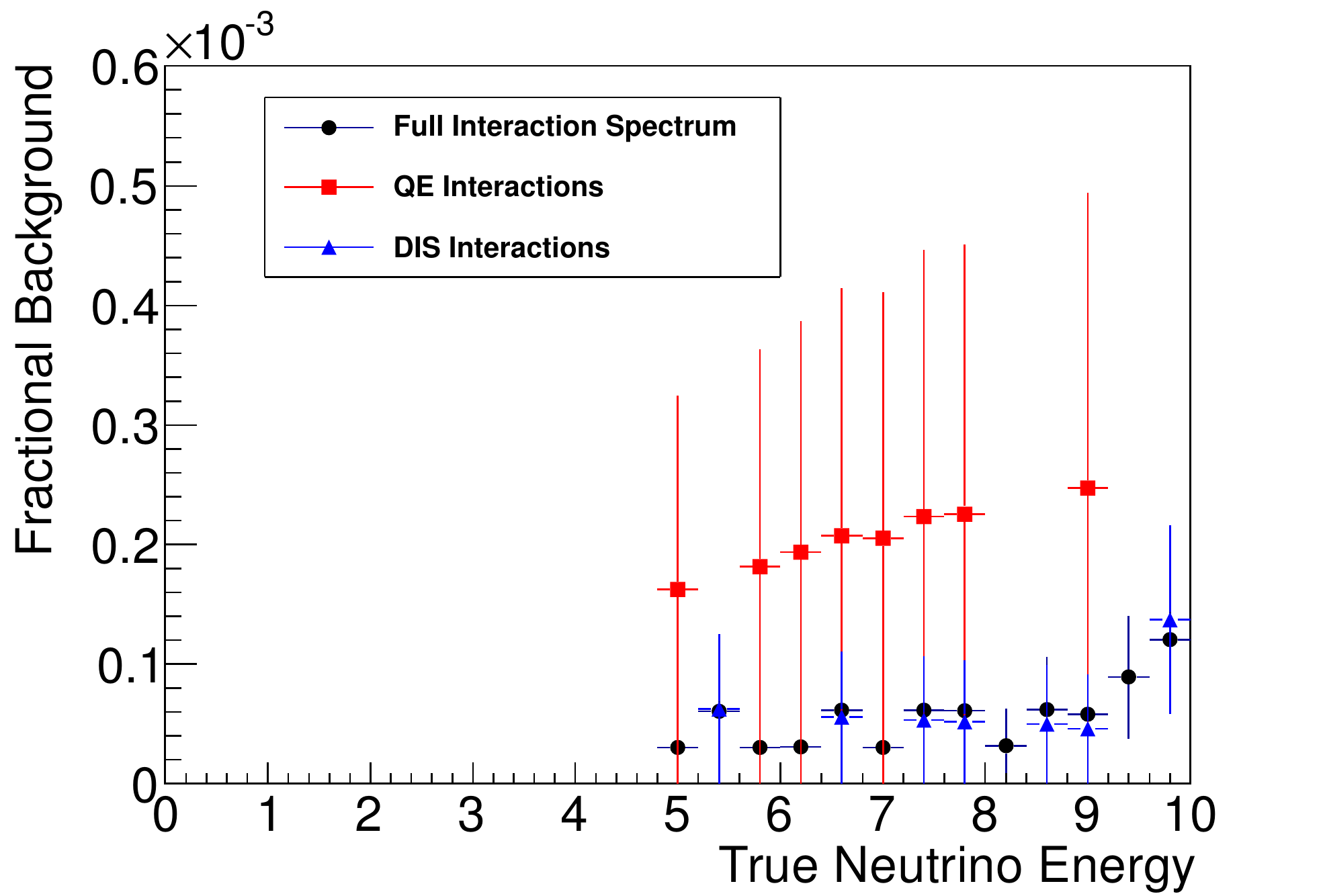} &
      \includegraphics[width=7.5cm, height=4.5cm]{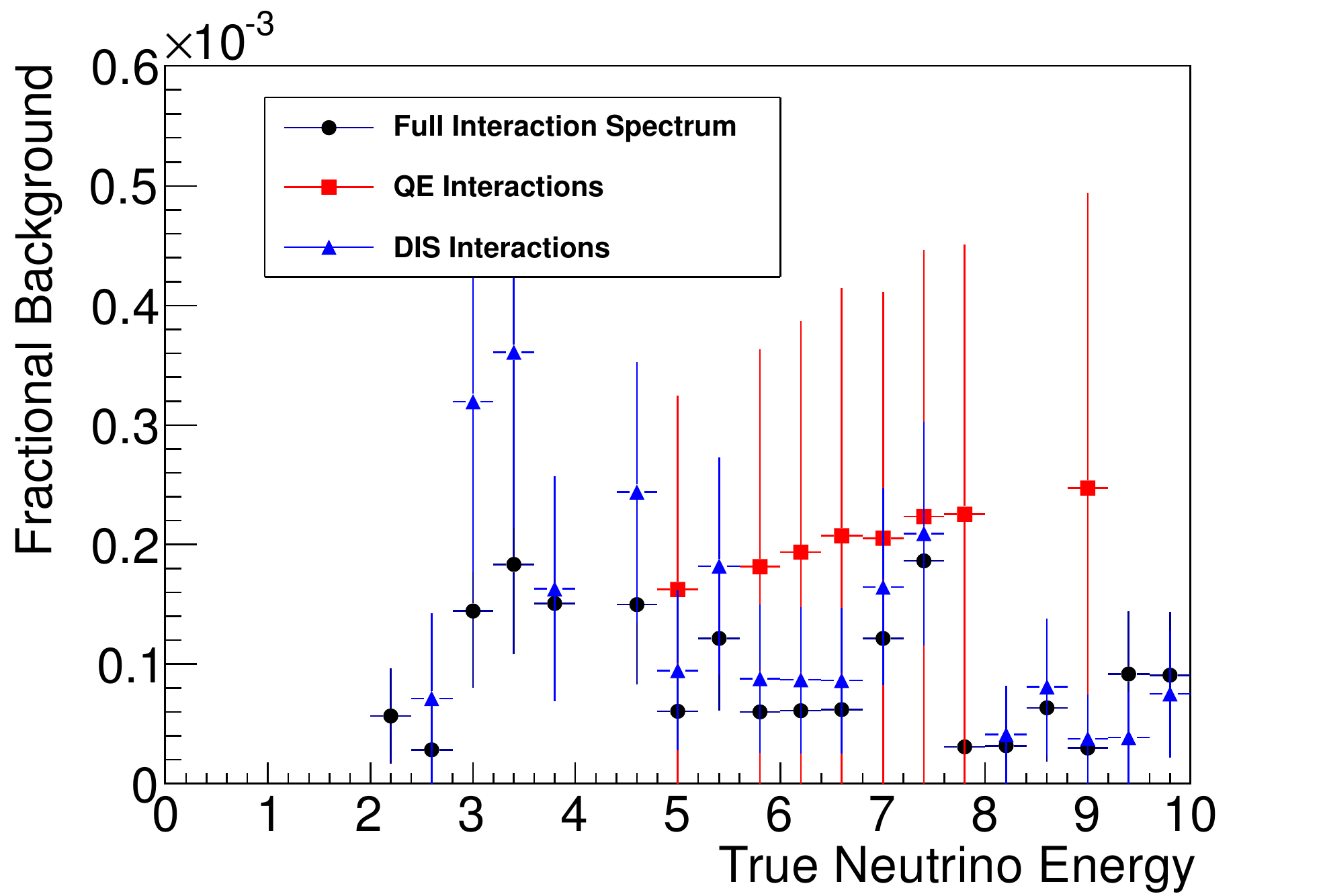}\\
      \includegraphics[width=7.5cm, height=4.5cm]{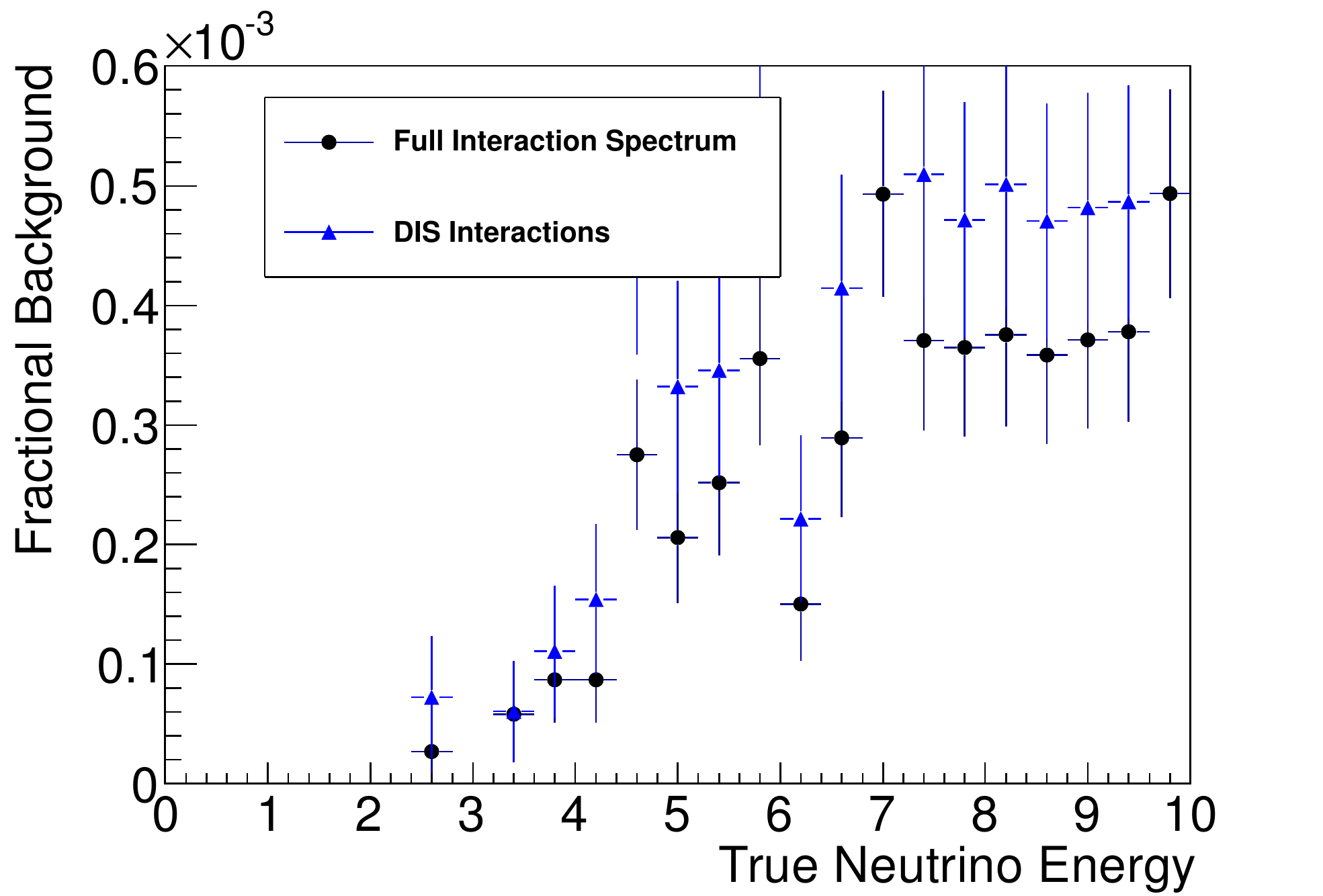} &
      \includegraphics[width=7.5cm, height=4.5cm]{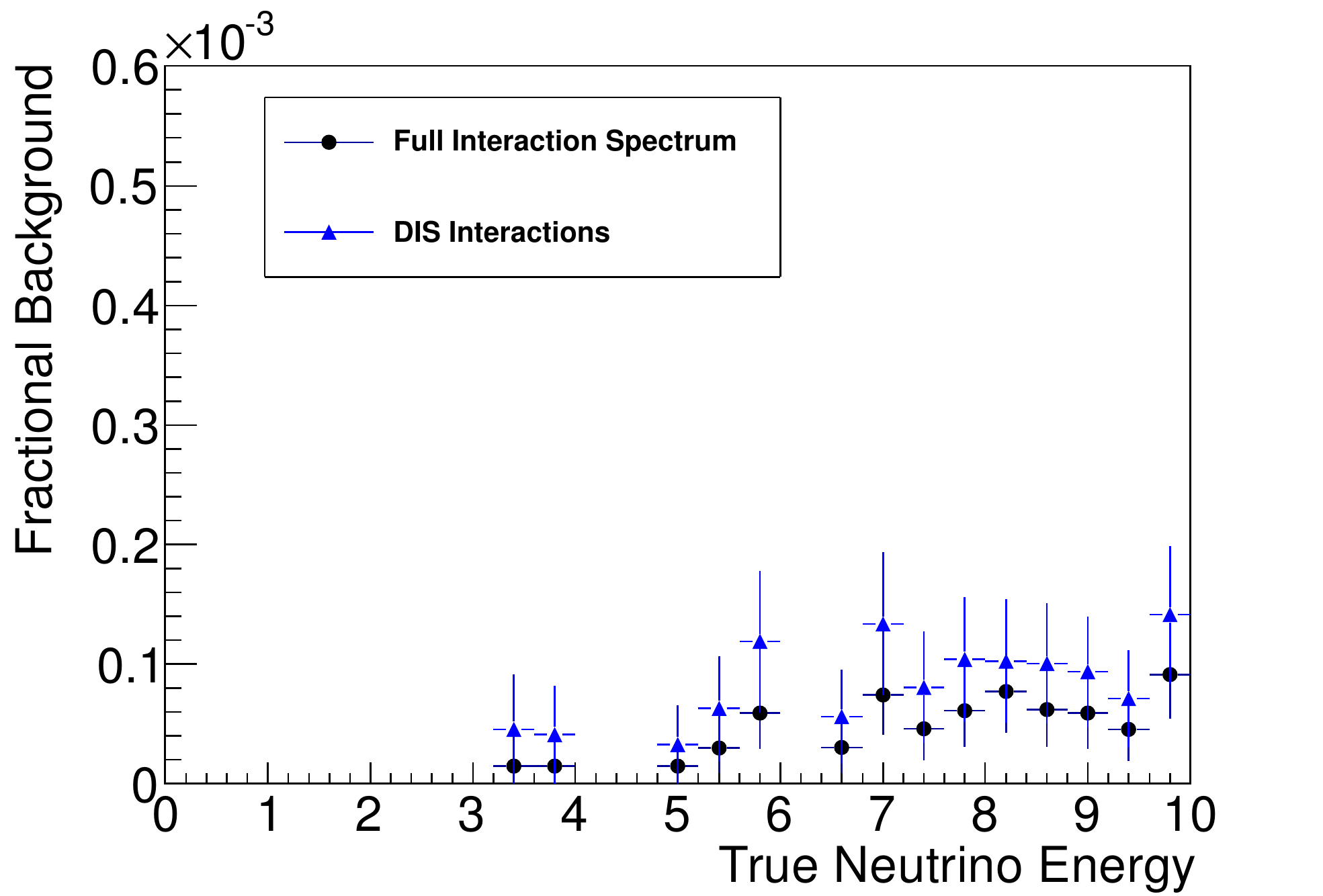}\\
      \includegraphics[width=7.5cm, height=4.5cm]{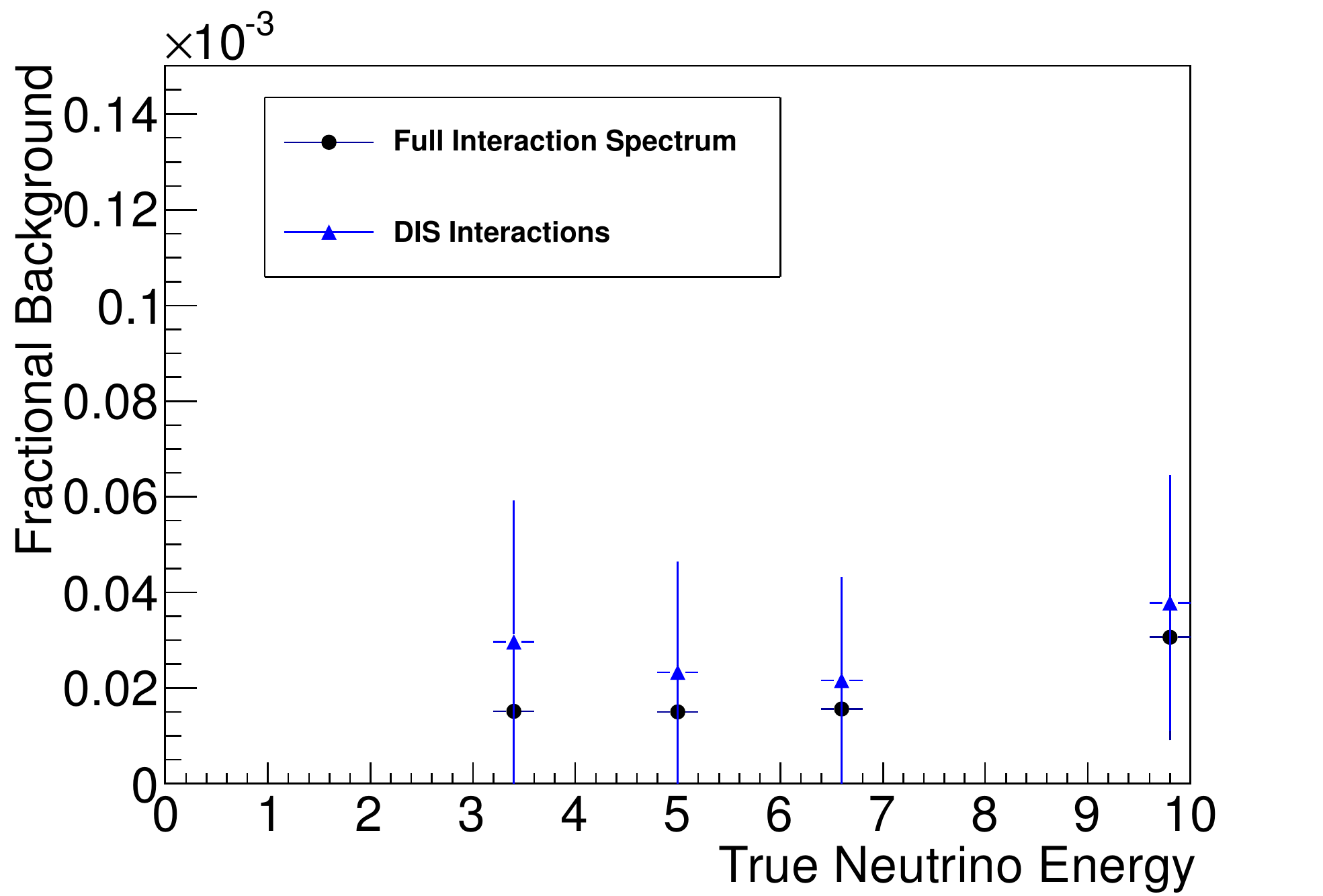} &
      \includegraphics[width=7.5cm, height=4.5cm]{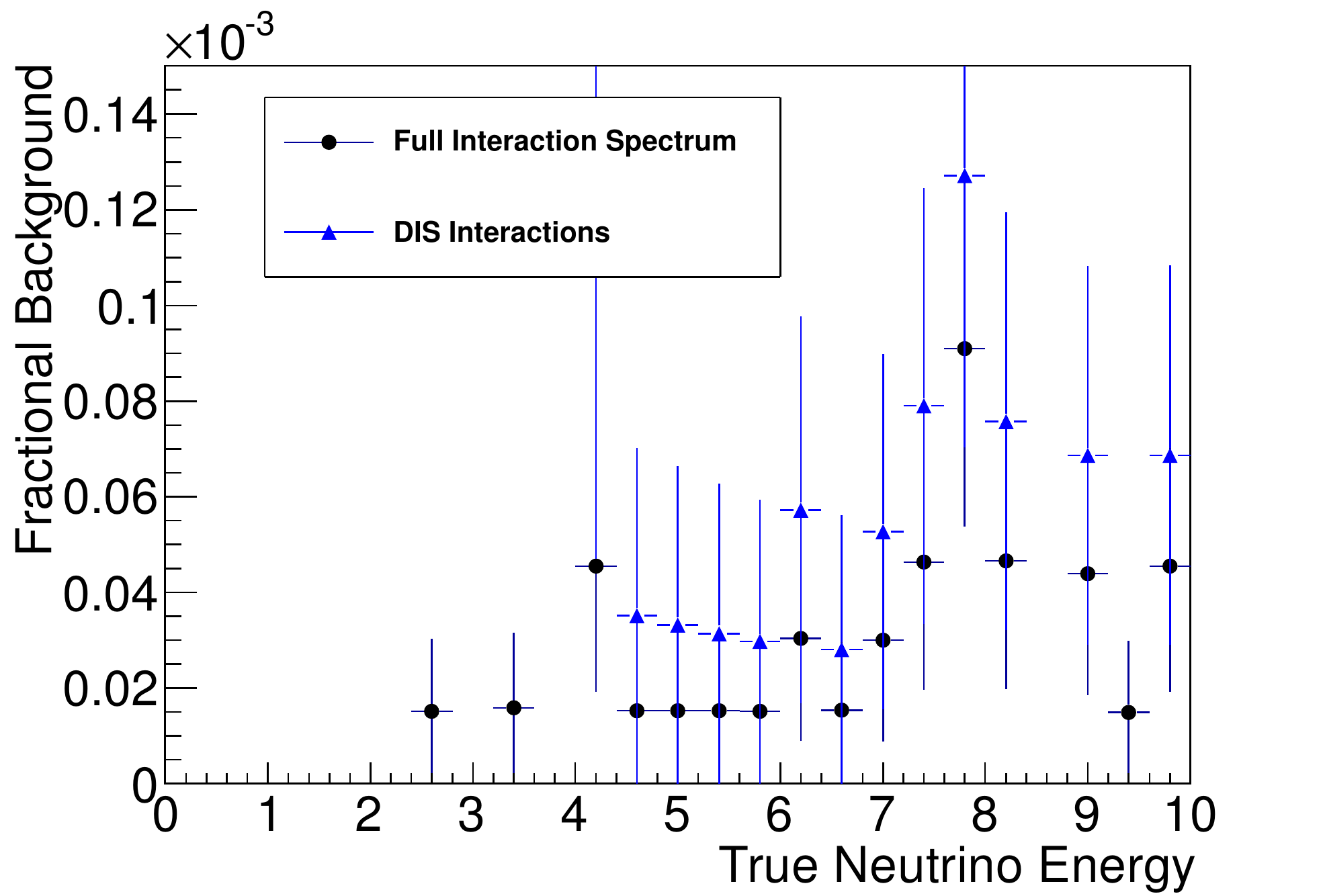}
    \end{array}$
  \end{center}
  \caption{Efficiencies for a pure DIS sample compared to the nominal
    case. (top) Signal efficiency, (second line)
    $\nu_\mu~(\overline{\nu}_\mu)$ CC background, (third line) NC
    background and (bottom) $\nu_e~(\overline{\nu}_e)$ CC
    background. $\nu_\mu$ appearance on the left and
    $\overline{\nu}_\mu$ appearance on the right.}
  \label{fig:DISonly}
\end{figure}

\begin{figure}
  \begin{center}$
    \begin{array}{cc}
      \includegraphics[width=7.5cm]{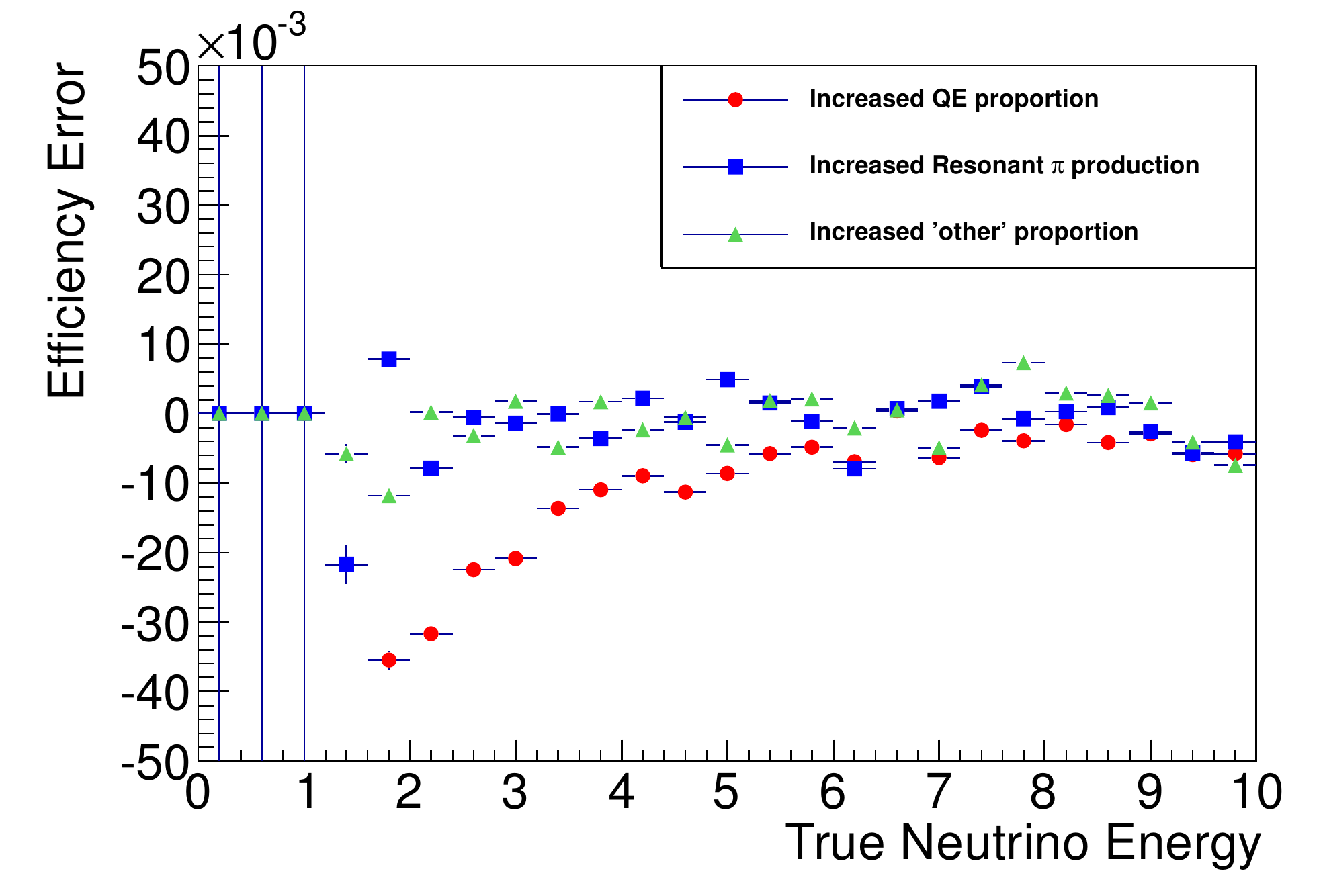} &
      \includegraphics[width=7.5cm]{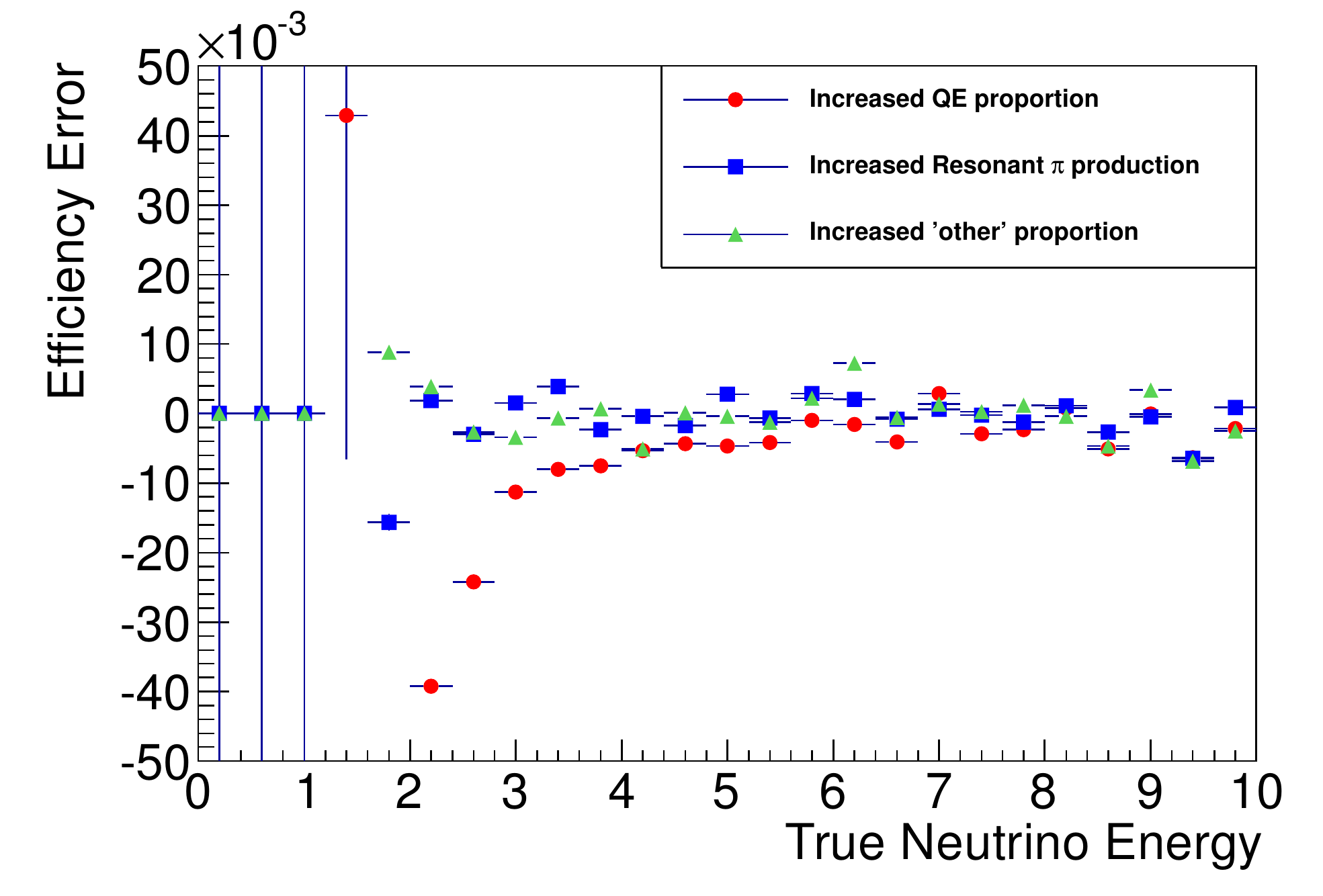}\\
      \includegraphics[width=7.5cm]{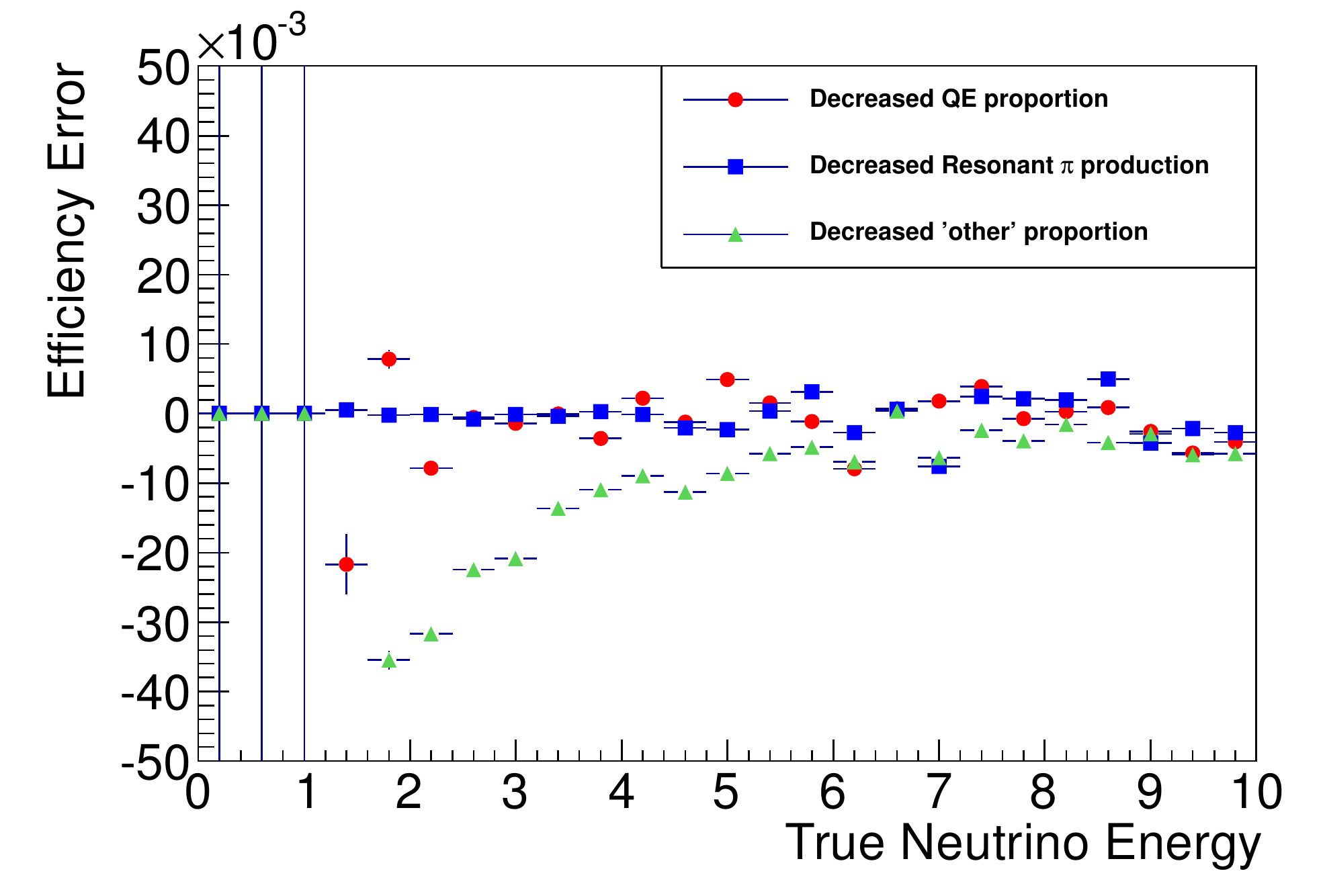} &
      \includegraphics[width=7.5cm]{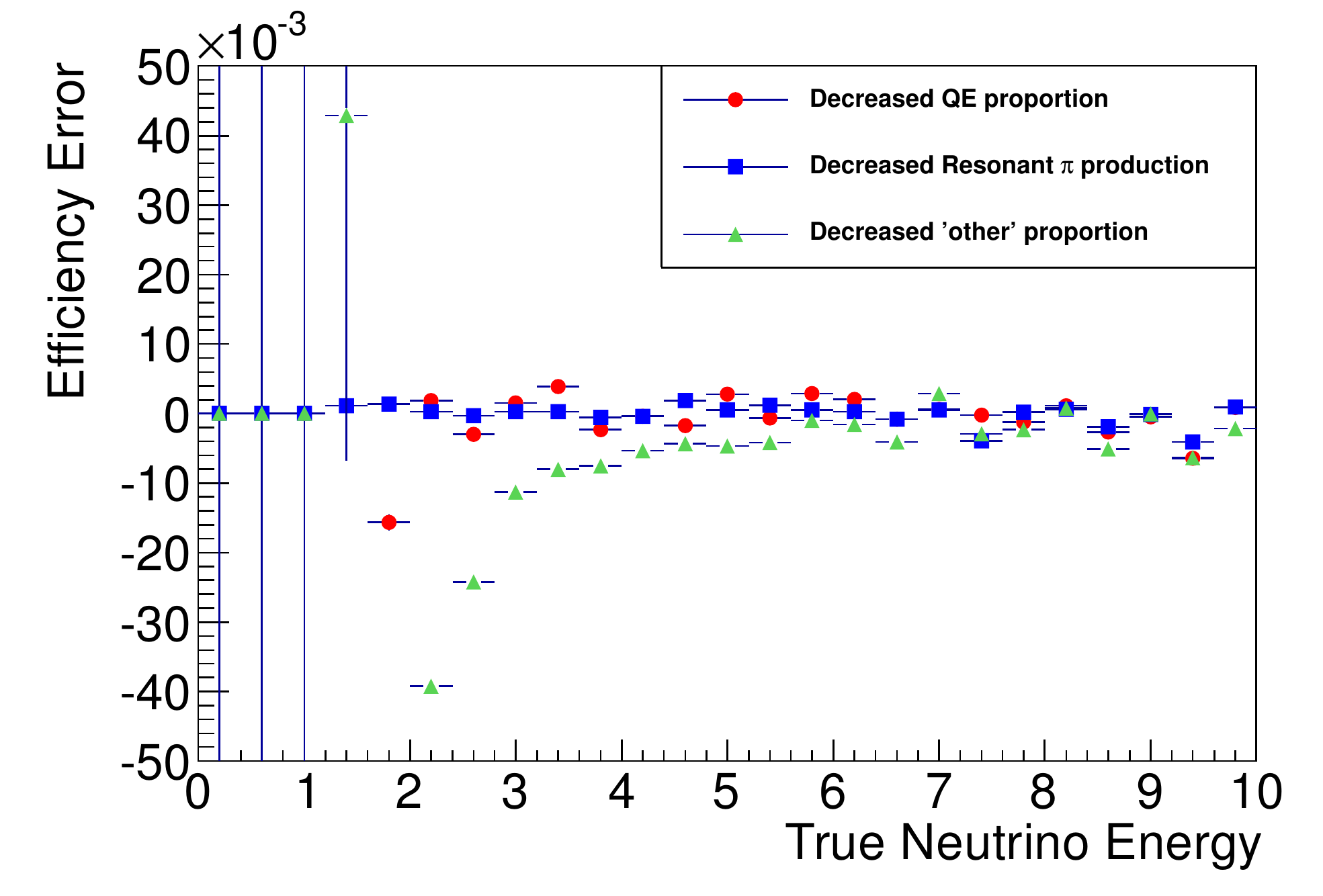}
    \end{array}$
  \end{center}
  \caption{Calculated error on signal efficiencies on increasing (top)
    and decreasing (bottom) the proportion of non-DIS interactions in
    the data-set. (left) Errors on true energy $\nu_\mu$ CC efficiency
    and (right) errors on true energy $\overline{\nu}_\mu$ CC
    efficiency}
  \label{fig:intvar}
\end{figure}

The exclusive QE, DIS and `other' cross-sections in the data sample
could have a significant effect on the signal efficiencies and
backgrounds. The efficiencies for the reconstruction of true QE and
true DIS interactions are compared to the nominal efficiency in
figure~\ref{fig:DISonly} where the dominance of DIS interactions in
the backgrounds is clear.
Although experimental data are available, confirming the presence of
non-DIS interactions in the energy region of interest, there are
significant errors in the transition regions (see for
example~\cite{Lyubushkin:2008pe,AguilarArevalo:2010zc}). These errors
lead to an uncertainty in the proportion of the different types of
interaction that can affect the efficiencies. In order to study the
systematic error associated with this effect, events of certain types
were randomly removed from the data-set and the mean effect
quantified. As an illustration of the method, consider the
contribution from QE interactions. Taking the binned errors on the
cross-section measurements
from~\cite{Lyubushkin:2008pe,AguilarArevalo:2010zc}, a run to reduce
the QE content would exclude a proportion of events in a bin so that
instead of contributing the proportion $\frac{N_{QE}}{N_{tot}}$, where
$N_{QE}$ and $N_{tot}$ are the number of QE interactions and the total
number of interactions in the bin of interest, it would instead
contribute: $\frac{N_{QE} - \sigma_{QE}N_{QE}}{N_{tot} -
  \sigma_{QE}N_{QE}}$, where $\sigma_{QE}$ is the proportional error
on the QE cross section for the bin. Since the data-set is finite and
an actual increase in the number of QE interactions is not possible,
the equivalent run to increase the QE contribution reduces the
contribution of the ``rest'' by an amount calculated to give the
corresponding proportional increase in QE interactions:
\begin{equation}
  \label{eq:incQE}
  \displaystyle\frac{N_{QE} + \sigma_{QE}N_{QE}}{N_{tot} + \sigma_{QE}N_{QE}} = \displaystyle\frac{N_{QE}}{N_{tot} - \epsilon N_{rest}} \, ;
\end{equation}
where $N_{rest}$ is the total number of non-QE interactions in the bin
and $\epsilon$ is the required proportional reduction in the `rest' to
simulate an appropriate increase in QE. Solving for $\epsilon$ yields
the required reduction:
\begin{equation}
  \label{eq:redRest}
  \epsilon = \displaystyle\frac{\sigma_{QE}}{1 + \sigma_{QE}} \, .
\end{equation}
The 1$\sigma$ systematic error can be estimated as the mean difference
between the nominal efficiency and the increase due to a higher QE
proportion or decrease due to exclusion. The errors in the true
$\nu_\mu$ and $\overline{\nu}_\mu$ efficiencies extracted using this
method that varies the contribution of QE, 1$\pi$ and other non-DIS
interactions are shown in figure \ref{fig:intvar}. Errors for 1$\pi$
resonant reactions are estimated to be $\sim$20\% below 5~GeV (as
measured by the K2K near detector~\cite{:2008eaa}) and at 30\%
above. Due to the large uncertainty, both theoretically and
experimentally, on the models describing other resonances, coherent,
diffractive and elastic processes, a very conservative error of 50\%
is taken when varying the contribution of the ``others''. As can be
seen in figure~\ref{fig:intvar}, the systematic effect is less than
1\% for neutrino energies between 3 GeV and 10 GeV increasing to 4\%
near 1 GeV, with increased QE and 1$\pi$ interactions
generally increasing the efficiency and increased contribution of the
``other'' interactions having the effect of decreasing
efficiency. This last result is likely to be predominantly due to
resonances producing multiple tracks. The effect on backgrounds is
expected to be minimal, as was also shown in figure~\ref{fig:DISonly}. At the
time of a Neutrino Factory, the cross-section uncertainties should be
much smaller than the ones assumed here, so we expect the systematic
error to be below 1\% for all energies.

\begin{figure}
  \begin{center}$
    \begin{array}{cc}
      \includegraphics[width=8cm, height=6cm]{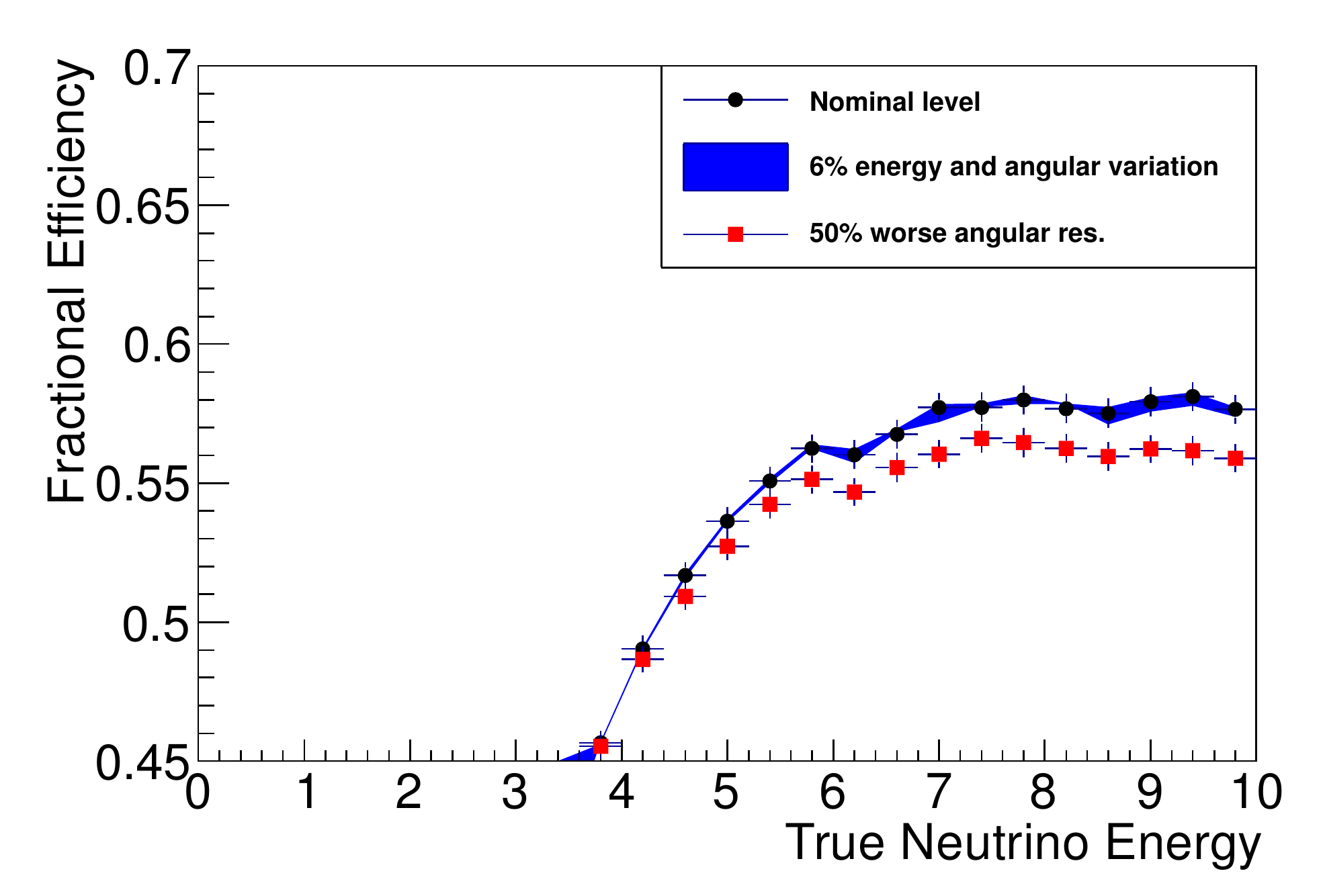} &
      \includegraphics[width=8cm, height=6cm]{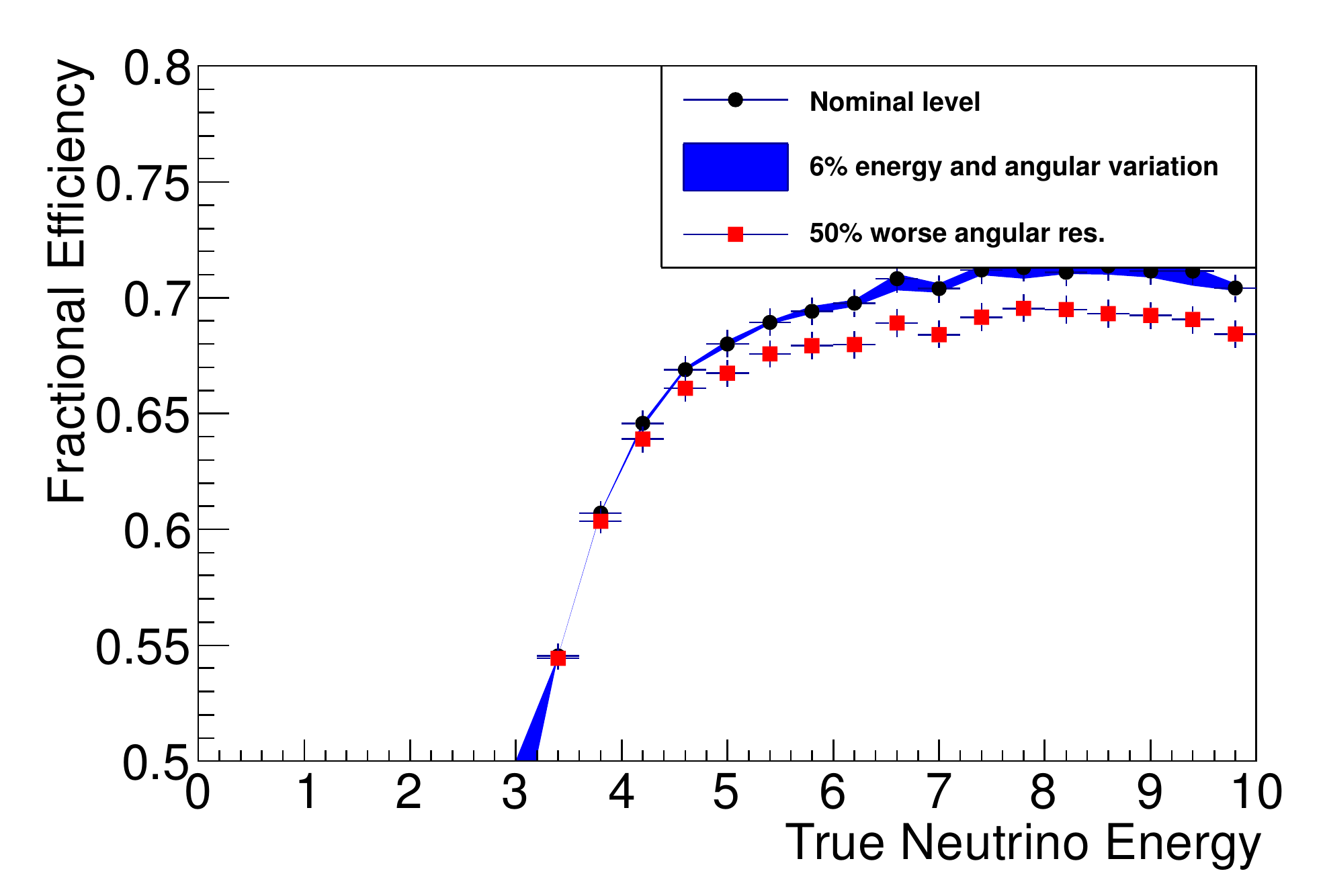}\\
      \includegraphics[width=8cm, height=6cm]{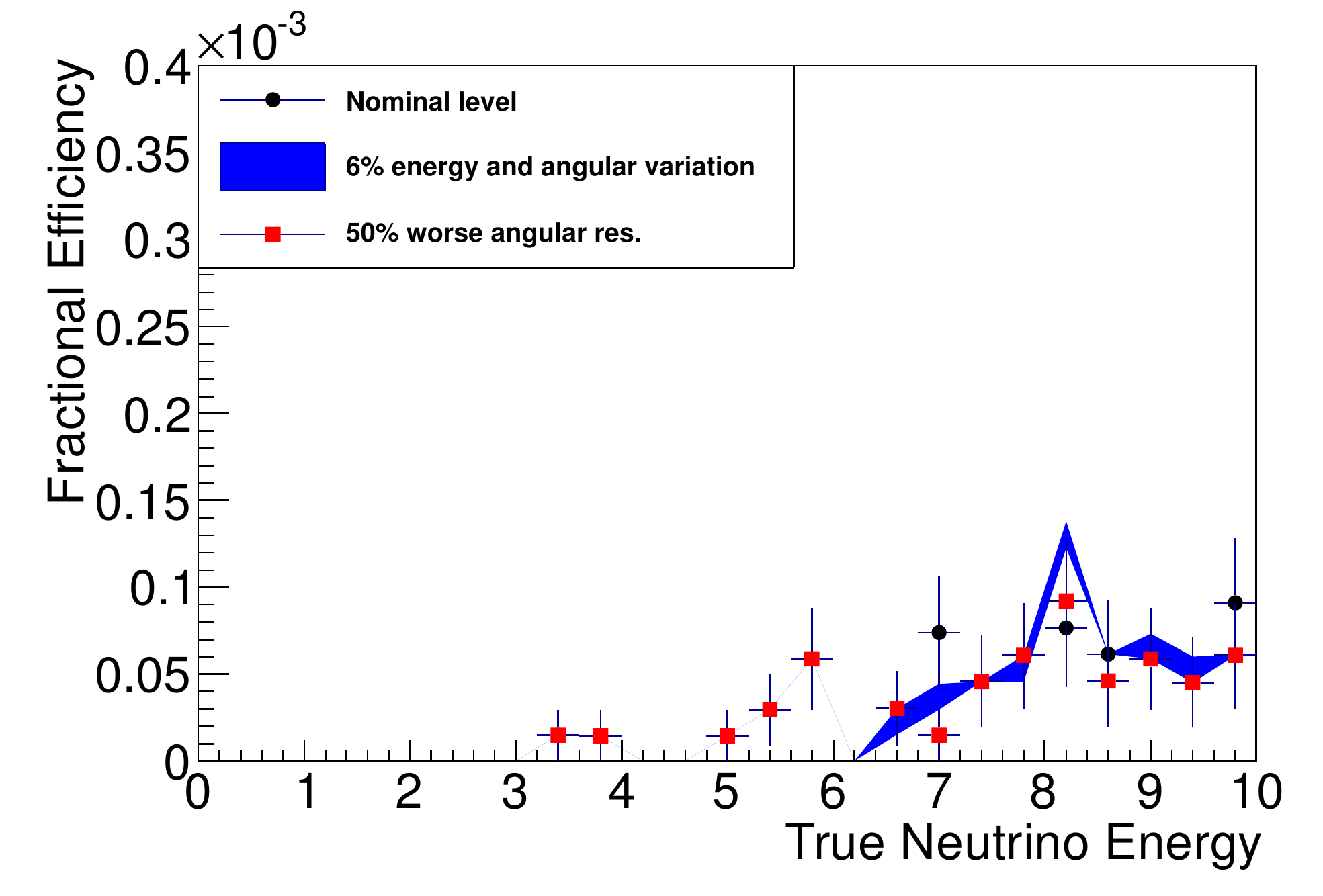} &
      \includegraphics[width=8cm, height=6cm]{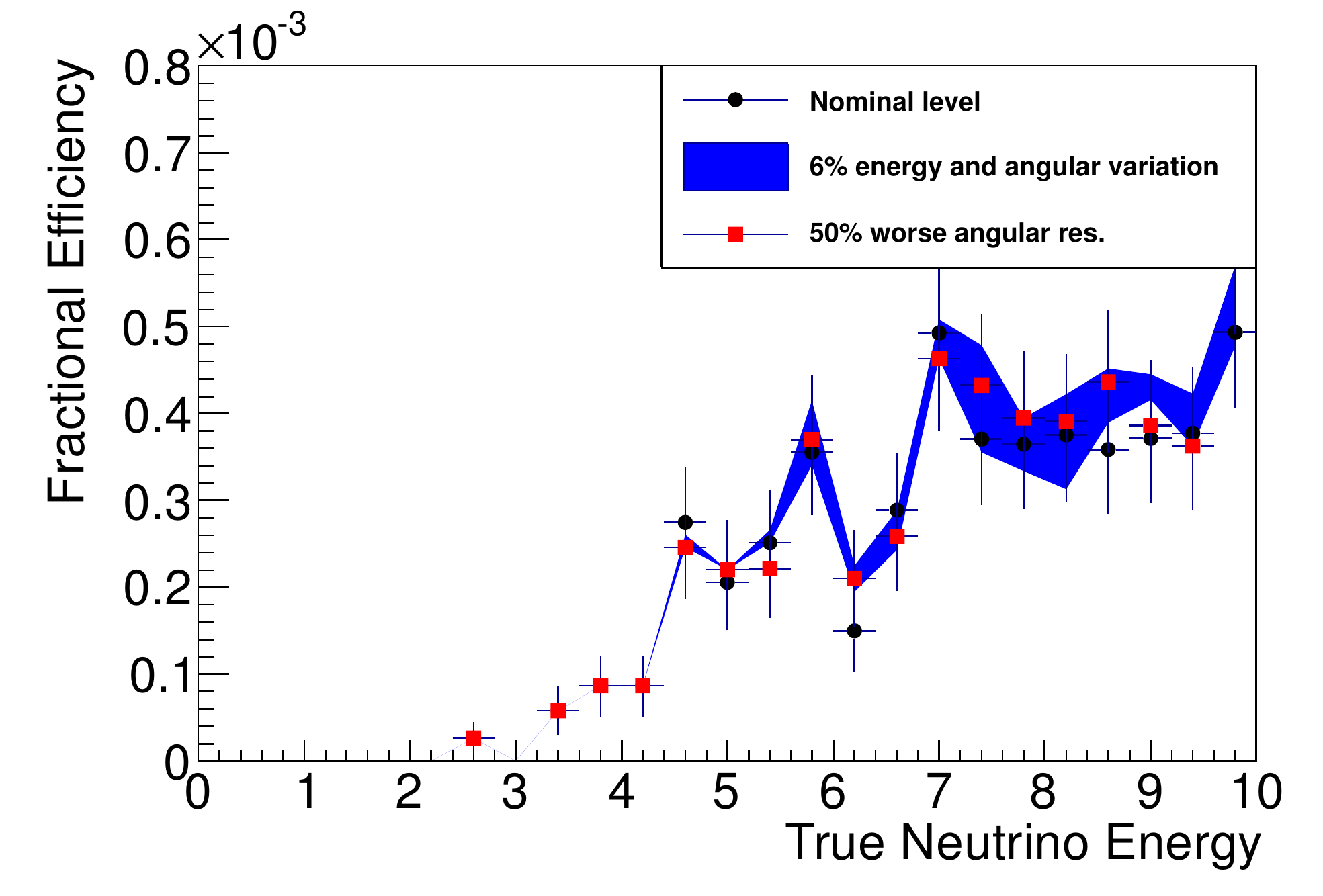}
    \end{array}$
  \end{center}
  \caption{Variation of signal efficiency (top) and NC backgrounds
    (bottom) due to a 6\% variation in the hadron shower energy and
    direction resolution and a more pessimistic 50\% reduction in
    angular resolution (focused on region of greatest variation).}
  \label{fig:hadresSyst}
\end{figure}

Another important source of systematic uncertainty is due to the error
in the muon momentum and the hadron shower energy used to reconstruct
the total neutrino energy. The muon is fully reconstructed, but the
hadron shower reconstruction is performed assuming a parametrisation
to the hadron energy and angular resolution. The particular choice of
this parametrisation introduces a systematic error on the resulting
signal and background efficiencies that should be studied. Taking a
6\% error as quoted for the energy scale uncertainty assumed by the
MINOS collaboration~\cite{Michael:2008bc} and varying the constants of
the energy and direction smears by this amount, it can be seen (blue
bands in figure \ref{fig:hadresSyst}) that, to this level, the
hadronic resolutions have little effect on the true neutrino-energy
efficiencies. However, the hadronic direction resolution is likely to
have far greater uncertainty and would be very sensitive to noise in
the readout electronics. Also shown in figure \ref{fig:hadresSyst} are
the efficiencies when the hadronic energy resolution parameters are
6\% larger but with a 50\% increase in the angular resolution
parameters. We expect in a real detector to measure the hadronic
angular resolution with a precision of better than 50\%, even though
we cannot quantify this precision yet. However, even at this level,
the observed difference in efficiency is only at the level of 1\%
above 7-8~GeV. A combination of the exclusive cross-sections and
hadronic energy uncertainties implies a total systematic uncertainty
for the measurement efficiency of order 1\% over the neutrino
energy range above 2~GeV.

\section{Comparison between event generators}
\label{sec:evtgen}
\begin{figure}
  \subfloat[GENIE and NUANCE $\nu_{\mu}$CC efficiency]{
    \includegraphics[width=8cm]{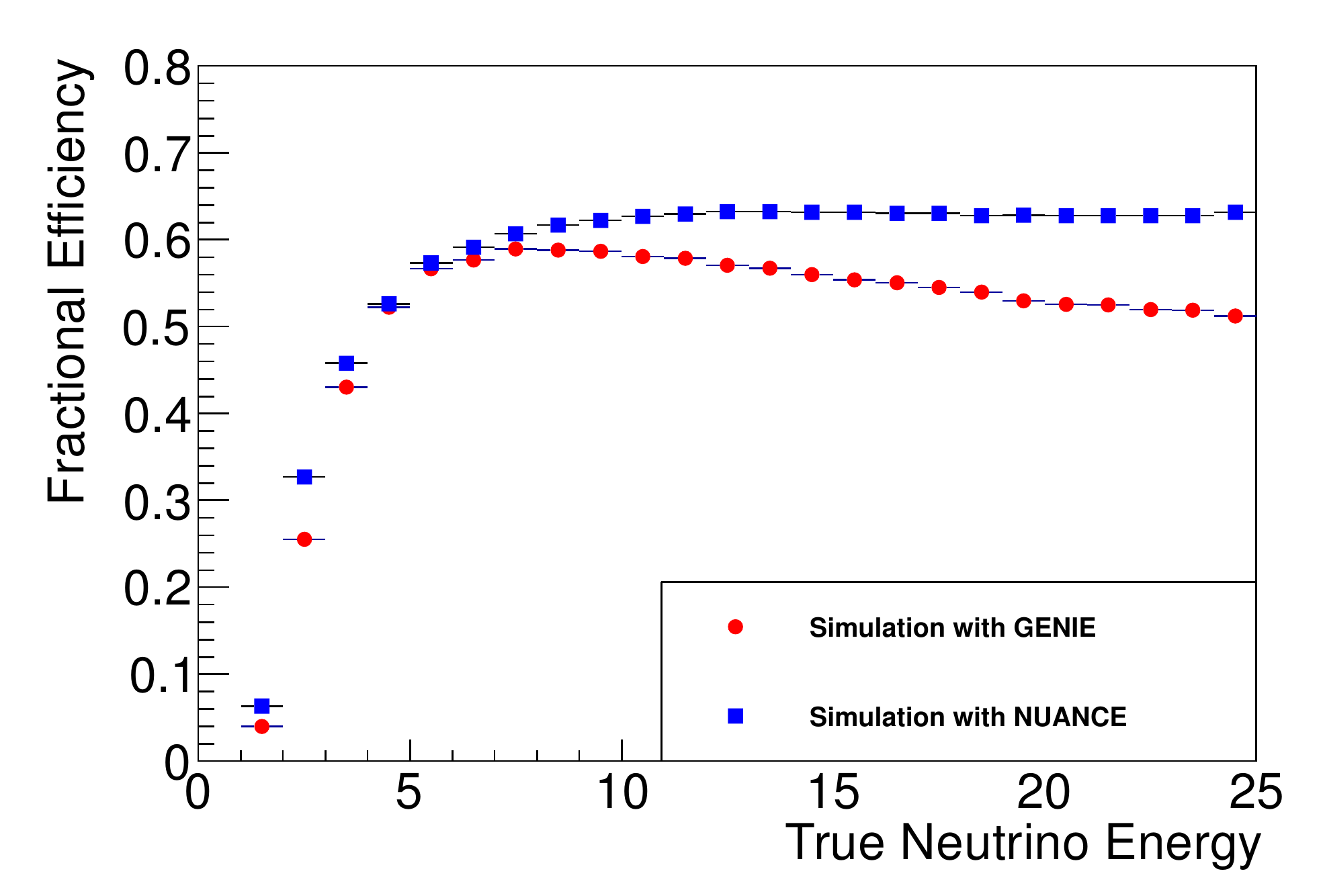}
  }
  \subfloat[GENIE and NUANCE $\bar{\nu}_{\mu}$CC efficiency]{
    \includegraphics[width=8cm]{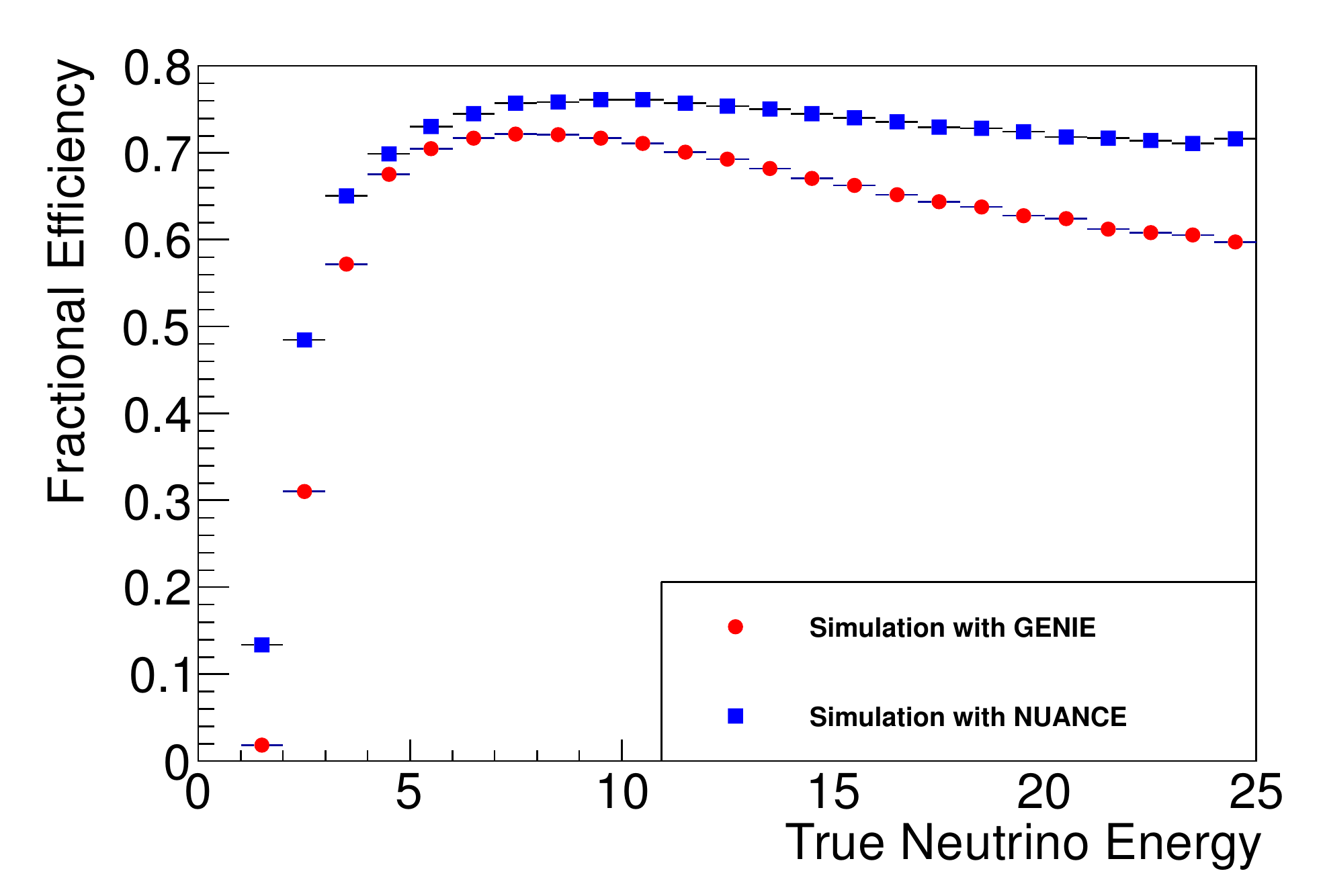}
  }\\
  \subfloat[Fractional difference between GENIE and NUANCE
    $\nu_{\mu}$ CC efficiencies]{
    \includegraphics[width=8cm]{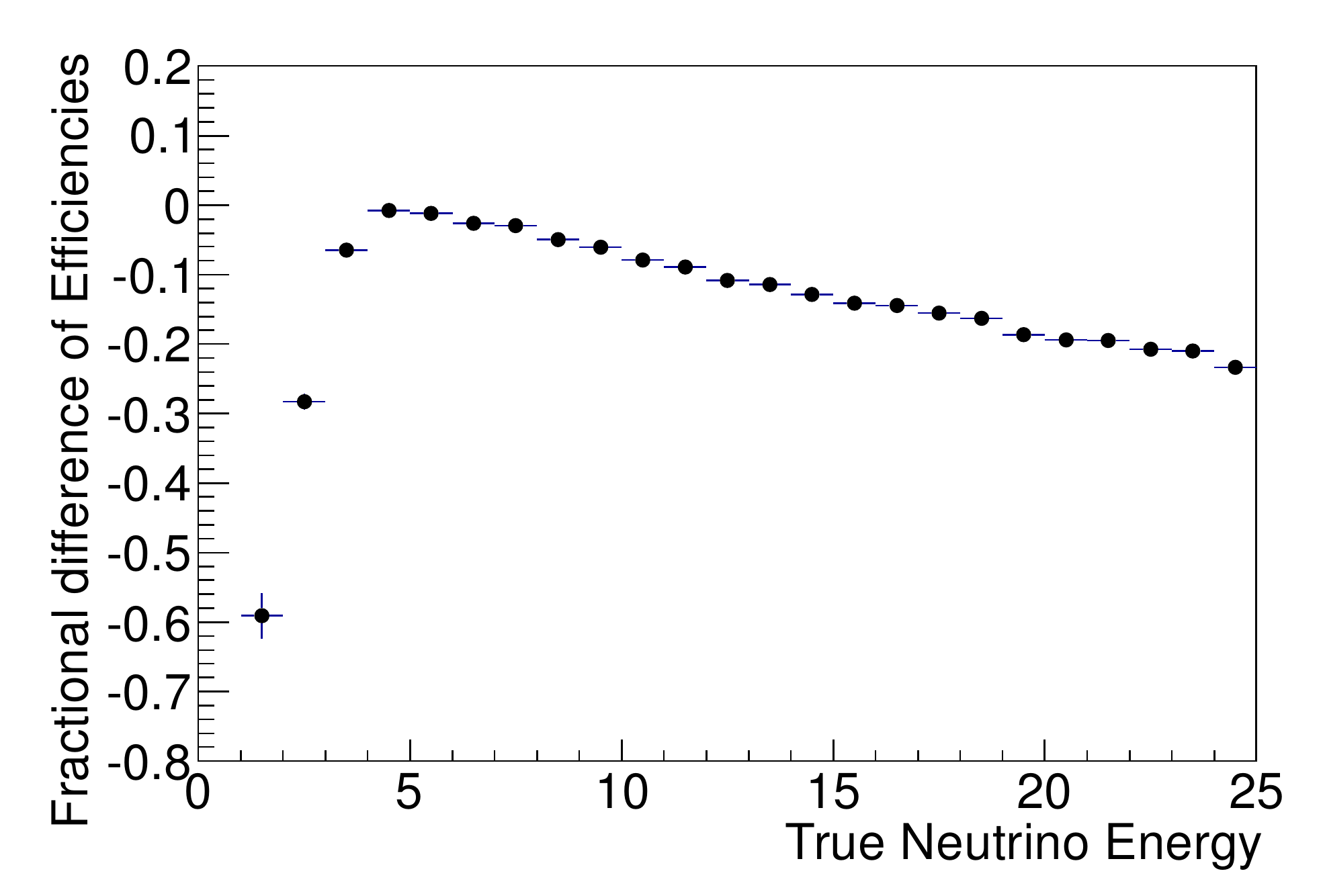}
  }
  \subfloat[Fractional difference between GENIE and NUANCE
    $\bar{\nu}_{\mu}$CC efficiencies] {
    \includegraphics[width=8cm]{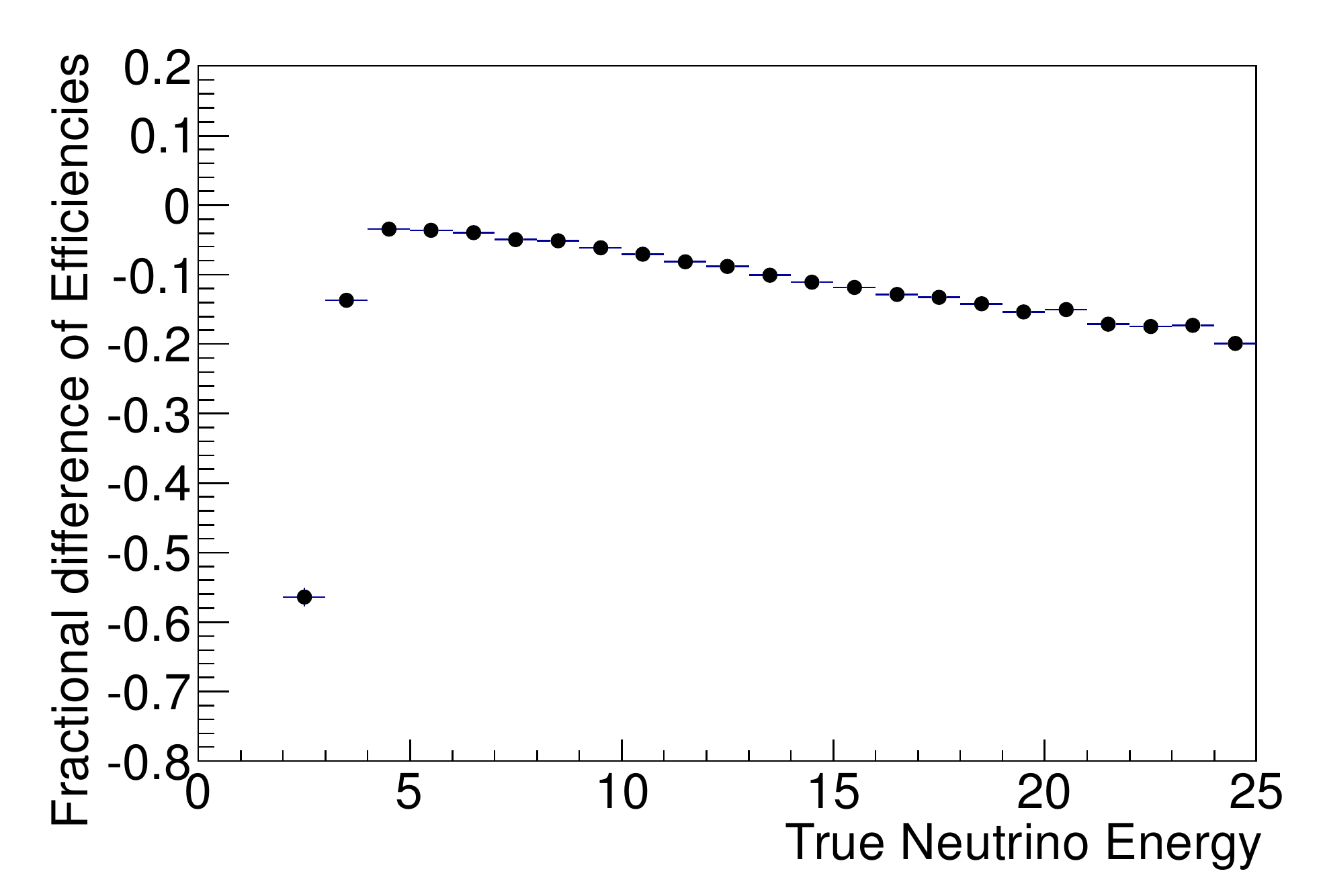}
  }
  \caption{The Golden channel analysis efficiencies from simulations
    of charge current neutrino events in a MIND detector generated
    using GENIE and NUANCE neutrino generators. The differences between
    the efficiencies as fractions of the GENIE efficiency are also shown.}
  \label{fig:gencomp}
\end{figure}

The above analysis uses the GENIE event generator to simulate the
interaction of events with matter in the detector. This generator was
assumed to bring this effort in line with current experiments,
such as MINOS, where good agreement with data has been achieved. 
It is useful to compare to a previous version of this
analysis with the NUANCE event generator \cite{NF:2011aa}. A
comparison of the neutrino charge current detection efficiencies
appears in figure~\ref{fig:gencomp}. This shows that the GENIE derived
analysis produces smaller positive identification of charge current
events. This loss of performance is linear with respect to neutrino
energy for energies greater than 5 GeV, so that the analysis of the
GENIE simulation is 20\% less efficient than that of the NUANCE
simulation. This difference is partially ascribed to differences in the parton 
distribution functions used by each generator and should not be
interpreted as a systematic error of the analysis, since the measured event rates
in a future experiment will be cross-checked against the simulations, so the efficiencies will
be determined much more accurately than the difference between generators.
We assume that GENIE, which has been benchmarked against recent neutrino
experiments, serves as a more realistic estimator for the efficiency of the analysis. 

\section{MIND Sensitivity}
\label{sec:MINDsens}
The Neutrino Factory is required to measure the CP-violating phase $\delta_{CP}$
simulaneously with $\theta_{13}$ while removing ambiguity caused by degenerate solutions.
Extracting the oscillation parameters from the observed signal at the
far detectors requires the accurate prediction of the expected flux
without oscillation, which is used along with the calculated
oscillation probabilities to fit the observed signal for the best
value of $\theta_{13}$ and $\delta_{CP}$. Due to the large distance to
the far detectors, the flux spectra expected from the decay rings of
the Neutrino Factory are accurately approximated by the flux from a
point source of muons travelling with appropriate Lorentz boost in the
direction of the detector. Using this flux or, alternatively, a
projection of the spectra observed in the near detector, the number of
true interactions expected in MIND as a function of energy can be calculated for each value of
$\theta_{13}$ and $\delta$. These spectra can then be multiplied by
the response matrices shown in the Appendix to
calculate the observed golden channel interaction spectrum expected
for some hypothetical values of the oscillation parameters.

A Neutrino Factory storing muons of energy 10~GeV is assumed, of which
5.0$\times 10^{20}$ per year of each species decay in the straight
sections pointing towards the MIND far detector of 100~ktonnes mass 
placed at a distance of 2000~km from the facility.

\subsection{The NuTS framework}
\label{sec:NUTS}
The Neutrino tool suite (NuTS) was developed for the studies presented
in~\cite{BurguetCastell:2001ez,BurguetCastell:2002qx,BurguetCastell:2005pa}. It
provides a framework for the generation of appropriate fluxes for
different neutrino accelerator facilities along with the necessary
infrastructure to calculate the true neutrino oscillation
probabilities for all channels. In addition, using the
parametrisation of the total interaction spectra calculated in
section~\ref{Sec:event_generation}, the expected number of events in a
given energy bin can be calculated. Using this framework and the
response matrices extracted for MIND, simulated data for an experiment
can be generated as
\begin{equation}
  \label{eq:simDat}
  Data^{i,j}_{sim} = smear\left( M_{sig}^iN_{sig}^{i,j} + \displaystyle\sum_{k}M_{bkg}^{i,k}N_{bkg}^{i,j,k}\right),
\end{equation}
for each polarity and detector baseline of interest, where
$M_{sig}^{i}$ is the response matrix for MIND for a particular signal
channel $i$ (stored $\mu^+$ or $\mu^-$), $N_{sig}^{i,j}$ is the 100\%
efficiency interaction spectrum in true $\nu$ energy bins for a
channel $i$ at a detector baseline $j$ (in this case, there is only one baseline at 2000~km),
$M_{bkg}^{i,k}$ is the response matrix for a background $k$
(mis-identification of CC interactions from other neutrino species or
from NC) to the appearance channel $i$, $N_{bkg}^{i,j,k}$ is the 100\%
expectation spectrum for a background $k$ to an appearance signal $i$
at a detector baseline $j$ and these expected values are used to
calculate an observed number of interactions following a Poisson
distribution.

\subsection{Fitting for $\theta_{13}$ and $\delta_{CP}$ simultaneously}
\label{sec:simFit}
Due to the correlation between $\theta_{13}$ and $\delta_{CP}$,
a simultaneous fit is necessary. Defining a grid of $\theta_{13}$ and 
$\delta_{CP}$ values the $\chi^2$ of a fit to
$Data^{i,j}_{sim}$ can be calculated using the function
\begin{eqnarray}
  \label{eq:chiPois}
    \chi^2 = \displaystyle\sum_{j} \left\{ 2\times\displaystyle\sum_{e}^{E_\mu}\right . &\left(A_jx_jN_{+,j}^e(\theta_{13},\delta_{CP}) - n_{+,j}^e + n_{+,j}^e\log\left(\frac{n_{+,j}^e}{A_jx_jN_{+,j}^e(\theta_{13},\delta_{CP})}\right)\right . \nonumber \\
      &\left . + A_jN_{-,j}^e(\theta_{13},\delta_{CP}) - n_{-,j}^e + n_{-,j}^e\log\left(\frac{n_{-,j}^e}{A_jN_{-,j}^e(\theta_{13},\delta_{CP})}\right)\right) \nonumber\\
    &\left . + \displaystyle\frac{(A_j-1)^2}{\sigma_A} + \displaystyle\frac{(x_j-1)^2}{\sigma_x} \right\},
\end{eqnarray}
where $n_{i,j}^e$ is the simulated data ($Data^{i,j}_{sim}$) for an energy
bin $e$, $N_{i,j}^e(\theta_{13},\delta_{CP})$ is the predicted
spectrum for the values of $\theta_{13}$ and $\delta_{CP}$ represented
by the grid point (calculated as in equation~\ref{eq:simDat} but
without a smear) and $j$ represents the baseline as in
equation~\ref{eq:simDat}. The uncertainty in the expected number of
interactions and expected ratio in interactions between neutrinos
and antineutrinos are represented by the additional free parameters
$A_j$ and $x_j$ respectively and their corresponding errors. 

\begin{figure}
  \begin{center}$
    \begin{array}{c}
    \includegraphics[width=10cm]{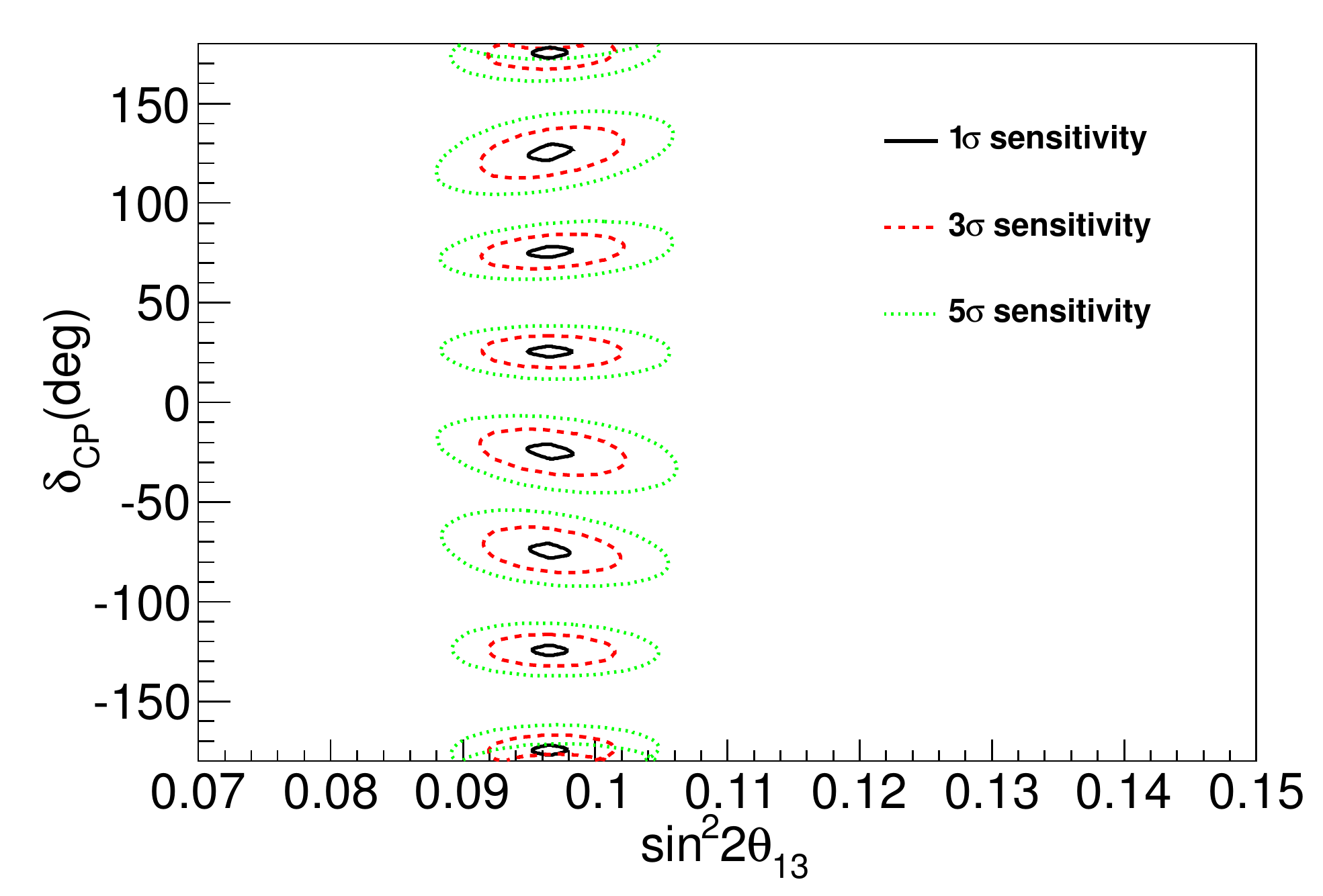} 
    \end{array}$
  \end{center}
  \caption{Examples of 1$\sigma$, 3$\sigma$ and 5$\sigma$ fits to simulated data 
    for the 10 GeV Neutrino Factory, for $\sin^2 2\theta_{13}=0.096$.}
  \label{fig:fits}
\end{figure}

We made two
assumptions regarding the overall event normalisation: one assumes 
a conservative error of  $\sigma_A = 0.025$ and the other assumes a more
optimistic (but realistic) assumption for a neutrino factory of $\sigma_A = 0.01$. 
The uncertainty in the ratio of cross sections between neutrinos
and antineutrinos is maintained fixed at $\sigma_x = 0.01$, which is the
level to which a near detector would seek to measure the interaction
cross-sections at the time of a neutrino factory. 
The minimisation of the parameters $A$ and $x$ is
performed analytically for each predicted dataset to leading
order. The contours at $\chi^2_{min}+9$ represent approximately the
3$\sigma$ level of understanding and those at $\chi^2_{min}+25$ represent 5$\sigma$. 
In such a fit, the experimentally
determined oscillation parameters \cite{Schwetz:2008er} are considered
fixed. While there would be some systematic error associated with
uncertainty in these parameters, systematics from the normalisation
and cross-section uncertainties are expected to dominate. 
Some examples for such fits, assuming $\sin^2 2\theta_{13}=0.096$, are shown in
figure~\ref{fig:fits}.

\subsection{Sensitivity to the mass hierarchy}
\label{sec:Masssens}
Distinguishing between the two possible mass hierarchies is 
important for the understanding of
the neutrino mass sector. The sensitivity to the true mass hierarchy
is defined here as when the true sign of $\Delta m^2_{13}$ can be
distinguished from the opposite sign to the appropriate $n\sigma$ level, that is
\begin{equation}
  \label{eq:Masssens}
  \chi^2_{min}(-\Delta m^2_{13}) - \chi^2_{min}(\Delta m^2_{13}) \geq n^2.
\end{equation}
Sensitivity to the mass hierarchy up to 5$\sigma$ covers
the whole region down to $\sin^2 2\theta_{13} < 10^{-4}$, so the Neutrino Factory would be able to cover easily the current value of  $\sin^2 2\theta_{13} \sim 0.1$. 

\subsection{Sensitivity to $\delta_{CP}$}
\label{sec:deltsens}
\begin{figure}
  \begin{center}$
    \begin{array}{cc}
      \includegraphics[width=0.49\linewidth]{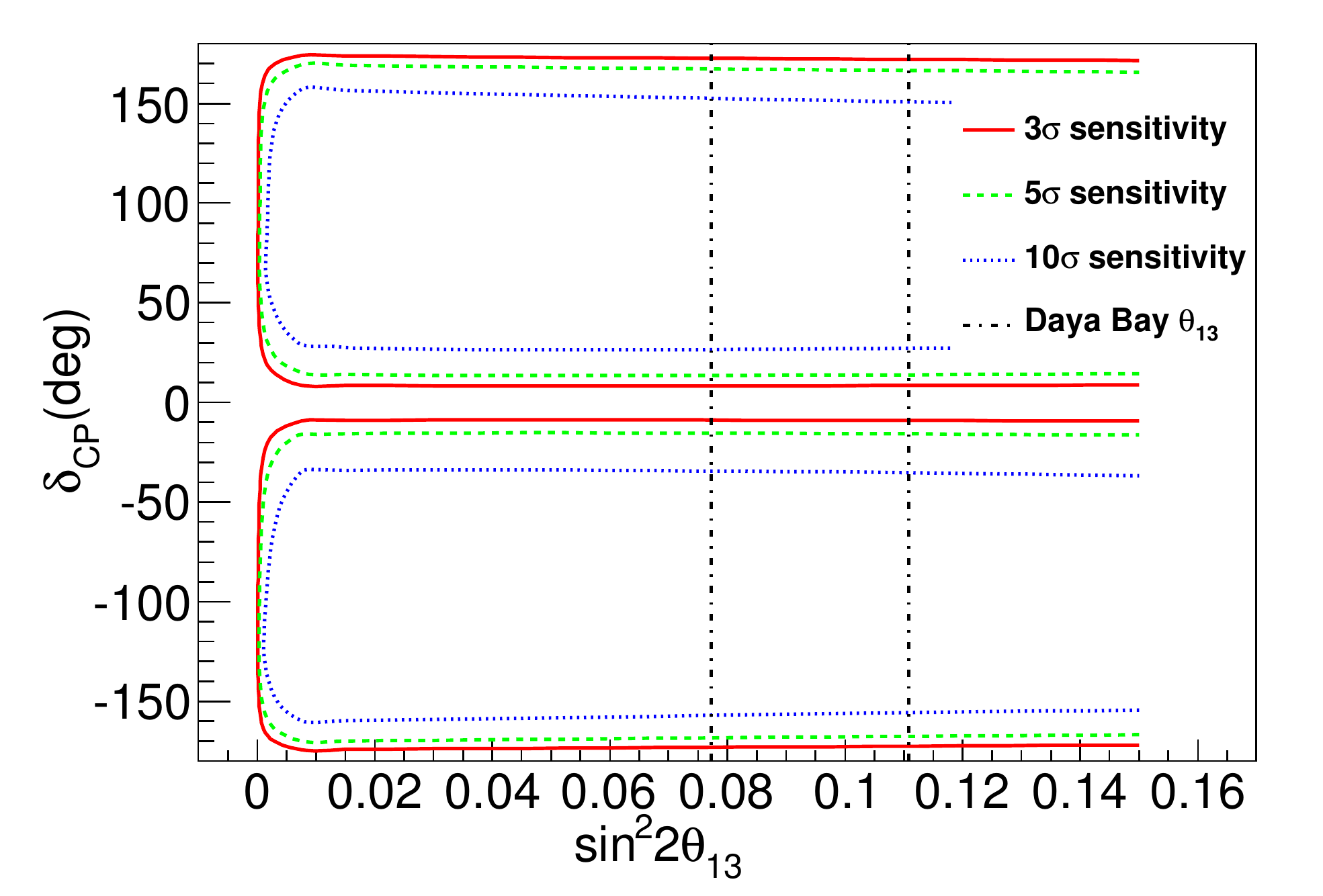} &
      \includegraphics[width=0.49\linewidth]{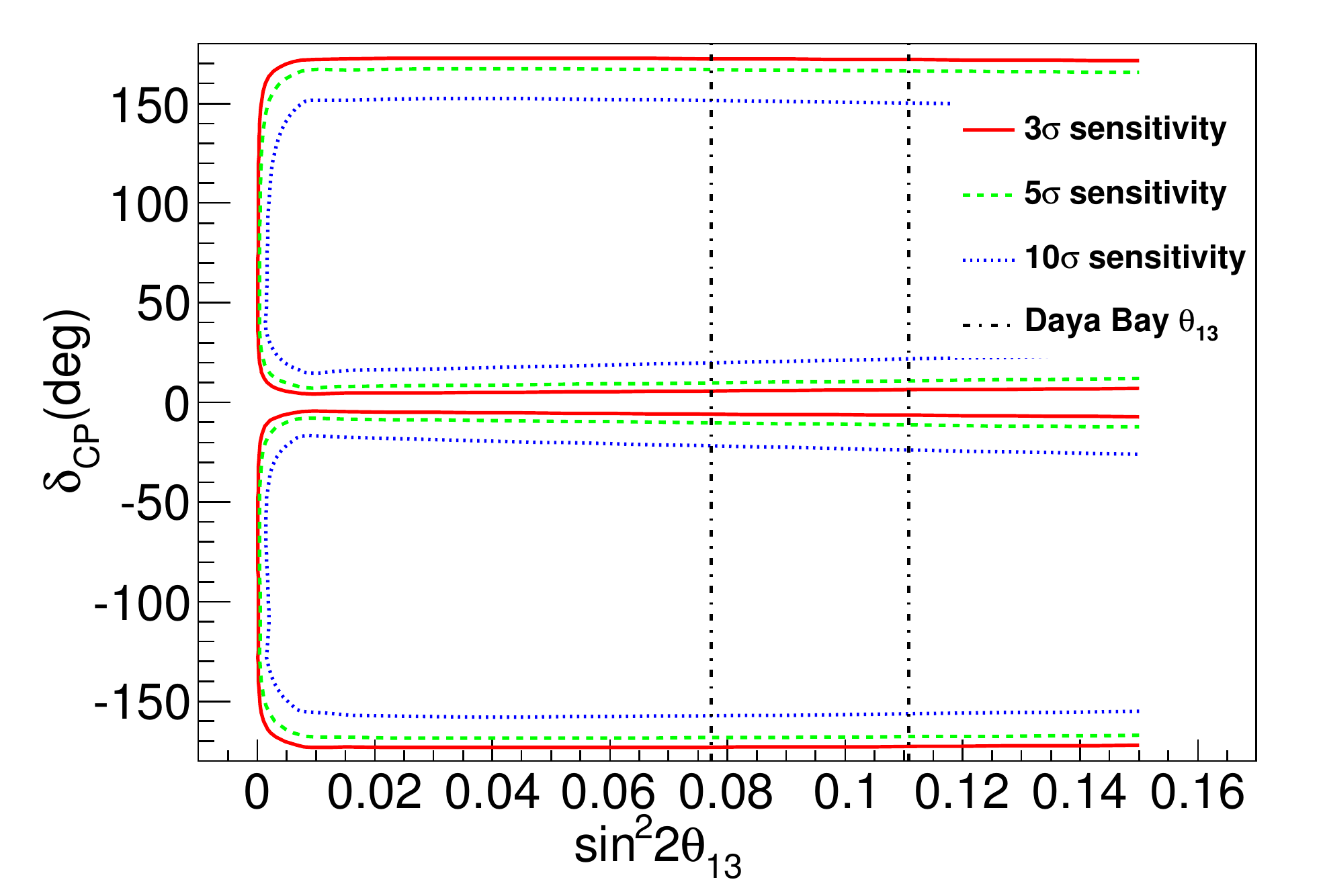}\\
      \includegraphics[width=0.49\linewidth]{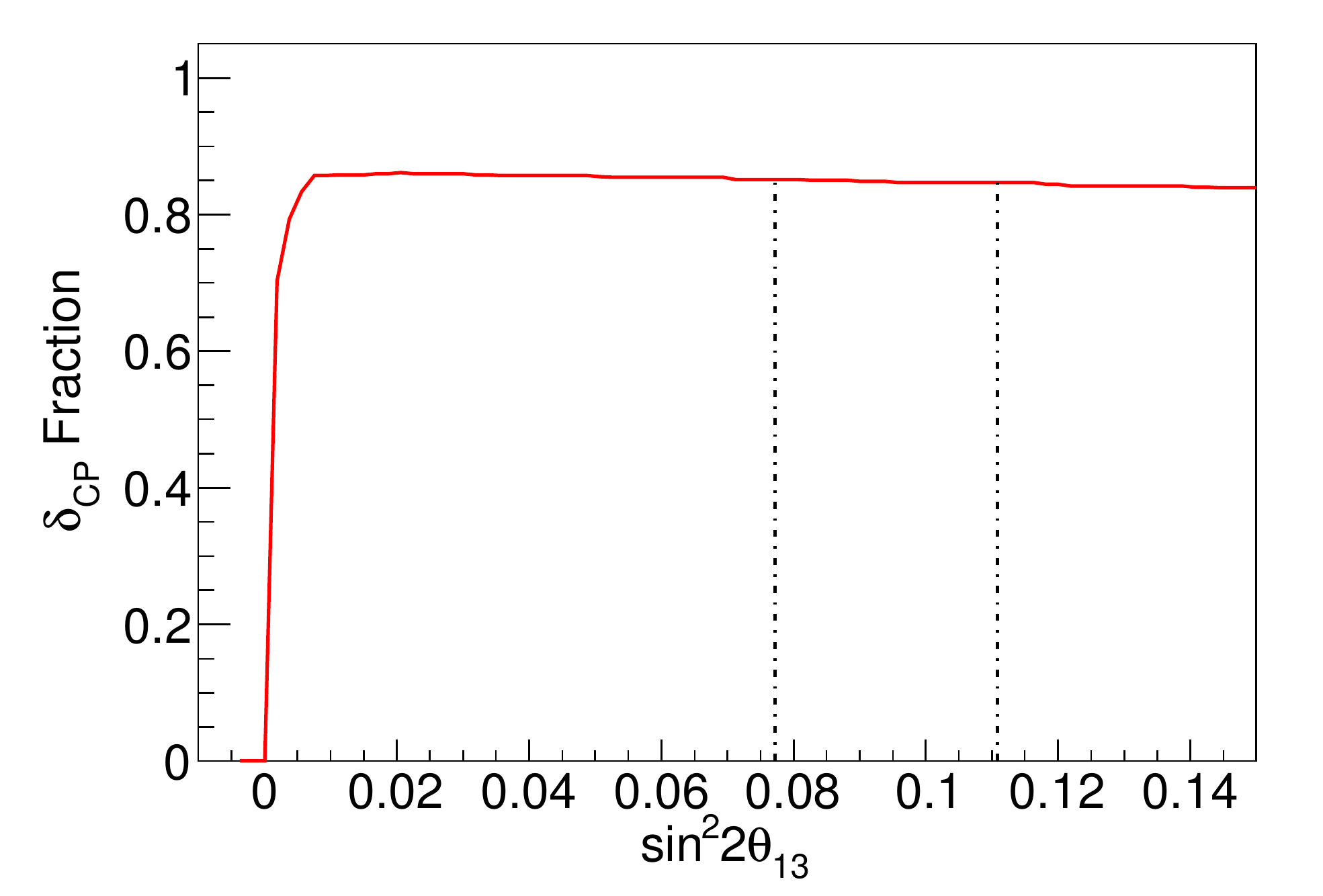} &
      \includegraphics[width=0.49\linewidth]{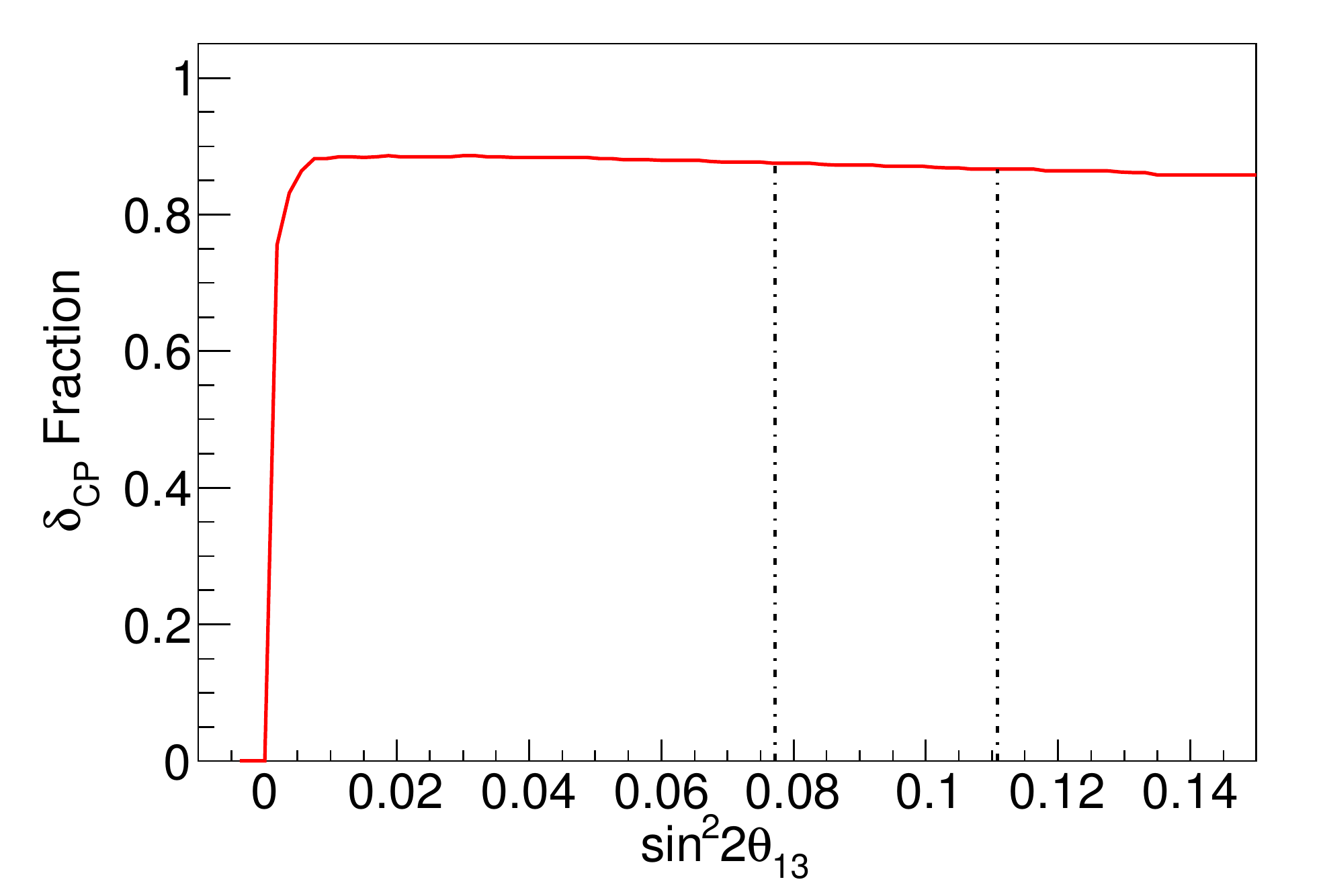}
    \end{array}$
  \end{center}
  \caption{\emph{$\delta_{CP}$ 3$\sigma$, 5$\sigma$ and 10$\sigma$ measurements (top)  and $\delta_{CP}$ 5$\sigma$ coverage to measure CP violation (bottom) as a function of $\sin^2 2\theta_{13}$  for true normal hierarchy (left) and true inverted hierarchy (right). The vertical lines represent the range of possible values of
$\sin^2 2\theta_{13}$. }}
  \label{fig:deltsens}
\end{figure}

Sensitivity to the measurement of $\delta_{CP}$ is defined 
as when the difference between the minimum $\chi^2$ with respect to the minimum
obtained by fitting with \emph{CP} conserving cases is greater than the appropriate $n\sigma$
level, that is
\begin{equation}
  \label{eq:deltsens}
  min(\chi^2(\delta_{CP} = 0), \chi^2(\delta_{CP} = 180), \chi^2(\delta_{CP} = -180)) - \chi^2_{min} \geq n^2.
\end{equation}
The $\delta_{CP}$ measurement and $\delta_{CP}$ coverage plots to measure CP violation by the 10 GeV Neutrino Factory are
shown in figure~\ref{fig:deltsens} for the normal mass hierarchy (left) and for the inverted mass
hierarchy (right).
The CP coverage to measure $\delta_{CP}$ at the 5$\sigma$ level for both normal and inverted mass hierarchy is $\sim$85\% in the range of the currently measured values of $\sin^2 2\theta_{13}$.  These results improve those presented by the Interim Design Report \cite{NF:2011aa}. 

\begin{figure}
  \begin{center}$
    \begin{array}{cc}
      \includegraphics[width=0.49\linewidth]{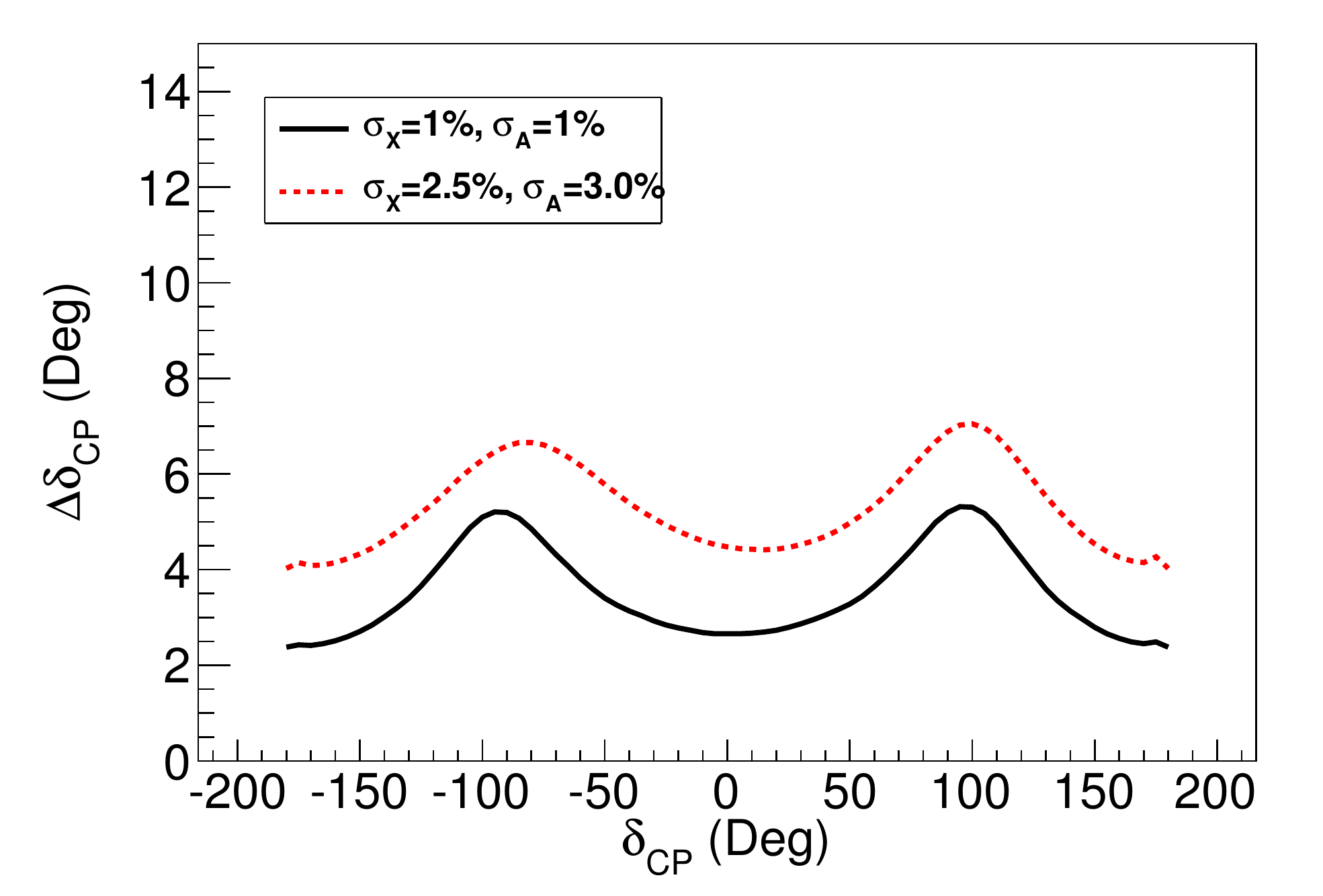} &
      \includegraphics[width=0.49\linewidth]{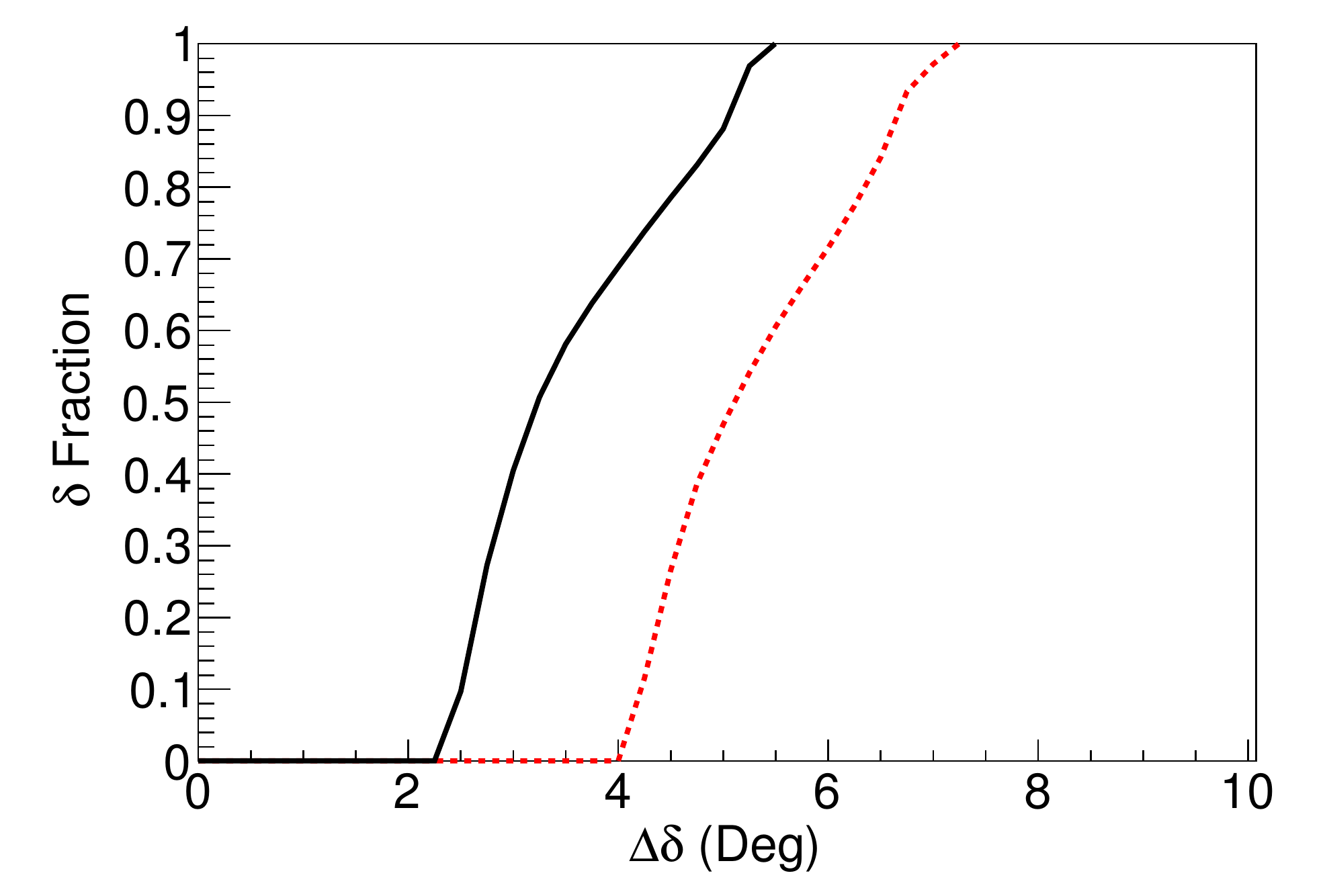}
    \end{array}$
  \end{center}
  \caption{Left: Precision in the value of $\delta_{CP}$ as a function of $\delta_{CP}$ under two
different assumptions: top line (3.0\% normalisation and 2.5\% cross-section error), bottom line (1\% normalisation and cross-section error). Right: $\delta_{CP}$ fraction coverage that can be
achieved above each value of $\Delta \delta_{CP}$ for 3.0\% and 2.5\%  errors (right curve) and for 1\% and 1\% errors (left curve).}
  \label{fig:deltadelta}
\end{figure}

The accuracy achieved in the measurement of $\delta_{CP}$ is an increasingly important
parameter to determine the performance of a facility \cite{Coloma:2012wq}. The $1\sigma$ error $\Delta \delta_{CP}$
for $\theta_{13}=9^\circ$ is shown in figure~\ref{fig:deltadelta} (left), under two assumptions of the overall normalisation and cross-section systematic errors: $(\sigma_A,\sigma_x)=$ (3.0\%,2.5\%) and  (1.0\%,1.0\%). Depending on the value of 
$\delta_{CP}$ and the level of systematic error, the accuracy in $\Delta \delta_{CP}$ is between 2.5\% and 5\%. 
Figure~\ref{fig:deltadelta} (right) shows the $\delta$ fraction coverage that can be
achieved above the value of $\Delta \delta_{CP}$ determined at a Neutrino Factory for each
of the two assumptions about the systematic error.

\section{Conclusions}
\label{sec:conclusions}
A new GEANT4 simulation of the Magnetised Iron Neutrino Detector (MIND), using GENIE as the neutrino event
generator, has been developed to determine the performance of MIND at
a Neutrino Factory. Considering the spectrum of neutrino interactions
in the energy region 0-25~GeV produced by the GENIE generator, this
simulation has been used to study the efficiency and the background
rejection of MIND. 
A detector of 100 ktonne at 2000 km was used to determine the expected
sensitivity to $\delta_{CP}$ and the mass hierarchy at a Neutrino Factory
for a value of $\sin^2\theta_{13}\sim 0.1$,
as determined recently by reactor and accelerator neutrino oscillation experiments. 

The proportions of quasi-elastic, single pion production and deep
inelastic events obtained from GENIE have been benchmarked with
experiments and a parametrisation of the total interaction spectrum,
in agreement with data, was used for the simulation.  Digitisation of
the events tracked through the simulation assumed read-out of the
scintillator using WLS fibres and electronics with 30\% QE and a
standard deviation on signal response of 6\%. Both assumptions should
be achievable and could, perhaps, turn out to be conservative if the
current trend in photon detectors and electronics performance
continues. These events have been used to study the efficiency and
background suppression of MIND in a search for wrong sign muons at a
Neutrino Factory storing both muon polarities. A re-optimised
reconstruction algorithm and a new analysis applied to data-sets
comprising several million $\nu_\mu~(\overline{\nu}_\mu)$ and
$\overline{\nu}_e~(\nu_e)$ CC and NC events, resulted in response
matrices describing the expected response of the detector. MIND showed
an efficiency plateau from 5~GeV in true neutrino energy at $\sim$60\%
for $\nu_\mu$ and $\sim$70\% for $\overline{\nu}_\mu$ and thresholds
at $\sim$2~GeV. All beam inherent backgrounds were simultaneously
suppressed at a level around $10^{-4}$. The difference in
efficiencies for the two polarities has been studied and found to be
predominantly due to the difference in the inelasticity spectrum
expected for neutrino and anti-neutrino interactions. We expect the dominant 
systematic errors to be due to the hadronic energy resolution and due to the 
different neutrino interaction types. Each contribute an expected systematic error 
on the signal efficiency of  around 1\%.

The response of MIND and a parametrisation of the total neutrino and
antineutrino cross-sections, calculated from the output of the GENIE
event generator, were used to study the sensitivity of two MIND
detectors at the Neutrino Factory. The obtained measurement of key parameters
of the PMNS matrix indicate that such an experiment would
determine the mass hierarchy, irrespective of the value of $\delta_{CP}$, at a level
better than 5$\sigma$, could perform measurements of $\delta_{CP}$ with an accuracy 
between 3\% and 5\% and would have an 85\% $\delta_{CP}$ coverage for the currently 
preferred value of  $\theta_{13} \sim 9.0^{\circ}$.

%
\begin{acknowledgments}
The authors acknowledge the support of the European Community under
the European Commission Framework Programme 7 Design Study: EUROnu,
Project Number 212372. 
The work was supported by the Science and Technology Facilities
Council (UK) and by the Spanish Ministry of Education and Science.  
\end{acknowledgments}
%
%
\
\bibliography{PRD_Golden}
%

\pagebreak
\section{Appendix}\label{app:response}

This appendix summarizes the response matrices of signal (wrong sign
$\nu_{\mu}$ and $\bar{\nu}_{\mu}$ appearance) and all
backgrounds. ($\bar{\nu}_{\mu}$ and $\nu_{\mu}$ CC, $\bar{\nu}_{e}$
and $\nu_{e}$ CC, $\bar{\nu}_{\tau}$ and $\nu_{\tau}$ CC, and NC) in
bins of true and reconstructed neutrino energy relevant to an
oscillation analysis. Each entry in the table is the survival
probability for each species. In all tables, columns represent the
true neutrino energy in GeV and rows the reconstructed energy, also in
GeV. The overflow bin in reconstructed energy represents all events
with a reconstructed energy greater than the known maximum.

\begin{table}[h!] 
\subsection{$\nu_{\mu}$ appearance}
\begin{tabular}{|c||c|c|c|c|c|c|c|c|c|c|c|}
\hline
 & 0.0-1.0 & 1.0-2.0 & 2.0-3.0 & 3.0-4.0 & 4.0-5.0 & 5.0-6.0 & 6.0-7.0 & 7.0-8.0 & 8.0-9.0 & 9.0-10.0 \\
 \hline 
0.0-1.0 & 0 & 1.392 & 3.940 & 2.052 & 2.402 & 1.323 & 0.3669 & 0.2460 & 0.3710 & 0.1218 \\
1.0-2.0 & 0 & 222.0 & 243.6 & 45.87 & 28.82 & 21.65 & 9.660 & 6.519 & 1.608 & 2.679 \\
2.0-3.0 & 0 & 149.4 & 1238 & 568.5 & 144.5 & 68.80 & 34.73 & 22.63 & 14.10 & 7.551 \\
3.0-4.0 & 0 & 14.13 & 807.9 & 1790 & 813.2 & 270.8 & 110.4 & 60.52 & 41.31 & 26.79 \\
4.0-5.0 & 0 & 1.392 & 162.0 & 1340 & 1848 & 859.0 & 318.7 & 146.9 & 81.87 & 57.24 \\
5.0-6.0 & 0 & 0.1070 & 14.64 & 375.5 & 1516 & 1695 & 805.7 & 339.7 & 152.2 & 83.54 \\
6.0-7.0 & 0 & 0.1070 & 1.914 & 62.40 & 593.3 & 1630 & 1616 & 841.6 & 369.3 & 178.8 \\
7.0-8.0 & 0 & 0 & 0.3378 & 7.725 & 123.1 & 744.8 & 1563 & 1462 & 783.8 & 347.1 \\
8.0-9.0 & 0 & 0 & 0 & 0.7242 & 18.37 & 200.4 & 836.3 & 1547 & 1360 & 771.6 \\
9.0-10.0 & 0 & 0 & 0.1126 & 0.3621 & 2.402 & 35.24 & 277.6 & 891.3 & 1454 & 1289 \\
10.0-11.0 & 0 & 0 & 0 & 0 & 0.6004 & 8.059 & 83.52 & 474.8 & 1520 & 3014 \\
\hline
\end{tabular}
\caption{Golden channel $\nu_{\mu}$ appearance signal efficiency. All values $\times 10^{-4}$.}
\end{table}

\begin{table}[h!] 
\begin{tabular}{|c||c|c|c|c|c|c|c|c|c|c|c|}
\hline
 & 0.0-1.0 & 1.0-2.0 & 2.0-3.0 & 3.0-4.0 & 4.0-5.0 & 5.0-6.0 & 6.0-7.0 & 7.0-8.0 & 8.0-9.0 & 9.0-10.0 \\
 \hline 
0.0-1.0 & 0 & 0 & 0 & 0 & 0 & 0 & 0 & 0 & 0 & 0 \\
1.0-2.0 & 0 & 0 & 0 & 0 & 0 & 0 & 0 & 0 & 0 & 0 \\
2.0-3.0 & 0 & 0 & 0 & 0 & 0 & 0.1203 & 0.1218 & 0 & 0 & 0.1198 \\
3.0-4.0 & 0 & 0 & 0 & 0 & 0 & 0.2406 & 0 & 0 & 0.1220 & 0.1198 \\
4.0-5.0 & 0 & 0 & 0 & 0 & 0 & 0 & 0.2437 & 0.2439 & 0 & 0.1198 \\
5.0-6.0 & 0 & 0 & 0 & 0 & 0 & 0 & 0 & 0 & 0.1220 & 0 \\
6.0-7.0 & 0 & 0 & 0 & 0 & 0 & 0 & 0 & 0.1220 & 0.1220 & 0.2396 \\
7.0-8.0 & 0 & 0 & 0 & 0 & 0 & 0.1203 & 0 & 0.1220 & 0 & 0.2396 \\
8.0-9.0 & 0 & 0 & 0 & 0 & 0 & 0 & 0 & 0 & 0 & 0.1198 \\
9.0-10.0 & 0 & 0 & 0 & 0 & 0 & 0 & 0.1218 & 0 & 0 & 0 \\
10.0-11.0 & 0 & 0 & 0 & 0 & 0 & 0 & 0 & 0 & 0 & 0.1198 \\
\hline
\end{tabular}
\caption{$\mu^{-}$ background from charge mis-identified $\bar{\nu}_{\mu}$ CC events All values $\times 10^{-4}$.}
\end{table}

\begin{table}[h!] 
\begin{tabular}{|c||c|c|c|c|c|c|c|c|c|c|c|}
\hline
 & 0.0-1.0 & 1.0-2.0 & 2.0-3.0 & 3.0-4.0 & 4.0-5.0 & 5.0-6.0 & 6.0-7.0 & 7.0-8.0 & 8.0-9.0 & 9.0-10.0 \\
 \hline 
0.0-1.0 & 0 & 0 & 0 & 0 & 0 & 0 & 0 & 0 & 0 & 0 \\
1.0-2.0 & 0 & 0 & 0 & 0 & 0 & 0 & 0 & 0 & 0 & 0 \\
2.0-3.0 & 0 & 0 & 0 & 0 & 0 & 0 & 0 & 0 & 0 & 0 \\
3.0-4.0 & 0 & 0 & 0 & 0 & 0.0596 & 0 & 0 & 0 & 0 & 0 \\
4.0-5.0 & 0 & 0 & 0 & 0.0596 & 0 & 0 & 0.0614 & 0 & 0 & 0 \\
5.0-6.0 & 0 & 0 & 0 & 0 & 0 & 0 & 0 & 0 & 0 & 0.0610 \\
6.0-7.0 & 0 & 0 & 0 & 0 & 0 & 0 & 0 & 0 & 0 & 0 \\
7.0-8.0 & 0 & 0 & 0 & 0 & 0 & 0 & 0 & 0 & 0 & 0 \\
8.0-9.0 & 0 & 0 & 0 & 0 & 0 & 0 & 0 & 0 & 0 & 0 \\
9.0-10.0 & 0 & 0 & 0 & 0 & 0 & 0 & 0 & 0 & 0 & 0 \\
10.0-11.0 & 0 & 0 & 0 & 0 & 0 & 0 & 0 & 0 & 0 & 0.0610 \\
\hline
\end{tabular}
\caption{$\mu^{-}$ background from $\nu_{e}$ CC events. All values $\times 10^{-4}$.}
\end{table}

\begin{table}[h!] 
\begin{tabular}{|c||c|c|c|c|c|c|c|c|c|c|c|}
\hline
 & 0.0-1.0 & 1.0-2.0 & 2.0-3.0 & 3.0-4.0 & 4.0-5.0 & 5.0-6.0 & 6.0-7.0 & 7.0-8.0 & 8.0-9.0 & 9.0-10.0 \\
 \hline 
0.0-1.0 & 0 & 0 & 0 & 0 & 0 & 0 & 0 & 0 & 0 & 0 \\
1.0-2.0 & 0 & 0 & 0 & 0.1729 & 0.0581 & 0.3538 & 0.0602 & 0.1221 & 0 & 0.0612 \\
2.0-3.0 & 0 & 0 & 0.1068 & 0 & 0.3488 & 0.2359 & 0.1807 & 0.4272 & 0.3680 & 0.1224 \\
3.0-4.0 & 0 & 0 & 0 & 0.2305 & 0.8138 & 0.8844 & 0.8431 & 0.6713 & 0.6133 & 0.8566 \\
4.0-5.0 & 0 & 0 & 0 & 0.1152 & 0.5232 & 0.7076 & 1.144 & 1.098 & 0.9812 & 0.9178 \\
5.0-6.0 & 0 & 0 & 0 & 0.0576 & 0.0581 & 0.5307 & 0.3011 & 0.5492 & 0.6746 & 0.6119 \\
6.0-7.0 & 0 & 0 & 0 & 0 & 0 & 0.1769 & 0.3011 & 0.4272 & 0.5519 & 0.4895 \\
7.0-8.0 & 0 & 0 & 0 & 0 & 0 & 0 & 0.0602 & 0.4272 & 0.2453 & 0.4283 \\
8.0-9.0 & 0 & 0 & 0 & 0 & 0 & 0 & 0 & 0.0610 & 0.1840 & 0.4283 \\
9.0-10.0 & 0 & 0 & 0 & 0 & 0 & 0 & 0 & 0 & 0 & 0.1836 \\
10.0-11.0 & 0 & 0 & 0 & 0 & 0 & 0 & 0 & 0 & 0 & 0.1836 \\
\hline
\end{tabular}
\caption{$\mu^{-}$ background from $\bar{\nu}_{\mu}$ NC events. All values $\times 10^{-4}$.}
\end{table}

\begin{table}[h!] 
\begin{tabular}{|c||c|c|c|c|c|c|c|c|c|c|c|}
\hline
 & 0.0-1.0 & 1.0-2.0 & 2.0-3.0 & 3.0-4.0 & 4.0-5.0 & 5.0-6.0 & 6.0-7.0 & 7.0-8.0 & 8.0-9.0 & 9.0-10.0 \\
 \hline 
0.0-1.0 & 0 & 0 & 0 & 0 & 0.5833 & 0 & 0 & 0.3747 & 0 & 0.1426 \\
1.0-2.0 & 0 & 0 & 0 & 14.70 & 8.750 & 8.569 & 12.33 & 8.992 & 9.264 & 6.845 \\
2.0-3.0 & 0 & 0 & 0 & 49.97 & 51.33 & 54.27 & 65.98 & 67.44 & 52.33 & 45.35 \\
3.0-4.0 & 0 & 0 & 0 & 38.21 & 51.92 & 58.08 & 71.91 & 76.06 & 78.66 & 79.71 \\
4.0-5.0 & 0 & 0 & 0 & 11.76 & 18.08 & 38.08 & 47.49 & 57.32 & 61.27 & 70.16 \\
5.0-6.0 & 0 & 0 & 0 & 0 & 5.833 & 16.19 & 28.77 & 32.41 & 33.32 & 42.78 \\
6.0-7.0 & 0 & 0 & 0 & 0 & 0 & 3.808 & 12.33 & 17.23 & 18.69 & 25.67 \\
7.0-8.0 & 0 & 0 & 0 & 0 & 0 & 1.587 & 4.109 & 10.30 & 10.89 & 13.69 \\
8.0-9.0 & 0 & 0 & 0 & 0 & 0 & 0 & 1.142 & 2.810 & 5.526 & 7.986 \\
9.0-10.0 & 0 & 0 & 0 & 0 & 0 & 0 & 0.2283 & 0.3747 & 2.113 & 4.278 \\
10.0-11.0 & 0 & 0 & 0 & 0 & 0 & 0 & 0 & 0 & 1.788 & 2.139 \\
\hline
\end{tabular}
\caption{$\mu^{-}$ reconstructed from $\nu_{\tau}$ CC events. All values $\times 10^{-4}$.}
\end{table}

\begin{table}[h!] 
\begin{tabular}{|c||c|c|c|c|c|c|c|c|c|c|c|}
\hline
 & 0.0-1.0 & 1.0-2.0 & 2.0-3.0 & 3.0-4.0 & 4.0-5.0 & 5.0-6.0 & 6.0-7.0 & 7.0-8.0 & 8.0-9.0 & 9.0-10.0 \\
 \hline 
0.0-1.0 & 0 & 0 & 0 & 0 & 0 & 0 & 0 & 0 & 0 & 0 \\
1.0-2.0 & 0 & 0 & 0 & 0 & 0 & 0 & 0 & 0 & 0 & 0 \\
2.0-3.0 & 0 & 0 & 0 & 0 & 0 & 0 & 0 & 0 & 0 & 0 \\
3.0-4.0 & 0 & 0 & 0 & 0 & 0 & 0 & 0 & 0.1637 & 0.4287 & 0 \\
4.0-5.0 & 0 & 0 & 0 & 0 & 0 & 0 & 0 & 0.1637 & 0.2858 & 0 \\
5.0-6.0 & 0 & 0 & 0 & 0 & 0 & 0 & 0 & 0 & 0 & 0 \\
6.0-7.0 & 0 & 0 & 0 & 0 & 0 & 0 & 0 & 0 & 0 & 0 \\
7.0-8.0 & 0 & 0 & 0 & 0 & 0 & 0 & 0 & 0 & 0 & 0 \\
8.0-9.0 & 0 & 0 & 0 & 0 & 0 & 0 & 0 & 0 & 0 & 0 \\
9.0-10.0 & 0 & 0 & 0 & 0 & 0 & 0 & 0 & 0 & 0 & 0 \\
10.0-11.0 & 0 & 0 & 0 & 0 & 0 & 0 & 0 & 0 & 0 & 0 \\
\hline
\end{tabular}
\caption{$\mu^{-}$ background from $\bar{\nu}_{\tau}$ CC events. All values $\times 10^{-4}$.}
\end{table}

\begin{table}[h!] 
\subsection{$\bar{\nu}_{\mu}$ appearance}
\begin{tabular}{|c||c|c|c|c|c|c|c|c|c|c|c|}
\hline
 & 0.0-1.0 & 1.0-2.0 & 2.0-3.0 & 3.0-4.0 & 4.0-5.0 & 5.0-6.0 & 6.0-7.0 & 7.0-8.0 & 8.0-9.0 & 9.0-10.0 \\
 \hline 
0.0-1.0 & 0 & 4.606 & 22.28 & 14.26 & 9.296 & 3.729 & 1.340 & 0.3659 & 0 & 0.7188 \\
1.0-2.0 & 0 & 85.70 & 262.9 & 93.41 & 54.43 & 26.11 & 18.15 & 9.025 & 4.757 & 1.318 \\
2.0-3.0 & 0 & 84.87 & 1565 & 656.4 & 116.1 & 51.85 & 39.48 & 28.05 & 20.61 & 12.94 \\
3.0-4.0 & 0 & 5.304 & 990.1 & 2364 & 795.4 & 164.5 & 63.85 & 49.15 & 46.11 & 38.57 \\
4.0-5.0 & 0 & 0.1396 & 161.9 & 1904 & 2513 & 880.2 & 226.6 & 89.76 & 66.35 & 54.63 \\
5.0-6.0 & 0 & 0 & 10.52 & 477.7 & 2145 & 2218 & 851.9 & 246.6 & 97.94 & 64.45 \\
6.0-7.0 & 0 & 0 & 0 & 60.02 & 801.6 & 2221 & 2032 & 842.4 & 283.0 & 111.5 \\
7.0-8.0 & 0 & 0 & 0.1238 & 3.751 & 152.5 & 1022 & 2181 & 1896 & 840.2 & 307.4 \\
8.0-9.0 & 0 & 0 & 0 & 0.6252 & 19.33 & 263.2 & 1169 & 2108 & 1738 & 832.3 \\
9.0-10.0 & 0 & 0 & 0 & 0.1250 & 1.223 & 47.28 & 355.9 & 1230 & 1996 & 1633 \\
10.0-11.0 & 0 & 0 & 0 & 0.1250 & 0.3670 & 8.301 & 98.33 & 603.4 & 2024 & 4033 \\
\hline
\end{tabular}
\caption{Golden channel $\bar{\nu}_{\mu}$ appearance signal efficiency. All values $\times 10^{-4}$.}
\end{table}

\begin{table}[h!] 
\begin{tabular}{|c||c|c|c|c|c|c|c|c|c|c|c|}
\hline
 & 0.0-1.0 & 1.0-2.0 & 2.0-3.0 & 3.0-4.0 & 4.0-5.0 & 5.0-6.0 & 6.0-7.0 & 7.0-8.0 & 8.0-9.0 & 9.0-10.0 \\
 \hline 
0.0-1.0 & 0 & 0 & 0 & 0 & 0 & 0 & 0 & 0 & 0 & 0 \\
1.0-2.0 & 0 & 0 & 0 & 0 & 0 & 0 & 0 & 0 & 0 & 0 \\
2.0-3.0 & 0 & 0 & 0 & 0 & 0 & 0 & 0 & 0 & 0 & 0 \\
3.0-4.0 & 0 & 0 & 0.3378 & 0.6035 & 0 & 0.2406 & 0.1223 & 0.1230 & 0 & 0 \\
4.0-5.0 & 0 & 0 & 0.2252 & 0.8449 & 0.3602 & 0.2406 & 0 & 0 & 0 & 0.2436 \\
5.0-6.0 & 0 & 0 & 0 & 0.1207 & 0.2402 & 0 & 0.1223 & 0.3690 & 0.1237 & 0.1218 \\
6.0-7.0 & 0 & 0 & 0 & 0 & 0 & 0.1203 & 0.2446 & 0 & 0.1237 & 0 \\
7.0-8.0 & 0 & 0 & 0 & 0 & 0 & 0.2406 & 0.1223 & 0.1230 & 0 & 0 \\
8.0-9.0 & 0 & 0 & 0 & 0.1207 & 0 & 0 & 0 & 0 & 0 & 0 \\
9.0-10.0 & 0 & 0 & 0 & 0 & 0 & 0 & 0 & 0.1230 & 0.1237 & 0.1218 \\
10.0-11.0 & 0 & 0 & 0 & 0 & 0 & 0.1203 & 0.2446 & 0.2460 & 0 & 0.3654 \\
\hline
\end{tabular}
\caption{$\mu^{+}$ background from charge mis-identified $\nu_{\mu}$ CC events All values $\times 10^{-4}$.}
\end{table}

\begin{table}[h!] 
\begin{tabular}{|c||c|c|c|c|c|c|c|c|c|c|c|}
\hline
 & 0.0-1.0 & 1.0-2.0 & 2.0-3.0 & 3.0-4.0 & 4.0-5.0 & 5.0-6.0 & 6.0-7.0 & 7.0-8.0 & 8.0-9.0 & 9.0-10.0 \\
 \hline 
0.0-1.0 & 0 & 0 & 0 & 0 & 0 & 0 & 0 & 0 & 0 & 0 \\
1.0-2.0 & 0 & 0 & 0 & 0 & 0 & 0 & 0 & 0 & 0 & 0 \\
2.0-3.0 & 0 & 0 & 0 & 0 & 0 & 0.0605 & 0.0608 & 0 & 0 & 0 \\
3.0-4.0 & 0 & 0 & 0.0613 & 0 & 0.0609 & 0 & 0 & 0.1221 & 0.1215 & 0.0602 \\
4.0-5.0 & 0 & 0 & 0 & 0.0627 & 0.1218 & 0.1211 & 0.2431 & 0.1832 & 0.1215 & 0.1204 \\
5.0-6.0 & 0 & 0 & 0 & 0 & 0 & 0 & 0 & 0.0611 & 0 & 0 \\
6.0-7.0 & 0 & 0 & 0 & 0 & 0.0609 & 0 & 0 & 0.1221 & 0 & 0 \\
7.0-8.0 & 0 & 0 & 0 & 0 & 0 & 0 & 0 & 0 & 0 & 0 \\
8.0-9.0 & 0 & 0 & 0 & 0 & 0 & 0 & 0 & 0 & 0 & 0.0602 \\
9.0-10.0 & 0 & 0 & 0 & 0 & 0 & 0 & 0 & 0.0611 & 0.0608 & 0.0602 \\
10.0-11.0 & 0 & 0 & 0 & 0 & 0 & 0 & 0 & 0 & 0 & 0 \\
\hline
\end{tabular}
\caption{$\mu^{+}$ background from $\bar{\nu}_{e}$ CC events. All values $\times 10^{-4}$.}
\end{table}

\begin{table}[h!] 
\begin{tabular}{|c||c|c|c|c|c|c|c|c|c|c|c|}
\hline
 & 0.0-1.0 & 1.0-2.0 & 2.0-3.0 & 3.0-4.0 & 4.0-5.0 & 5.0-6.0 & 6.0-7.0 & 7.0-8.0 & 8.0-9.0 & 9.0-10.0 \\
 \hline 
0.0-1.0 & 0 & 0 & 0 & 0 & 0 & 0 & 0 & 0 & 0 & 0 \\
1.0-2.0 & 0 & 0 & 0 & 0 & 0 & 0.0586 & 0 & 0.0607 & 0 & 0 \\
2.0-3.0 & 0 & 0 & 0 & 0.1167 & 0 & 0.1173 & 0 & 0.0607 & 0 & 0.0606 \\
3.0-4.0 & 0 & 0 & 0 & 0 & 0 & 0.1759 & 0.0597 & 0.1213 & 0.1210 & 0.2422 \\
4.0-5.0 & 0 & 0 & 0 & 0 & 0 & 0.0586 & 0.1194 & 0.2427 & 0.1814 & 0.0606 \\
5.0-6.0 & 0 & 0 & 0 & 0 & 0 & 0 & 0.0597 & 0.0607 & 0.1210 & 0.1817 \\
6.0-7.0 & 0 & 0 & 0 & 0 & 0 & 0 & 0 & 0.0607 & 0.1814 & 0.0606 \\
7.0-8.0 & 0 & 0 & 0 & 0 & 0 & 0 & 0 & 0 & 0.0605 & 0 \\
8.0-9.0 & 0 & 0 & 0 & 0 & 0 & 0 & 0 & 0 & 0 & 0 \\
9.0-10.0 & 0 & 0 & 0 & 0 & 0 & 0 & 0 & 0 & 0 & 0.0606 \\
10.0-11.0 & 0 & 0 & 0 & 0 & 0 & 0 & 0 & 0 & 0 & 0 \\
\hline
\end{tabular}
\caption{$\mu^{+}$ background from $\nu_{\mu}$ NC events. All values $\times 10^{-4}$.}
\end{table}

\begin{table}[h!] 
\begin{tabular}{|c||c|c|c|c|c|c|c|c|c|c|c|}
\hline
 & 0.0-1.0 & 1.0-2.0 & 2.0-3.0 & 3.0-4.0 & 4.0-5.0 & 5.0-6.0 & 6.0-7.0 & 7.0-8.0 & 8.0-9.0 & 9.0-10.0 \\
 \hline 
0.0-1.0 & 0 & 0 & 0 & 15.53 & 1.498 & 2.763 & 3.066 & 2.128 & 1.857 & 2.020 \\
1.0-2.0 & 0 & 0 & 0 & 0 & 22.47 & 33.16 & 31.88 & 27.66 & 25.29 & 23.99 \\
2.0-3.0 & 0 & 0 & 0 & 62.11 & 64.41 & 77.99 & 74.19 & 75.28 & 67.44 & 59.34 \\
3.0-4.0 & 0 & 0 & 0 & 54.35 & 63.67 & 68.47 & 66.62 & 61.70 & 73.59 & 69.82 \\
4.0-5.0 & 0 & 0 & 0 & 15.53 & 28.46 & 42.99 & 54.36 & 55.15 & 55.15 & 53.91 \\
5.0-6.0 & 0 & 0 & 0 & 0 & 5.243 & 19.65 & 36.17 & 36.99 & 41.29 & 43.43 \\
6.0-7.0 & 0 & 0 & 0 & 0 & 3.745 & 5.220 & 11.44 & 20.62 & 27.01 & 26.39 \\
7.0-8.0 & 0 & 0 & 0 & 0 & 0 & 0.6141 & 4.292 & 12.27 & 16.72 & 18.94 \\
8.0-9.0 & 0 & 0 & 0 & 0 & 0 & 0 & 1.635 & 2.619 & 5.430 & 10.48 \\
9.0-10.0 & 0 & 0 & 0 & 0 & 0 & 0 & 0.4087 & 1.964 & 3.572 & 5.555 \\
10.0-11.0 & 0 & 0 & 0 & 0 & 0 & 0.3070 & 0.2044 & 0.4910 & 1.286 & 5.176 \\
\hline
\end{tabular}
\caption{$\mu^{+}$ reconstructed from $\bar{\nu}_{\tau}$ CC events. All values $\times 10^{-4}$.}
\end{table}

\begin{table}[h!] 
\begin{tabular}{|c||c|c|c|c|c|c|c|c|c|c|c|}
\hline
 & 0.0-1.0 & 1.0-2.0 & 2.0-3.0 & 3.0-4.0 & 4.0-5.0 & 5.0-6.0 & 6.0-7.0 & 7.0-8.0 & 8.0-9.0 & 9.0-10.0 \\
 \hline 
0.0-1.0 & 0 & 0 & 0 & 0 & 0 & 0 & 0 & 0 & 0 & 0 \\
1.0-2.0 & 0 & 0 & 0 & 0 & 0 & 0 & 0.6849 & 0.1873 & 0.6501 & 0.2852 \\
2.0-3.0 & 0 & 0 & 0 & 0 & 0 & 0.3174 & 0.2283 & 0.9367 & 0.4876 & 0.9982 \\
3.0-4.0 & 0 & 0 & 0 & 0 & 0.5833 & 0 & 0.2283 & 0.3747 & 0.9751 & 1.426 \\
4.0-5.0 & 0 & 0 & 0 & 0 & 0 & 0 & 0 & 0.3747 & 0.4876 & 0.7130 \\
5.0-6.0 & 0 & 0 & 0 & 0 & 0 & 0 & 0 & 0 & 0.3250 & 0.8556 \\
6.0-7.0 & 0 & 0 & 0 & 0 & 0 & 0 & 0 & 0 & 0 & 0.1426 \\
7.0-8.0 & 0 & 0 & 0 & 0 & 0 & 0 & 0 & 0 & 0 & 0 \\
8.0-9.0 & 0 & 0 & 0 & 0 & 0 & 0 & 0 & 0 & 0 & 0 \\
9.0-10.0 & 0 & 0 & 0 & 0 & 0 & 0 & 0 & 0 & 0 & 0 \\
10.0-11.0 & 0 & 0 & 0 & 0 & 0 & 0 & 0 & 0 & 0 & 0 \\
\hline
\end{tabular}
\caption{$\mu^{+}$ background from $\nu_{\tau}$ CC events. All values $\times 10^{-4}$.}
\end{table}

\end{document}